\pdfoutput=1 
\documentclass[
 aps,%
 twocolumn,%
 prb,%
 groupedaddress,%
 superscriptaddress,%
  showpacs,%
 letterpaper,%
 amsfonts,%
 10pt,%
 floatfix,%
 noeprint%
]{revtex4-1}

\RequirePackage{amsmath,amssymb,bm}
\RequirePackage{graphicx}
\RequirePackage[FIGTOPCAP,nooneline]{subfigure}
\RequirePackage{hyperref}
\hypersetup{%
  breaklinks = {true},
  citecolor = {blue},
  colorlinks = {true},
  linkcolor = {red},
  pdfauthor = {\textcopyright\ F\'{a}bio Hip\'{o}lito },
  pdfcreator = {\LaTeX\ and \flqq hyperref\frqq},
  pdffitwindow = {true},
  pdfmenubar = {true},
  pdfpagelayout = {SinglePage},
  pdfstartview = {Fit},
  pdftoolbar = {true},
  plainpages = {false},
}
\usepackage[usenames,dvipsnames]{xcolor}

\newcommand{\ie}{\emph{i.e.\ }}

\newcommand{\br}{\mathbf{r}}
\newcommand{\bE}{\mathbf{E}}
\newcommand{\bk}{\mathbf{k}}

\newcommand{\eps}{\epsilon}
\def\ket#1{|#1\rangle}
\def\bra#1{\langle#1|}
\def\braket#1#2{\langle#1|#2\rangle}

\newcommand{\Fref}[1]{Fig.~\ref{#1}}
\newcommand{\Eqref}[1]{equation~(\ref{#1})}
\newcommand{\bAcal}{\boldsymbol{\mathcal{A}}}
\newcommand{\bardelta}{\bar{\delta}}
\newcommand{\sigmadc}{\sigma^{(2)}_{\textrm{dc}}(\omega)}
\newcommand{\bs}[1]{\boldsymbol{#1}}
\newcommand{\teq}{{\,=\,}}
\newcommand{\tequiv}{{\,\equiv\,}}
\newcommand{\tsim}{{\,\sim\,}}

\DeclareMathOperator{\Tr}{Tr}

\RequirePackage[normalem]{ulem} 

\graphicspath{ {./Figs/} } 

\begin{document}

\title{Nonlinear photocurrents in two-dimensional systems based on graphene 
and boron nitride}

\author{F. Hipolito}
\affiliation{NUS Graduate School for Integrative Sciences and Engineering,
Centre for Life Sciences, Singapore 117456}
\affiliation{%
Centre for Advanced 2D Materials, National University of Singapore,
6 Science Drive 2, Singapore 117546%
}

\author{Thomas G. Pedersen}
\affiliation{Department of Physics and Nanotechnology,
Aalborg University, DK-9220 Aalborg {\O}st, Denmark}
\affiliation{Center for Nanostructured Graphene (CNG), 
DK-9220 Aalborg {\O}st, Denmark}

\author{Vitor M. Pereira}
\thanks{Corresponding author: vpereira@nus.edu.sg}
\affiliation{%
Department of Physics, National University of Singapore,%
2 Science Drive 3, Singapore 117542%
}
\affiliation{%
Centre for Advanced 2D Materials, National University of Singapore,
6 Science Drive 2, Singapore 117546%
}

\date{\today}

\begin{abstract}
The dc photoelectrical currents can be generated purely as a non-linear effect 
in uniform media lacking inversion symmetry without the need for a material 
junction or bias voltages to drive it, in what is termed photogalvanic effect. 
These currents are strongly dependent on the polarization state of the 
radiation, as well as on topological properties of the underlying Fermi surface 
such as its Berry curvature. 
In order to study the intrinsic photogalvanic response of gapped graphene, 
biased bilayer graphene (BBG), and hexagonal boron nitride (hBN), we compute 
the non-linear current using a perturbative expansion of the density matrix. 
This allows a microscopic description of the quadratic response to an 
electromagnetic field in these materials, which we analyze as a function of 
temperature and electron density.
We find that the intrinsic response is robust across these systems and allows 
for currents in the range of pA\,cm/W to nA\,cm/W. At the independent-particle 
level, the response of hBN-based structures is significant only in the 
ultra-violet due to their sizable band-gap. However, when Coulomb interactions 
are accounted for by explicit solution of the Bethe-Salpeter equation, we find 
that the photoconductivity is strongly modified by transitions involving 
exciton levels in the gap region, whose spectral weight dominates in the 
overall frequency range. 
Biased bilayers and gapped monolayers of graphene have a strong 
photoconductivity in the visible and infrared window, allowing for photocurrent 
densities of several nA\,cm/W. We further show that the richer electronic 
dispersion of BBG at low energies and the ability to change its band-gap on 
demand allows a higher tunability of the photocurrent, including not only its 
magnitude but also, and significantly, its polarity.
\end{abstract}

\pacs{78.67.-n,78.67.Wj,81.05.ue,42.65.An}

\maketitle

%
\section{\label{sec:Intro}Introduction}

Non-linear photocurrents consist of dc electric currents induced in a material 
by non-linear interaction with an external electromagnetic field. This 
is frequently termed \emph{photogalvanic}, or \emph{optical rectification} 
effect \cite{Bass1962,Ivchenko2005}.
It belongs to a vast class of processes arising from the non-linear 
response of matter to light fields, such as higher harmonic generation, 
sum/difference frequency generation, optical rectification, parametric 
oscillation, two-photon absorption, or stimulated Raman scattering.
These processes have a broad range of applications, from non-linear optical 
microscopy, optical switching, tunable high-frequency lasers, surface analysis 
with non-linear optics, to parametric down conversion for generation of 
entangled photons \cite{Ghosh1987, Shih1988, Boyd2008, Shen2002} and 
photocurrents \cite{Bass1962, Shen2002}.
The photogalvanic effect is particularly interesting for optoelectronic 
applications as it opens the door for light-induced injection and steering of 
electric or spin currents \cite{Ivchenko2005} without electrical contacts or the 
need of electric fields to separate photo-excited electron-hole pairs, as 
commonly happens in most applications of photo-generated currents.

In metallic structures, the non-linear response to external fields is 
frequently dominated by surface states as a result of the large absorption in 
metals. This is both interesting and convenient in bulk crystals with 
inversion symmetry because elastic even-order interactions with electromagnetic 
fields are suppressed by symmetry in the bulk, but might be possible at the 
surface where the inversion symmetry is broken\cite{Boyd2008, Haussuhl2007}.
 
The wealth of strictly two-dimensional systems that emerged in the wake of the 
success of graphene, with a range of different intrinsic opto-electronic 
properties that continues to grow, provide a particularly fertile setting to 
explore non-linear optical properties. Two of their important characteristics 
in this context are the intrinsically low absorption as light traverses only 
one or a few atomic layers, and the ease to fabricate planar heterostructures 
based on combining different functionalities that are important for 
photoelectric and photogalvanic devices. Moreover, in many of them the 
electronic density and/or gap can be tuned externally by field effect 
\cite{Zhang2009, Castro2007} or electrolytic gating \cite{Das2008}, thereby 
allowing control over two key parameters that determine the optical response.

In this paper, we analyze the characteristics of the second-order electrical 
conductivity tensor with a focus on the intrinsic photoconductivity of 
graphene-based systems and boron-nitride using a perturbative expansion of the 
density matrix. 
We start with a brief analysis of symmetry, followed by an overview of the 
formalism proposed in Ref. \onlinecite{Aversa1995} to compute the relevant 
response functions. 
In the main sections, we analyze in detail the features of the 
photoconductivity in monolayers of ``gapped graphene'', of which the canonical 
example is hBN (hexagonal boron nitride), as well as their bilayer counterparts 
which can display a richer response when the inter-layer bias and chemical 
potential are independently controlled.
The photogalvanic reponse is a second order non-linear effect and, 
therefore, requires broken inversion symmetry in the dipole limit. We 
systematically investigate the influence of stacking sequence on the inversion 
symmetry of various bilayer structures.
It has previously been established that the second order non-linear optical 
response, in the form of second harmonic generation, in bilayer graphene is 
very large \cite{Brun2015}. Moreover, the response is highly tunable by 
external gating or variations in doping level. We find that similar conclusions 
hold for the photogalvanic response.

In the last section we address the role of excitons in the photoconductivity of 
these two-dimensional insulators, and establish that, similarly to the large 
renormalization that they cause in the linear optical conductivity, the 
excitonic response is crucial for an accurate characterization of the quadratic 
response.
The important role of excitons in second harmonic generation from 
two-dimensional materials has previously been demonstrated for MoS$_2$ and 
hBN\cite{Trolle2013,Pedersen2015}. Hence, in the last part of this paper, we 
apply the excitonic intraband formalism developed in Ref. 
\onlinecite{Pedersen2015} to include Coulomb interactions in the non-linear 
photoconductivity of hBN. In this manner, we show that excitons lead to a large 
red-shift of the response as well as a renormalization of the spectrum above 
the band gap.

\section{Symmetry considerations}

The essence of response theory applied to photocurrent generation is that, 
when an electromagnetic wave with electric field 
\begin{equation}
  \bE(t) = \frac{1}{2} \sum_{\omega_n} E_{\omega_n}^\alpha e^{-i\omega_n t}
    \, \hat{\textbf{e}}_\alpha
  \label{eq:E-field}
\end{equation}
impinges on an electronic system, the induced current density can be, quite 
generically, expressed as a sum of contributions proportional to increasing 
powers of the field magnitude,
\begin{equation}
\label{eq:J:Nord}
  J_{\lambda}(t) = \sum_{N=1}^{\infty} \;
    \sum_{\{\omega_k\}} 
    \sigma^{(N)}_{\lambda\alpha\dots\beta}(\omega_1,\dots,\omega_N) 
    \frac{ E_{\omega_1}^{\alpha} \dots E_{\omega_N}^{\beta} }
    { e^{i(\omega_1+\dots+\omega_N)} }
  .
\end{equation}
This defines the $N$-th order frequency-dependent conductivity, a tensor of 
rank $N+1$ that determines the Cartesian component $\lambda$ of the 
current, $J_\lambda(t)$, as a function of time. A perennial question in this 
context is to determine 
$\sigma^{(N)}_{\lambda\alpha\dots\beta}(\omega_1,\dots,\omega_N)$ at a given 
order from the microscopic details of the target system (symmetry, electronic 
structure, interactions, etc.).

The dc photocurrent arising in the intrinsic photogalvanic effect results from 
the existence of a non-zero quadratic response $\sigmadc \equiv 
\sigma^{(2)}_{\lambda\alpha\beta}(\omega,-\omega)$ in systems that fulfill 
certain symmetry criteria.
Foremost is the fact that, being an odd-rank tensor, it can only be sustained 
in the absence of inversion symmetry. In addition to this, the most basic 
threefold symmetry common to all honeycomb-based lattices considerably reduces 
the non-zero independent components of $\sigma^{(2)}$. 
Under the $C_3$ point group symmetry, we have only two independent in-plane 
components, say $ \sigma_{111}^{(2)} $ and $ \sigma_{222}^{(2)} $, as well as 
the constraints $ \sigma_{211}^{(2)} = \sigma_{121}^{(2)} = \sigma_{112}^{(2)} = 
-\sigma_{222}^{(2)} $, $ \sigma_{122}^{(2)} = \sigma_{212}^{(2)} = 
\sigma_{221}^{(2)} = -\sigma_{111}^{(2)} $.
In Table \ref{tab:sym:threefold}, we identify the symmetry and non-vanishing 
components of rank-3 tensors for mono and bilayers of graphene and hBN that 
are appropriate for the target systems in this paper.
It contains the relevant space (SG) and point groups (PG) for the lattices 
under consideration and highlights the presence or absence of inversion 
symmetry in each case \cite{Dresselhaus2008, Haussuhl2007}.

%
\begin{table}[t]
\centering
\begin{tabular}{cccccc}\hline
  material & stacking & SG       & PG & i & $ t_{ijk} \neq 0 $ \\ \hline
 graphene & SL  & $ P6/mmm $   & $6/mmm \equiv D_{6h}$    & yes &none \\ 
          & AB  & $P\bar{3}m1$ & $\bar{3}m \equiv D_{3d}$ & yes & none \\ 
(biased)  & AB  & $ P3m1 $     & $3m \equiv C_{3v} $      & no  & $ t_{222} $ \\
\hline
hBN & SL  & $ P\bar{6}m2 $ & $ \bar{6}m2  \equiv D_{3h}$ & no  & $ t_{222} $ \\ 
    & AA  & $ P\bar{6}m2 $ & $ \bar{6}m2  \equiv D_{3h}$ & no  & $ t_{222} $ \\ 
    & AA' & $ P\bar{3}m1 $ & $ \bar{3}2/m \equiv D_{3d}$ & yes & none \\ 
    & AB  & $ P3m1       $ & $ 3m         \equiv C_{3v}$ & no  & $ t_{222} $ \\ 
    & A'B & $ P\bar{3}m1 $ & $ \bar{3}2/m \equiv D_{3d}$ & yes & none \\ 
    & AB' & $ P\bar{3}m1 $ & $ \bar{3}2/m \equiv D_{3d}$ & yes & none \\ 
\hline
\end{tabular}
\caption{Summary of basic symmetry properties for monolayers and bilayers of 
graphene and hBN. SG: space group, PG: point group, i: inversion. The last 
column indicates the independent, non-vanishing elements of a rank 3 tensor.
The labeling of stacking order in hBN follows Ref. \onlinecite{Ribeiro2011}, 
where AA indicates two pairs of vertically aligned atoms in the unit cell, 
while only one pair is vertically aligned in the AB stacking.}
\label{tab:sym:threefold}
\end{table}

Among the cases selected to analyze in detail here, the restrictions imposed 
by lattice symmetry are most stringent in graphene, where both free standing 
monolayer and AB bilayers have inversion symmetry. However, second order 
response is possible in biased AB graphene bilayers because the potential 
difference between the two layers naturally breaks that symmetry\cite{Brun2015}.
In contrast, due to the presence of two distinct elements in the unit cell, an 
hBN monolayer does not have an inversion center and, consequently, its 
quadratic response can be finite. In addition, the stacking order plays an 
important role in hBN bilayers since a majority of possible stackings results 
in inversion-symmetric lattices (namely, AA', A'B, and AB'). 
In the remainder of this paper we will compute and analyze the intrinsic 
photoconductivity of the systems listed in the table whose response is not 
suppressed by symmetry.

\section{Calculation of non-linear response functions}
\label{sec:IntraResp}
We shall be interested in the interaction of the electronic system in a crystal 
with light, for which we can neglect the position dependence of the
electromagnetic field in a first (dipole) approximation, and write the total 
Hamiltonian of the system as
\begin{equation}
  \hat H = \hat H_0 + \hat{V}(t)
  ,\qquad \hat{V}(t) = e\, \hat{\br}\cdot\bE(t)
  .
  \label{eq:H}
\end{equation}
Here, $\hat{H}_0$ represents the Hamiltonian of the electrons in the 
periodic crystal and $\mathbf{E}(t)$ the explicitly time-dependent 
external field and $e > 0$ is the elementary charge. The field is taken to be 
monochromatic and parametrized as in Eq. \ref{eq:E-field}.
The eigenvalues of the unperturbed Hamiltonian $\hat{H}_0$ define the band 
energies $\eps_n(\bk)$, and its eigenstates, $\ket{n\bk}$, are the 
corresponding Bloch waves. In the derivations below, electronic 
interactions are neglected and, hence, collective effects arise simply on 
account of the Fermi-Dirac statistics. In the last section, however, we 
investigate the effect of Coulomb interactions on the excited electronic 
states, i.e. excitons. These effects are studied only for the simple case of 
monolayer hBN, though.

\subsection{Perturbative expansion of the density matrix}

The time evolution of a system governed by the Hamiltonian \eqref{eq:H} is 
entirely determined if one computes the time-dependent density operator
\begin{equation}
  \hat{\rho}(t) \equiv \sum_{mn} \rho_{mn}(t) | m \rangle\langle n | 
  ,
  \label{eq:rho-def}
\end{equation}
that obeys the dynamical (quantum Liouville) equation 
$ i \hbar \, \partial\hat\rho / \partial t = \big[ \hat H, \hat \rho 
\big]$ or, more explicitly, 
\begin{equation}
  \frac{ \partial \rho_{mn} }{ \partial t } = 
  \frac{ \epsilon_{mn} \, \rho_{mn} }{ i\hbar } +
  \sum_l \frac{ V_{ml} \, \rho_{ln} \!-\!\rho_{ml} \, V_{ln} }
  { i\hbar }
  .
  \label{eq:EqMotion}
\end{equation}
Knowledge of $\hat{\rho}(t)$ permits one to readily quantify and characterize 
the electric current density in terms of the polarization and intensity of the 
field $\bE(t)$. The current density is given in terms of the single-particle 
velocity operator,
\begin{equation}
  \hat{\bm{v}} \equiv \frac{i}{\hbar}[\hat{H},\,\hat{\br}]
  \label{eq:v-def}
  ,
\end{equation}
by
\begin{align}
  \mathbf{J}(t) & \equiv
  \Tr\!\big[ \hat{\rho}(t) \, \mathbf{\hat j} \big] =
  -\frac{ g e }{ \Omega } \sum_\mathbf{k} \sum_{mn} \bm{v}_{nm} \, 
\rho_{mn}(t) 
  ,
  \label{eq:jObs}
\end{align}
where $g \teq 2$ accounts for the spin degeneracy and $\Omega$ is the 
$D$-dimensional volume of the system. 

The formal integration of \eqref{eq:EqMotion} is facilitated in the interaction 
picture with respect to the external perturbation, and leads to the 
conventional perturbative expansion of the density operator in powers of the 
perturbation $\hat{V}(t)$ \cite{Sakurai}. In a crystal, however, such 
straightforward expansion cannot be directly used for the perturbation defined 
in \eqref{eq:H} because it involves matrix elements of the position between 
Bloch states such as $\bra{m\bk}\hat{\br}\ket{n\bk'}$, which are notably ill 
defined \cite{Blount1962}. The integration of Eq. \ref{eq:EqMotion} thus 
requires a more careful treatment of these matrix elements. Moreover, the 
potential alternative of effecting a gauge transformation to describe the 
external field in the minimal coupling scheme $\mathbf{p} \to \mathbf{p} + e 
\mathbf{A}$ is plagued by its own difficulties in the non-linear response 
functions it generates, most notably the appearance of non-physical divergences 
in the dc limit of an insulator at zero temperature \cite{Moss1990}. 
To navigate these difficulties, we follow the systematic approach proposed 
in Ref. \onlinecite{Aversa1995} to handle the external field perturbation 
as expressed in \Eqref{eq:H}. In order to provide here a self-contained 
account of our calculations we briefly review the key aspects of that approach.

An important step is to express matrix elements of the position operator 
between Bloch states as \cite{Blount1962}
\begin{equation}
  \langle m\mathbf{k} | \hat{\mathbf{r}} | n \mathbf{k'} \rangle \teq
  i \delta_{mn}\! \boldsymbol{\nabla_\mathbf{k}} \delta_{\mathbf{k,k'} }
  +
  \delta_{\mathbf{k,k'}} \, \boldsymbol{\mathcal{A}}_{mn}(\mathbf{k})
  ,
  \label{eq:r-matrix}
\end{equation}
where $\boldsymbol{\mathcal{A}}_{mn}(\bk)$ is the so-called Berry connection
\cite{Berry1984,Xiao2010,Aversa1995} and 
\begin{equation}
\delta_{\mathbf{k,k'}} \equiv
\frac{ \Omega_C }{ \Omega } \sum_\mathbf{R} e^{ i ( \mathbf{k -k}') 
\cdot \mathbf{R} } 
\end{equation}
is the Kronecker delta with $\Omega_C$ representing the volume of the unit
cell. We normalize the Bloch states in the finite volume 
$\Omega$ as
\begin{equation}
  \psi_{n\bk}(\br) \equiv \braket{\br}{n\bk} = 
  \frac{e^{i\bk\cdot\br}}{\sqrt{\Omega}} u_{n\bk}(\br)
  ,
\end{equation}
with $u_{n\bk}(\br)$ a cell-periodic function. The Berry connection then has
the explicit form
\begin{equation}
  \label{eq:BerryConn-def}
  \boldsymbol{\mathcal{A}}_{mn} \equiv
  \frac{ i }{ \Omega_c} \int_{\Omega_C} \!\!\!\mathrm{d}\mathbf{r} 
  \, u_{m\mathbf{k}}^*( \mathbf{r} )
  \boldsymbol{\nabla}_\mathbf{k} u_{n\mathbf{k}}( \mathbf{r} )
  .
\end{equation}
For brevity, we will frequently omit the explicit $\bk$ dependence in 
$\bAcal_{mn}$, $\rho_{mn}$ and other quantities when there is no risk of 
confusion. Equation \eqref{eq:r-matrix} suggests a natural identification of two 
types of matrix elements, interband and intraband, respectively given by
\begin{subequations}
  \label{eq:r-inter-intra-def}
  \begin{align}
    \mathbf{r}_{mn}^{(e)} & \equiv
    \langle m\, \mathbf{k} | \mathbf{ \hat r}^{(e)} | n\, \mathbf{k}' \rangle=
    \bardelta_{mn}\ \delta_{\mathbf{k,k'}}
    \boldsymbol{\mathcal{A}}_{mn}
    ,
    \\
    \mathbf{r}_{mn}^{(i)} & \equiv
    \delta_{mn} \Big[ \delta_{\mathbf{k,k'}}
    \boldsymbol{\mathcal{A}}_{mn} 
    +i\boldsymbol{\nabla}_\mathbf{k} \delta_{\mathbf{k,k'}} \Big]
    .
  \end{align}
\end{subequations}
In this expression, $\delta_{mn}$ is the usual Kronecker delta, and 
$\bardelta_{mn}\tequiv 1 - \delta_{mn}$. This allows a decomposition of the 
position operator into a purely interband component and another purely 
intraband, $\hat{\br} = \hat{\br}^{(i)} + \hat{\br}^{(e)}$, with
\begin{align}
  \hat{\br}^{(e)} & \equiv
  \sum_{ \bk \bk' } \sum_{ m\ne n } \br^{(e)}_{mn} \ \ket{m\bk}\bra{n\bk'},
  \\
  \hat{\br}^{(i)} & \equiv
  \sum_{ \bk \bk' } \sum_{ m } \br^{(i)}_{mm}  \ \ket{m\bk}\bra{m\bk'}.
\end{align}
With these definitions, Eq. \ref{eq:EqMotion} can be recast as \cite{Aversa1995}
\begin{align}
  i \hbar \frac{ \partial \rho_{mn} }{ \partial t } & =
  \hbar\omega_{mn} \, \rho_{mn}
  + ie\, \big( \rho_{mn} \big)_{\!;\mathbf{k}} \cdot \mathbf{E}(t)
  \nonumber\\
  & +e \, \sum_l \bigg[
  \bardelta_{ml}\boldsymbol{\mathcal{A}}_{ml} \, \rho_{ln}
  -\rho_{ml} \bardelta_{ln} \boldsymbol{\mathcal{A}}_{ln} \bigg] \!\!
  \cdot \mathbf{E}(t)
  ,
  \label{eq:EqMotion:3}
\end{align}
where $\hbar\omega_{mn}(\bk) \tequiv \eps_m(\bk) - \eps_n(\bk)$ and the second 
term contains the generalized (gauge invariant) gradient \cite{Aversa1995}
\begin{equation}
  \big( \rho_{mn} \big)_{\!;\mathbf{k}} \equiv 
  \boldsymbol{\nabla}_\mathbf{k} 
  \rho_{mn} -i \rho_{mn} \big( \boldsymbol{\mathcal{A}}_{mm} - 
  \boldsymbol{\mathcal{A}}_{nn} \big) 
  .
  \label{eq:GenD}
\end{equation}
The purpose of writing Eq. \ref{eq:EqMotion} as \eqref{eq:EqMotion:3} is that, 
now, all the matrix elements appearing in \eqref{eq:EqMotion:3} are well 
defined and non-singular, which would not be the case if we had generated an 
equivalent expansion in terms of matrix elements $\br_{mn}$ directly from 
\eqref{eq:EqMotion}. 

Equation \eqref{eq:EqMotion:3} can be straightforwardly integrated 
recursively yielding a series in increasing powers of the electric field,
\begin{equation}
  \rho_{mn}(t) = \sum_{N=0}^{\infty} \rho_{mn}^{(N)}(t) 
  .
  \label{eq:rho-series}
\end{equation}
Each term $\rho_{mn}^{(N)}(t)$ is associated with the $(N-1)$-th order 
response function. Since we are ultimately interested in the electrical 
currents induced by the external radiation field, we will compute the 
non-linear conductivity which is the natural response function for this case.
In the absence of the light field, the system is in equilibrium and its 
effective single-particle density matrix reduces to 
$\rho_{mn}(t)\bigr|_{\bE=0} \!\teq \rho_{mn}^{(0)} \teq \delta_{mn} 
f[\eps_m(\bk)]$, a simple Fermi-Dirac distribution. This unperturbed density 
matrix begins the iterative solution of Eq. \ref{eq:EqMotion:3} which is then 
straightforward and we obtain, in first order in the electric field,
\begin{subequations}
\label{eq:rho}
\begin{equation}
  \label{eq:rho:1}
  \frac{ \rho_{mn}^{(1)}( t ) }{ 2\hbar / e } \teq 
  \!\!\sum_{\omega_1} \bigg[
  \bar\delta_{mn} \mathcal{A}_{mn}^\alpha f_{nm}
  \!-\!i \delta_{mn} \frac{ \partial f_n }{ \partial k_\alpha }
  \bigg]
  \frac{ E_{\omega_1} ^\alpha e^{-i ( \omega_1 +i\eta ) t}
  }{ \omega_1 -\omega_{mn} + i\eta }
  ,
\end{equation}
where we introduced $f_{n} \equiv f[\eps_n(\bk)]$, $ f_{nm} \equiv 
f[\eps_n(\bk)] - f[\eps_m(\bk)] $, and the frequency summation $ 
\sum_{\omega_1} $ should be interpreted as including the two possible Fourier 
components, \ie  $ \sum_{\omega_1} g( \omega_1) \equiv g( -\omega_1) +g( 
\omega_1) $.
Also, $i\eta$ with $\eta$ a positive infinitesimal is added to each frequency
to ensure adiabatic turn-on of the field. However, to improve numerical
stability and account for broadening in realistic spectra, we will keep $\eta$
finite but small throughout.
In addition, in \eqref{eq:rho:1} and henceforth, Greek 
superscripts denote Cartesian components and are implicitly summed over when 
they appear repeated. 
The second-order contribution to the density matrix reads as
\begin{align}
  \label{eq:rho:2}
  \rho _{mn} ^{ (2) } &( t ) =
  \frac{ e^2 }{ 4\hbar^2 } \sum_{\omega_2,\omega_1} \sum_{l}
  \frac{ E_{\omega_2}^\beta E_{\omega_2}^\alpha 
  e^{ -i( \omega_2 +\omega_1 + 2i\eta ) t }
  }{ \omega_2 +\omega_1 -\omega_{mn} +2i\eta } \times
  %
  \nonumber \\ &
  %
  \Bigg[
  %
  %
  \bar\delta_{lm} \bar\delta_{ln} \bigg( 
  \frac{ \mathcal{A}_{ml}^\beta \mathcal{A}_{ln}^\alpha f_{nl}
  }{ \omega_1 -\omega_{ln} +i\eta }
  -\frac{ f_{lm} \mathcal{A}_{ml}^\alpha \mathcal{A}_{ln}^\beta
  }{ \omega_1 -\omega_{ml} +i\eta }  \bigg)
  %
  \nonumber \\ &
  %
  %
  -\frac{ \delta_{lm} \delta_{mn} }{ \omega_1 +i\eta }
  \frac{ \partial^2 f_n }{ \partial k_\beta \partial k_\alpha }
  %
  %
  -i \, \delta_{lm} \bar\delta_{mn} \frac{\mathcal{A}_{mn} }{ \omega_1 +i\eta }
  \frac{ \partial f_n }{ \partial k_\alpha }
  %
  \nonumber \\ &
  %
  %
  %
  -i\, \delta_{lm} \bar\delta_{mn} \bigg( 
  \frac{ \mathcal{A}_{mn}^\alpha f_{nm} }{ \omega_1 -\omega_{mn} +i\eta }
  \bigg)_{;k_\beta}
  \Bigg] ,
\end{align}
\end{subequations}
where the generalized derivative introduced in Eq. \ref{eq:GenD} appears 
explicitly in the last term.

\subsection{Linear and quadratic response functions}
\label{sec:IntraResp:JandD}

The current response at any desired order is obtained by substituting the 
perturbative expansion \eqref{eq:rho-series} in \Eqref{eq:jObs},
\begin{equation}
  \mathbf{J}(t) = \sum_{N=0}^{\infty} \mathbf{\hat J}^{(N)}(t) 
  ,
\end{equation}
where
\begin{equation}
  \mathbf{\hat J}^{(N)}(t) \equiv -\frac{ge}{\Omega} \sum_\mathbf{k} 
  \sum_{mn} \bm{v}_{nm} \, \rho_{mn}^{(N)}(t)
  .
\end{equation}
According to the earlier definition in \Eqref{eq:J:Nord}, 
\begin{equation}
  J_{\lambda}^{(N)}(t) = \sum_{\{\omega_k\}} 
    \sigma^{(N)}_{\lambda\alpha\dots\beta}(\omega_1,\dots,\omega_N)
    \frac{ E_{\omega_1}^{\alpha} \dots E_{\omega_N}^{\beta} }
    { e^{-i(\omega_1+\dots+\omega_N)t} }
  ,
  \label{eq:J-N-harmonic}
\end{equation}
which defines the $N$-th order optical conductivity 
$\sigma^{(N)}_{\lambda\alpha\dots\beta}(\omega_1,\dots,\omega_N)$. In this 
expression, $\Sigma_{\{\omega_k\}}$ means a summation that runs over all 
$\omega_1,\dots,\omega_N$, and each $\omega_k$ takes all the values that define 
the harmonic content of the external field. In the case of a single source of 
monochromatic light as in \eqref{eq:E-field}, we have simply 
$\omega_k=\pm\omega$. Equation \eqref{eq:J-N-harmonic} transparently shows 
that, in quadratic and higher orders, the time dependence of the response is 
richer than that of the external field, with the characteristic appearance of 
up to $N$ higher order harmonics of the input frequency. In particular, in 
second order we see that it is possible to induce a dc contribution (constant 
in time) to $J_{\lambda}^{(N)}(t)$ that is determined by 
$\sigma^{(2)}_{\lambda\alpha\beta}(\omega,-\omega)$. \emph{This particular 
response function --- the photoconductivity --- is our main focus in this 
paper}.

\begin{figure}[t]
\centering
\includegraphics[width=\columnwidth]{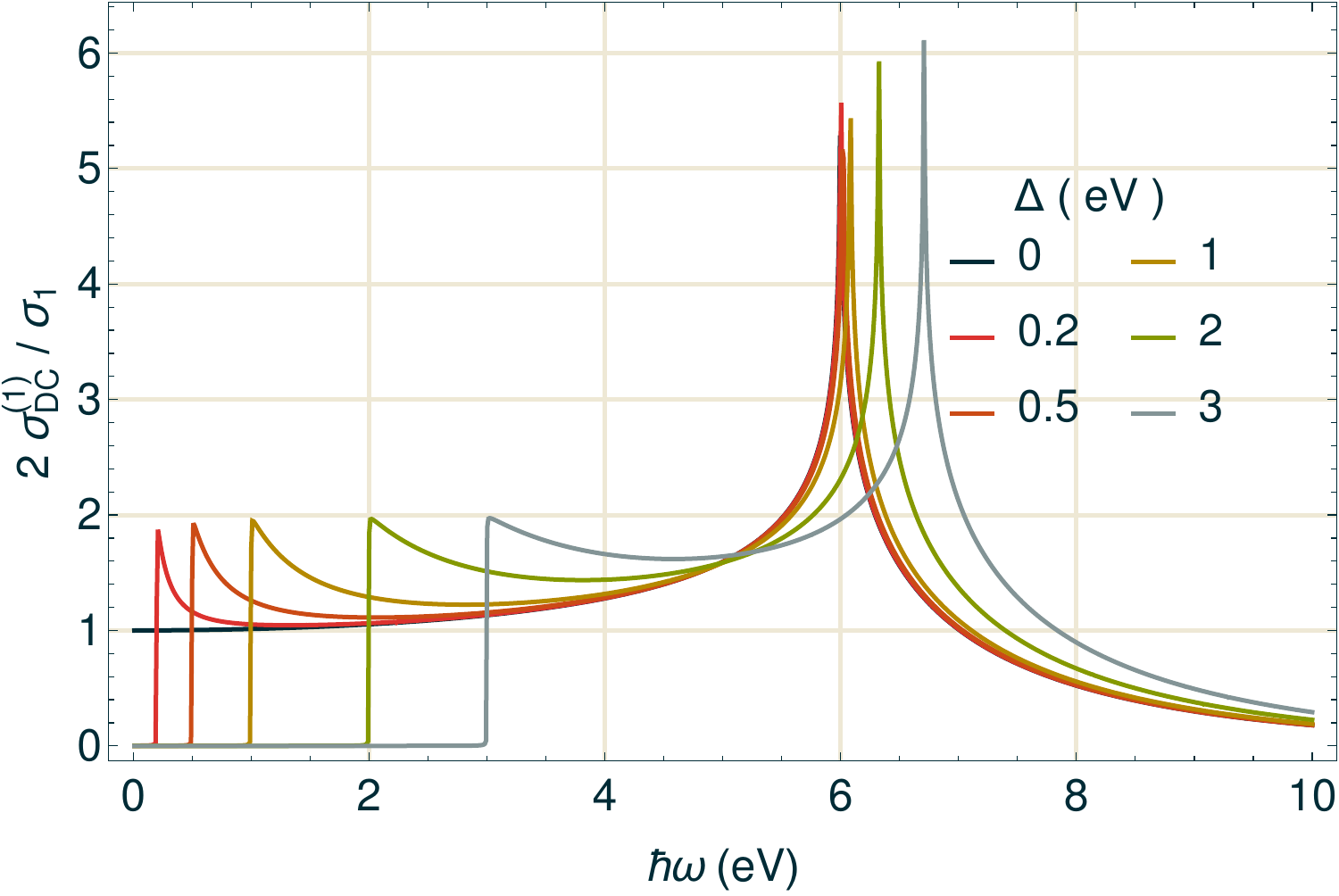}
\caption{
Diagonal components of $\text{Re}\,\sigma^{(1)}_{\alpha\beta}(\omega)$ in 
gapped graphene.
The factor of 2 in the vertical axis stems from the definitions 
\eqref{eq:E-field} and \eqref{eq:J-N-harmonic}, and $\sigma_1\equiv e^2/4\hbar$.
We used the TB model (\ref{eq:monolayer:Hamil}) $ \big[ \gamma_0 = 
3 \, \mathrm{eV}, \, \Gamma = 0.5 \,\mathrm{meV}, \, \mu = 0 \, 
\mathrm{eV} \big]$. 
}
  \label{fig:graph:OpCond}
\end{figure}

As a preliminary illustration, one obtains the linear optical conductivity by 
directly combining the results in Eq. \ref{eq:rho} with the definitions above:
\begin{align}
\label{eq:sigma1}
\sigma_{\lambda\alpha}^{(1)}(\omega) = &
\frac{ 2 i g \hbar^2 \, \sigma_1 }{ \Omega } \sum_\mathbf{k}
\sum_{mn} \sum_{\omega} \frac{ v_{nm}^\lambda }{ \hbar\omega -\epsilon_{mn} 
+i\Gamma }
\nonumber \\ &
\times\bigg[ -\frac{ \delta_{mn} }{ \hbar } \frac{ \partial f_n }{ \partial 
k_\alpha }
+\bar\delta_{mn} \frac{ v_{mn}^\alpha f_{nm} }{ \epsilon_{mn} } 
\bigg]
,
\end{align}
with $ \sigma_1 \tequiv e^2 / 4\hbar $, $ \omega_{mn} \tequiv 
\epsilon_{mn}/\hbar $, and $ \eta \tequiv \Gamma / \hbar $.
We have written $ \sigma_{\lambda\alpha} ^{(1)} ( \omega ) $ explicitly in 
terms of the matrix elements of the velocity operator $ v_{mn} $ using the 
relation $ \mathcal{A}_{ mn }^\alpha \teq i\hbar v_{ mn }^\alpha /\epsilon_{ mn 
}$ that follows from Eqs.~(\ref{eq:v-def} and \ref{eq:r-inter-intra-def}) when 
$m\neq n$.
Note that, in this form, the two terms that make up the result above directly, 
and individually, reflect the interband ($\propto \delta_{mn}$) and intraband 
($\propto\bardelta_{mn}$) contributions to the overall optical conductivity. 
For 
example, the Drude component $\propto (\hbar\omega_1 + i\Gamma)^{-1}$ can 
be read from the first term in \eqref{eq:sigma1} due to the constraint $m=n$. 
In 
fact, one can readily see that
\begin{equation}
  \text{Re}\Bigl[\sigma_{\alpha\alpha}^{(1),{\rm intra}}(\omega,T=0)\Bigr]
  = D^\alpha(\mu)\, \delta(\omega)
  ,
\end{equation}
where the Drude weight is $D^\alpha(\mu) \tequiv 
2\pi\hbar\sigma_{1}N(\mu)\sum_{m} \big\langle\left|v_{mm}^{\alpha} 
(\bk)\right|^{2}\big\rangle_{\mu}$, $N(\mu)$ is the density of states (DOS) at 
the Fermi level, and $\langle\cdot\rangle_\mu$ denotes an average over the 
Fermi surface \cite{Gusynin2006}.
This component is zero in the presence of a band gap for a clean system (as 
happens in most of the situations we consider throughout this paper) as long as 
the chemical potential remains in the gapped region and the temperature is low.
In such cases, that include the peculiar situation of undoped graphene, the 
linear conductivity is uniquely determined by the interband component 
\cite{Peres2006a, Gusynin2006, Stauber2008, Mak2008a}.
Several examples are shown in \Fref{fig:graph:OpCond}.

\begin{figure}
\centering
\subfigure[]{%
\label{fig:lattice:triangular}
\includegraphics[height=0.14\textheight]{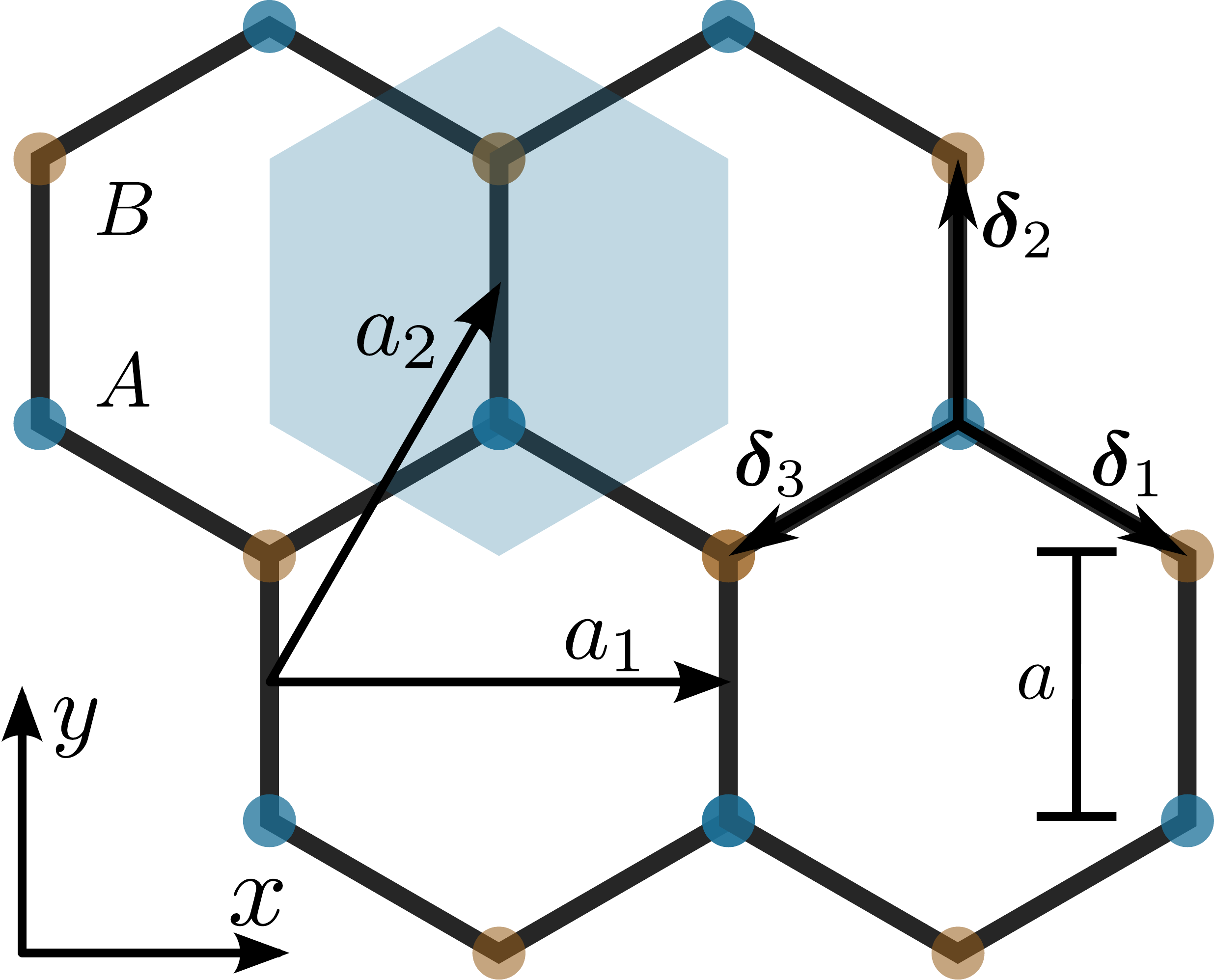}}
\subfigure[]{%
\label{fig:lattice:triangular:reciprocal}
\includegraphics[height=0.14\textheight]{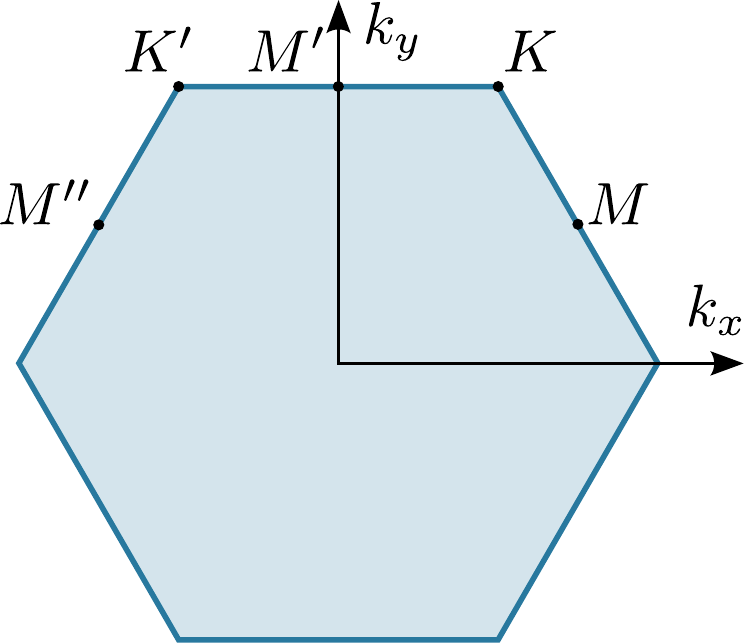}}
\caption%
{The 2D honeycomb lattice and respective first Brillouin zone (BZ).
The real lattice (a) contains two distinct elements (light and dark disks) and 
the respective Wigner-Seitz (WS) cell is represented by the shaded hexagon.
In (b) we draw the conventional representation of the associated BZ.}
\label{fig:lattice}
\end{figure}

The clear distinction between inter and intraband terms in the final 
expressions for the conductivity is a direct result of the earlier 
decomposition of the position matrix elements \eqref{eq:r-inter-intra-def}, and 
propagates to higher orders \cite{Aversa1995}. In particular, the quadratic 
conductivity can be decomposed into four distinct contributions:
\begin{equation}
  \label{eq:sigma2:full}
  \sigma _{\lambda\alpha\beta} ^{(2)} ( \omega_1, \omega_2 ) = 
  \sigma _{\lambda\alpha\beta} ^{(2,ee)} +
  \sigma _{\lambda\alpha\beta} ^{(2,ie)} +
  \sigma _{\lambda\alpha\beta} ^{(2,ei)} +
  \sigma _{\lambda\alpha\beta} ^{(2,ii)}
  ,
\end{equation}
where
\begin{subequations}
\begin{align}
  \sigma _{\lambda\alpha\beta} ^{(2,ee)} ( \omega_1, \omega_2 ) & \equiv
  \sigma_2 g \frac{ \hbar^3 \gamma_0 }{ \Omega a }
  \sum_\mathbf{k} \sum_{lmn}
  \frac{ \bar{\delta}_{lm} \bar{\delta}_{ln} \, v_{nm}^\lambda
  \, / \,  \epsilon_{ml} \epsilon_{ln}
  }{ \hbar(\omega_2 \!+ \!\omega_1) \!- \!\epsilon_{mn} \!+ \!2i\Gamma }
\nonumber \\ &
  \hspace*{-3em}\times\bigg(
  \frac{ v_{ml}^\beta v_{ln}^\alpha f_{nl} }{
  \hbar\omega_1 -\epsilon_{ln} +i\Gamma }
  -\frac{ f_{lm} v_{ml}^\alpha v_{ln}^\beta }{
  \hbar\omega_1 -\epsilon_{ml} +i\Gamma }
  \bigg)
  \label{eq:sigma2:ee}
  ,
\\*[1em]
  \sigma _{\lambda\alpha\beta} ^{(2,ie)} ( \omega_1, \omega_2 ) & \equiv
  \sigma_2 g \frac{ \hbar^2 \gamma_0  }{ \Omega a }
  \sum_\mathbf{k} \sum_{mn}
  \frac{ -\bar{\delta}_{mn} \, v_{nm}^\lambda }{ 
  \hbar(\omega_2 \!+ \!\omega_1) \!- \!\epsilon_{mn} \!+ \!2i\Gamma }
  \nonumber \\ &\times
  \bigg( \frac{ v_{mn}^\alpha f_{nm}/ \epsilon_{mn} 
  }{ \hbar\omega_1 -\epsilon_{mn} +i\Gamma } \bigg)_{;k_\beta}
  \label{eq:sigma2:ie}
  ,
\\*[1em]
  \sigma _{\lambda\alpha\beta} ^{(2,ei)} ( \omega_1, \omega_2 ) & \equiv
  \sigma_2 g \frac{ \hbar^2 \gamma_0  }{ \Omega a }
  \sum_\mathbf{k} \sum_{mn}
  \frac{ -\bar{\delta}_{mn} / \epsilon_{mn} }{ 
  \hbar(\omega_2 \!+ \!\omega_1) \!- \!\epsilon_{mn} \!+ \!2i\Gamma }
  \nonumber \\ &\times
  \frac{ v_{nm}^\lambda v_{mn}^\beta }{ \hbar\omega_1 +i\Gamma }
  \frac{ \partial f_{nm} }{ \partial k_\alpha }
  \label{eq:sigma2:ei}
  ,
\\*[1em]
  \sigma _{\lambda\alpha\beta} ^{(2,ii)} ( \omega_1, \omega_2 ) & \equiv
  \sigma_2 g 
  \frac{ \hbar \gamma_0 }{ \Omega a }
  \frac{ ( \hbar\omega_1 +i\Gamma )^{ -1 } }{ 
  \hbar(\omega_2 \!+ \!\omega_1) \!- \!\epsilon_{mn} \!+ \!2i\Gamma }
  \nonumber \\ & \times
  \sum_\mathbf{k} \sum_{n}
  v_{nn}^\lambda 
  \frac{ \partial^2 f_n }{ \partial k_\beta \partial k_\alpha } 
  \label{eq:sigma2:ii}
  .
\end{align}
\end{subequations}
In these expressions, $ee$ refers to a contribution including 
\emph{interband} matrix elements only, $ii$ to that including purely 
\emph{intraband}, and $ie$, $ei$ to those that include one \emph{inter} and 
one \emph{intraband} matrix element. For later convenience, the constant 
$\sigma_2 \tequiv e^3 a/ 4 \gamma_0 \hbar$ is defined in terms of the in-plane 
nearest-neighbor lattice parameter $a$ and hopping integral $\gamma_0$ (cf. 
Fig. \ref{fig:lattice}).

The results \eqref{eq:sigma2:full} completely describe the quadratic response 
of the system for any combination of the pair of frequencies $\omega_1$ and 
$\omega_2$. Henceforth, we shall be interested only in the specific case of $ 
\sigma_{222}^{(2)}(\omega,-\omega)$ that characterizes the intrinsic 
photoconductivity of the system: how much current density is driven in the 
system for a given intensity and polarization of the incident electromagnetic 
radiation. This response function is associated with the effect known as 
photoconductivity. To ease the notation, we define 
\begin{equation}
  \sigmadc \equiv \sigma_{222}^{(2)}(\omega,-\omega)
  ,
\end{equation}
where the subscript ``dc'' emphasizes that the induced current is constant in 
time. Since there is no risk of ambiguity and, moreover, a system with $C_3$ 
point group symmetry has only one independent tensor component, $\sigmadc$ will 
be used from this point on.

\section{Photoconductivity of monolayer honeycomb lattices}
\label{sec:monolayers}

To actually compute the linear and quadratic conductivities in 
Eq. \ref{eq:sigma1} or \ref{eq:sigma2:full}, we must determine not only the 
electronic energy bands, but also the matrix elements of the velocity, 
$\bm{v}_{mn}$, and Berry connection, $\bAcal_{mn}$, involving any two bands. A 
simple one orbital tight-binding (TB) model provides an accurate, yet simple, 
description of such quantities in graphene and boron nitride (monolayers and 
bilayers). 

Consider the general case of a single layer of a honeycomb lattice where the 
atoms residing in the A and B sublattices can be different, a canonical example 
being a monolayer of hBN. The direct and reciprocal lattices of such a crystal 
are illustrated in \Fref{fig:lattice}. In a single 
orbital, nearest neighbor tight-binding modeling of the relevant electronic 
degrees of freedom, the Hamiltonian operator takes the form
\begin{equation}
  \hat{H} = -\gamma_0 \sum_{\bk} \Psi^\dagger_\bk\, h_\bk\, \Psi_\bk
  \label{eq:Htb-1}
  ,
\end{equation}
where $\Psi^\dagger \tequiv [a^\dagger_\bk\; b^\dagger_\bk]$ comprises the 
Fourier-transformed electron creation operators at sites of the A and B 
sublattices, and $h_\bk$ is the reduced Hamiltonian in the crystal momentum 
representation:
\begin{align}
\label{eq:monolayer:Hamil}
h_\bk \equiv 
\begin{pmatrix}
  -\Delta/2               & \phi( \mathbf{k} ) \\
  \phi^*( \mathbf{k} )  & +\Delta/2
\end{pmatrix}
.
\end{align}
Henceforth, we use units of energy such that $\gamma_0 \teq 1$.
Here, $\Delta$ quantifies the difference in the atomic energy of A- and B-type 
atoms, and $ \phi( \mathbf{k} ) \tequiv e^{i k_y a} +2 e^{-i k_y a/2 } \cos( 
\sqrt{3} k_x a /2 ) $, $a$ being the nearest neighbor distance.
This description yields the simple two-band energy dispersion
\begin{equation}
  \epsilon_{\pm} (\bk) = \pm\sqrt{|\phi(\bk)|^2 + \Delta^2/4}
  \label{eq:Ek-2band}
  .
\end{equation}
In this tight-binding parametrization, the velocity matrix elements and Berry 
connection \eqref{eq:monolayer:Hamil} are simply
\begin{equation}
\hbar \bm{v}_{mn} \!=\! \bra{m\bk} \bm{\nabla}_{\bk}h_\bk \ket{m\bk}  
,\quad
\bAcal_{mn} \!\! = i \bra{m\bk} \bm{\nabla}_{\bk} \ket{n\bk} 
,
\end{equation}
where $\ket{n\bk}$ are the normalized eigenstates of \eqref{eq:monolayer:Hamil}.

\subsection{Hexagonal boron nitride}
\label{sec:IntraResp:BNsl}
A system to which the Hamiltonian above is directly relevant is that of a 
monolayer of boron nitride, whose crystal lattice consists of a honeycomb 
structure where the B and N atoms occupy distinct sublattices.
We follow the parametrization of Ref.~\onlinecite{Pedersen2015} for the hopping 
and gap, namely $ \gamma_0 \teq 2.33 \, \mathrm{eV} $, $ \Delta \teq 7.80 
\,\mathrm{eV}$, and the nearest neighbor distance is $ a \teq 1.45 \times 
10^{-10} \, \mathrm{m}$.

\begin{figure}
\subfigure[][]{%
 \includegraphics[width=\columnwidth]{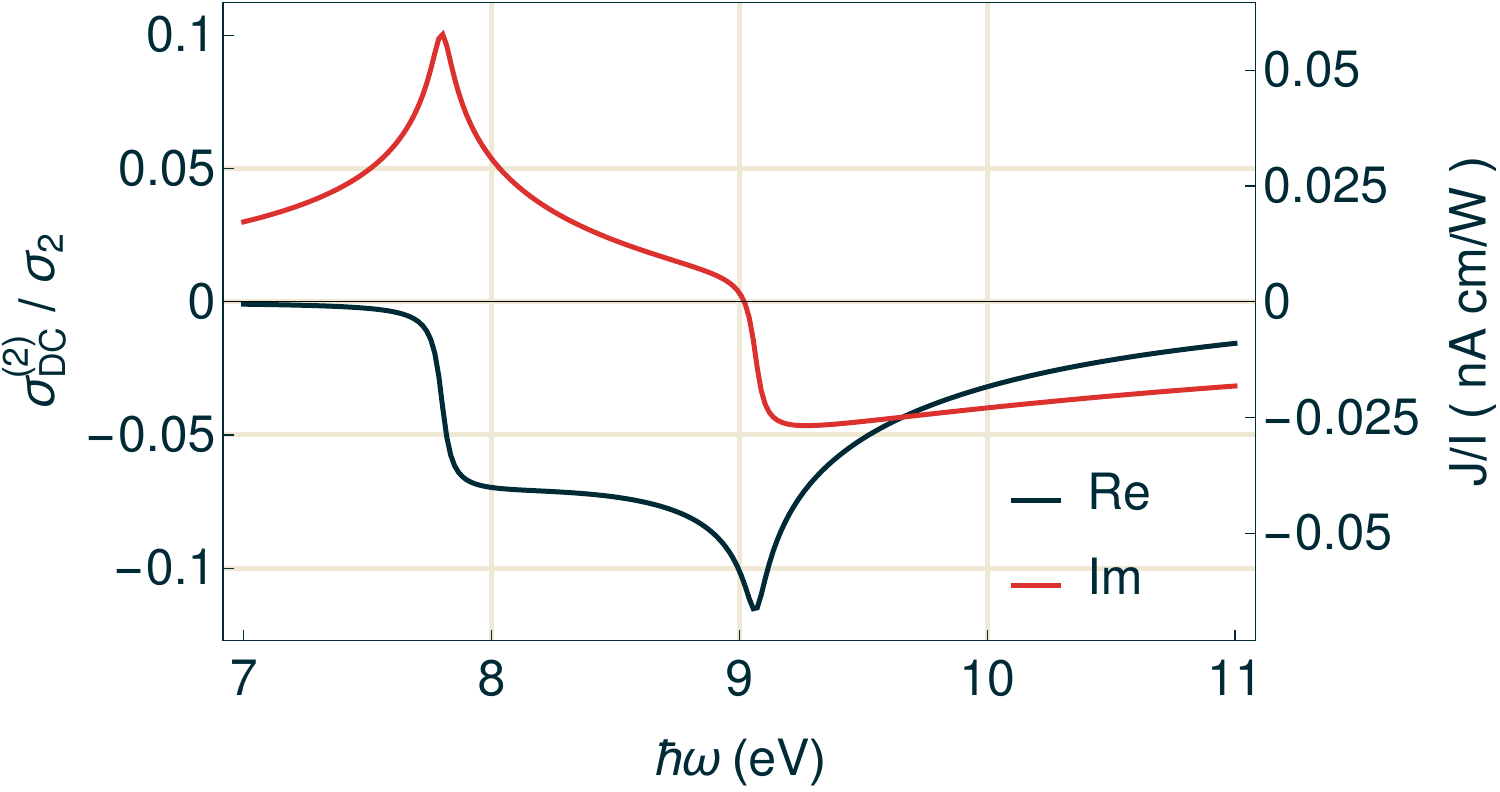}%
 \label{fig:BN:SL:s2}%
}
\subfigure[][]{%
 \includegraphics[width=\columnwidth]{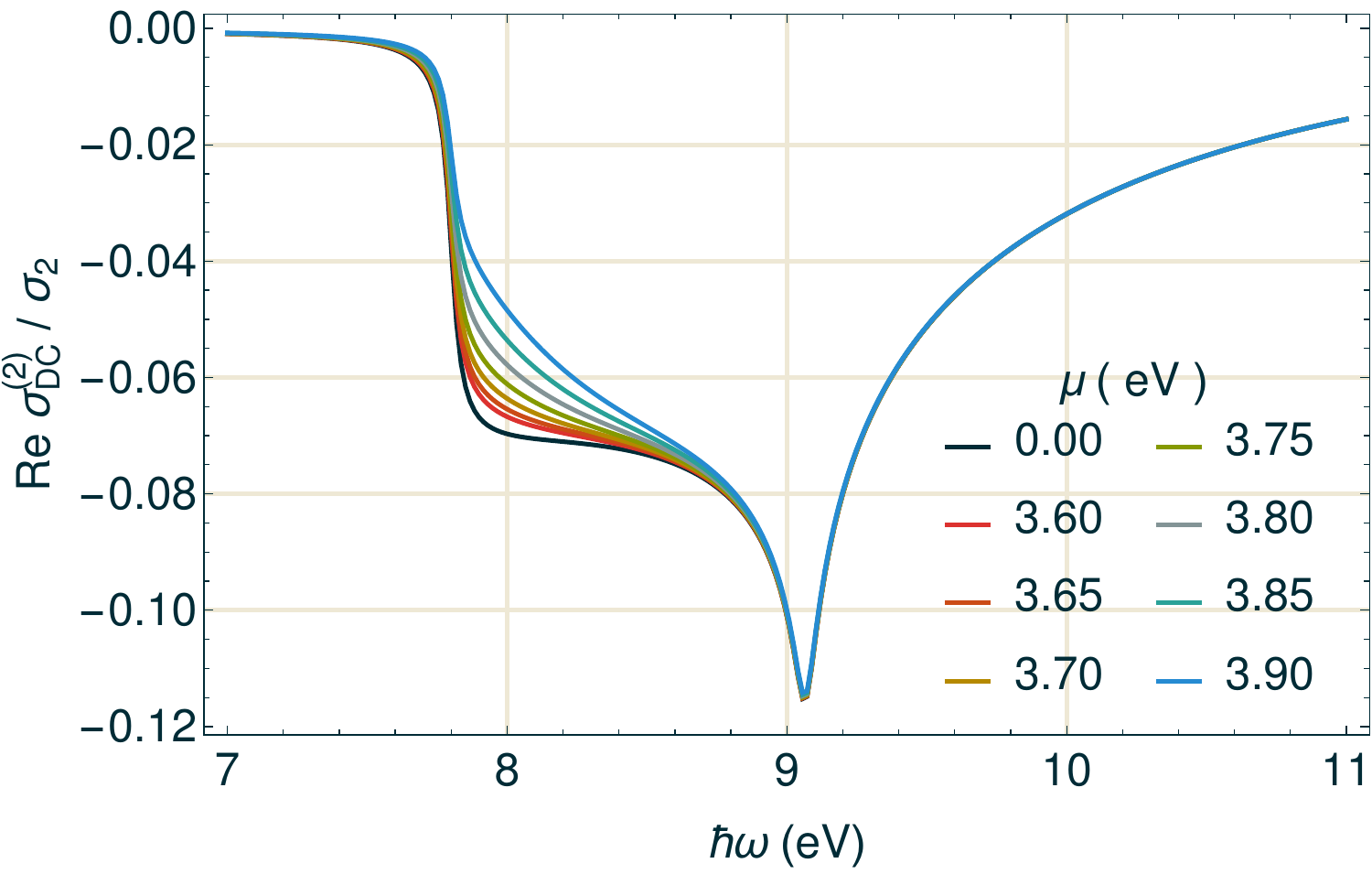}%
 \label{fig:BN:SL:mu}%
}
\caption{%
  The photoconductivity $\sigmadc$ for a hBN monolayer in units of $\sigma_2 
  \teq 3.79\times 10^{-15}$\,S\,m/V. 
  We consider the TB model (\ref{eq:monolayer:Hamil}) 
  $ ( \gamma_0 \teq 2.33 \, \mathrm{eV}$, \,
  $ \Delta \teq 7.8 \, \mathrm{eV}$, \,
  $ \Gamma \teq 0.03   \, \mathrm{eV}$, \,
  $ \mu \teq 0 \, \mathrm{eV} ) $.
  \subref{fig:BN:SL:s2} Real and imaginary parts of $ \sigmadc $ at 
  room-temperature.
  Right scale measures the electric current per laser intensity $( J/I 
  )$\cite{EndNote-1}.
  In \subref{fig:BN:SL:mu}, we analyze the effect of finite doping by showing 
  $ \mathrm{Re} \, \sigmadc $ at different chemical potential and high 
  temperature: $ T \teq 1500 \,\mathrm{K} $.
 }
 \label{fig:BN:SL}
\end{figure}

The symmetry constraints greatly facilitate the calculation of the quadratic 
conductivity because, as only $\sigma_{222}^{(2)}(\omega,-\omega)$ needs to be 
computed explicitly, the last two contributions \eqref{eq:sigma2:ei} and 
\eqref{eq:sigma2:ii} vanish identically as long as time-reversal symmetry is 
preserved.
Hence, quite generically only $\sigma_{222}^{(2,ee)}(\omega,-\omega)$ and 
$\sigma_{222}^{(2,ie)}(\omega,-\omega)$ need to be computed in a system with 
threefold plane rotational symmetry. Moreover, in any two-band model, 
$\sigma_{222}^{(2,ee)}$ is also identically zero for the same reason.

Figure \ref{fig:BN:SL}(a) shows the real and imaginary parts of 
$\sigmadc$ for a hBN monolayer computed directly from 
the results in eqs.~\eqref{eq:sigma2:full}. 
As anticipated from the nature of the frequency denominators in those 
expressions, the system mainly responds for photon energies at, or above, the 
band gap. The peak at $ \approx 9$\,eV is associated with virtual transitions 
between the van Hove singularities at the $M$ point in the Brillouin zone.

Unlike other two-dimensional (2D) crystals such as graphene or transition-metal 
dichalcogenides, the large gap in hBN makes it impossible to change the Fermi 
level by electrostatic gating. On the other hand, a small amount of impurities 
might introduce a shallow donor (acceptor) band and allow the chemical 
potential to be driven close to the edge of the conduction (valence) band. This 
scenario is explored in Fig. \ref{fig:BN:SL}(b), where we show the effect of 
varying the chemical potential in the vicinity of the band edge at high 
temperature.
High temperature is chosen here because the interaction with very intense laser 
light, as required to observe non-linear effects, generates hot carriers in the 
material. Experiments \cite{Sun2008, Ruzicka2010, Ruzicka2010a, Lui2010, 
Sun2012, Tielrooij2013} in graphene indicate hot carrier temperatures in the 
range 1000 to 3600 K.  

\subsection{Gapped graphene}
\label{sec:IntraResp:graphSL}

The case of hBN can be seen as an extreme limit of ``gapped graphene'' in 
the framework of the effective two-band tight-binding model introduced in 
\eqref{eq:monolayer:Hamil}. We use the designation ``gapped graphene'' to 
describe a nearest-neighbor tight-binding model like that of graphene, but where 
the sublattice symmetry is explicitly broken by introducing a potential energy 
that differs by an amount $\Delta$ between the two sublattices [see 
\eqref{eq:monolayer:Hamil}].
Second order non-linearities are not expected in pristine graphene, or any odd
numbered Bernal stacked multi-layers due to the presence of an inversion center 
\cite{Jorio2011a}. Breaking the sublattice symmetry in a monolayer, in addition 
to opening a band-gap, lifts this restriction.
It is of general interest to describe and understand the behavior of the 
photoconductivity as a function of gap magnitude in such a system: 
on the one hand, such sublattice symmetry breaking has been predicted to take 
place when graphene is grown or transferred to particular substrates 
\cite{Giovannetti2007, Zhou2007, Slawinska2010, Dean2010, Otrix2012, 
ZhiGuo2014, Bokdam2014, Woods2014, Yankowitz2014, Huang2014};
on the other hand, models such as \eqref{eq:monolayer:Hamil} are frequently 
used as minimal descriptions of the low-energy details in many transition metal 
dichalcogenides. We note that inversion symmetry in graphene can also be
broken by rolling the material into chiral nanotubes \cite{Pedersen2009}, for 
which the gapped graphene model may also be applied.

\begin{figure}
\includegraphics[width=1.\columnwidth]{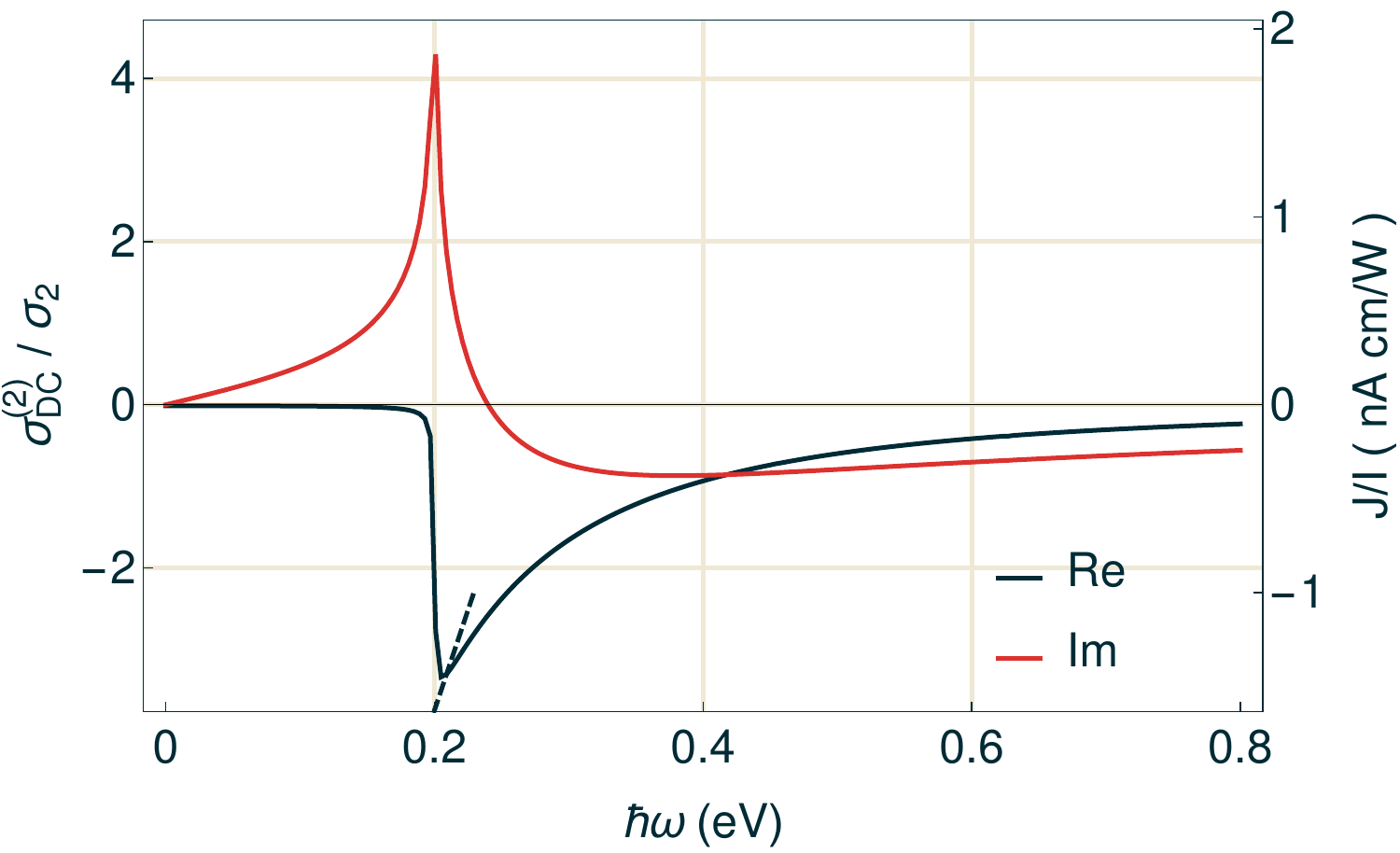}%
\caption{%
  Real and imaginary parts of the non-vanishing components of the 
  photoconductivity tensor $ \sigmadc $, $ -\sigma_{211}^{(2)} $, $ 
  -\sigma_{121}^{(2)} $ and  $ -\sigma_{112}^{(2)} $, whose curves perfectly   
  overlap as expected from symmetry.
  Right scale measures the electric current per laser intensity $( J/I 
  )$\cite{EndNote-1}.
  We consider the TB model \eqref{eq:monolayer:Hamil} 
  $ ( \gamma_0 \teq 3 \, \mathrm{eV}$, \,
  $ \Delta \teq 0.2 \, \mathrm{eV}$, \,
  $ \Gamma \teq 1 \,   \mathrm{meV}$, \,
  $ \mu \teq 0 \, \mathrm{eV}$, \,
  $ T \teq 1 \,\mathrm{K} ) $.
}
\label{fig:graphSLs2}
\end{figure}

\begin{figure}
\subfigure[][]{%
 \includegraphics[width=\columnwidth]{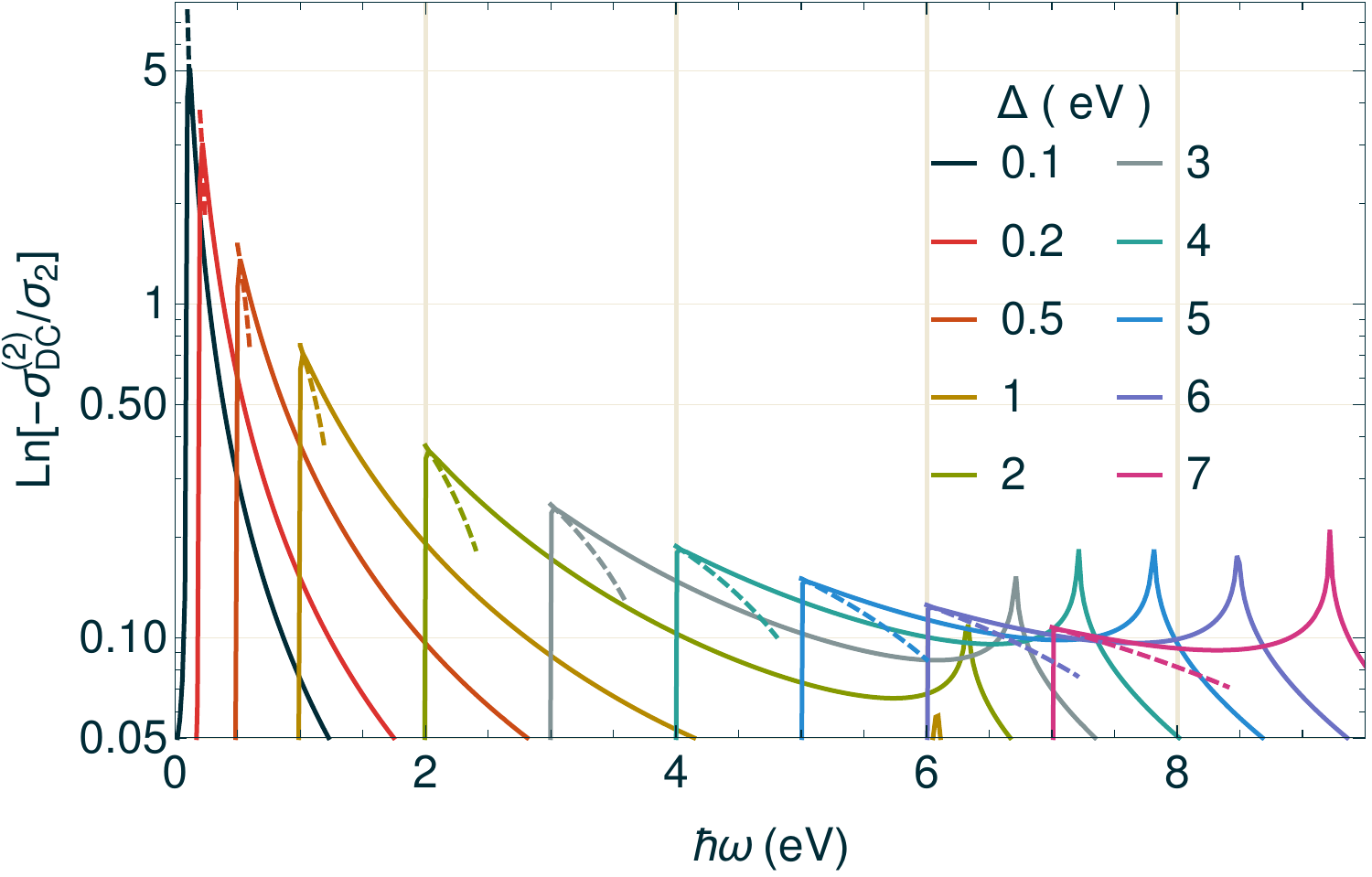}%
 \label{fig:graphSLDelta}%
}
%
%
\subfigure[][]{%
 \includegraphics[width=\columnwidth]{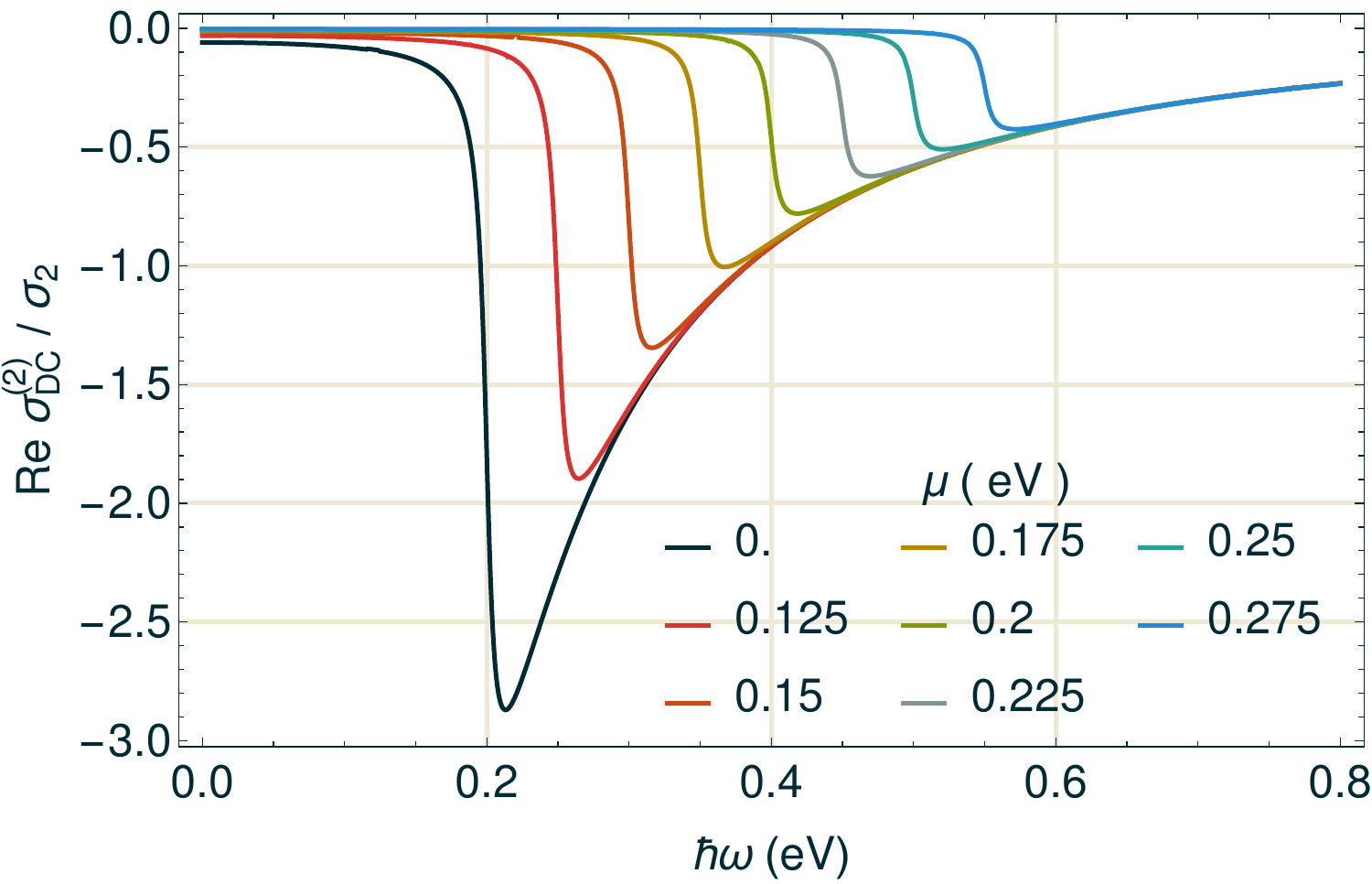}%
  \label{fig:graphSLmu}%
}
\caption{
Photocurrent of gapped graphene as a function of gap 
\subref{fig:graphSLDelta}, and chemical potential \subref{fig:graphSLmu} 
whose respective magnitudes are indicated in the legends. In 
\subref{fig:graphSLDelta}, dashed lines refer to the threshold behavior 
given in \Eqref{eq:sigma2-at-onset}, and the vertical scale is logarithmic.
Other parameters are as in \Fref{fig:graphSLs2}.
}
  \label{fig:graphSL}
\end{figure}

Using the parameters relevant for graphene to be definite, we computed 
explicitly the four non-vanishing elements of the photoconductivity 
tensor $\sigma^{(2)}_{\lambda\alpha\beta}(\omega,-\omega)$, which are shown in 
\Fref{fig:graphSLs2} when the gap $\Delta \teq 200 \, \mathrm{meV}$. 
That the curves for distinct components coincide and cannot be distinguished in 
the figure documents a correct implementation of our computation of the various 
terms in \eqref{eq:sigma2:full}.
%
%
%
The real part of $\sigmadc$ should share key features of the joint density of 
states for transitions between the valence and conduction bands.
In particular, it should display an onset of response at precisely $ 
\hbar\omega \teq \Delta $ (for $T\teq0$), and van Hove singularities at 
frequencies coinciding with transitions between locally flat portions of the 
band dispersion.
The band-gap feature can be clearly identified in the figure.
Note, however, an important difference in contrast to the case shown in Fig.
\ref{fig:BN:SL} for an actual realization of hBN: for small gap, the response 
is \emph{much stronger} at frequencies in the vicinity of the gap (compare the 
magnitude of $ \rm{Re}\, \sigmadc $ at $ \hbar\omega \teq \Delta $ in the two 
cases).
Since the low frequency features are governed by the nature of virtual 
transitions in the vicinity of the $K$ point of the Brillouin zone (BZ), we can understand this 
behavior from an exact analytical standpoint which is possible to establish 
after expanding \eqref{eq:sigma2:full} in the vicinity of $K$:
\begin{align}
  \rm{Re}\,\sigmadc & \approx \sigma_{2} 
  \bigg[ 
  -\frac{1}{4\Delta}+\bigg(\frac{21}{32\Delta^{2}}-\frac{25}{576} \bigg) 
  \big( \hbar\omega -\Delta \big) \bigg]
  \nonumber \\ & \times
  \theta \big( \hbar\omega -\Delta \big)
  .
  \label{eq:sigma2-at-onset}
\end{align}
This curve that defines the onset of photoconductivity at $\hbar\omega \gtrsim 
\Delta$ is shown in Figs. \ref{fig:graphSLs2} and \Fref{fig:graphSL}(a) as 
dashed 
lines.
%
It is clear that the singular behavior at $ \hbar\omega \teq \Delta$ should be 
more prominent the smaller the gap. Since the magnitude of 
$\rm{Re}\,\sigma^{(2)}$ there is exactly $ -\sigma_2 / 4\Delta$, the 
frequencies near the optical absorption edge will entirely dominate the 
photoconductivity response for gaps smaller than $1$\,eV. This is shown 
explicitly in \Fref{fig:graphSL}(a), where we plot the photoconductivity for 
different gaps at very low temperature.
The effect of varying the chemical potential is studied in 
\Fref{fig:graphSL}(b) for representative cases. Since all the virtual 
transitions that define the response in a translationally invariant system are 
vertical (i.e., conserving the crystal momentum $\bm{k}$), at $T\approx 0$ a 
chemical potential in the conduction band will block any response for 
$\hbar\omega < 2|\mu|$ due to Pauli exclusion, as seen in the figure.

\section{Photoconductivity of bilayer honeycomb lattices}
\label{sec:bilayers}

\subsection{Biased bilayer graphene}
\label{sec:graphAB}

If it is not clear within the current experimental landscape whether a 
realization of small $\Delta$ gapped graphene is a realistic prospect, the 
existence of a band-gap in biased bilayer graphene (BBG) is a well-established 
experimental fact \cite{Zhang2009}. 
Crucially, its gap is a function of the externally driven inter-layer bias 
voltage and, hence, tuneable \cite{Castro2007, Zhang2009}. From this 
perspective, a BBG is a more natural candidate to explore the quadratic 
response to light. 

A minimal tight-binding model that captures the electronic structure of 
Bernal-stacked bilayer graphene in the presence of a uniform electric field 
perpendicular to the plane, is given by \cite{Castro2007, McCann2006}
\begin{align}
\label{eq:bilayer:Hamils:graph}
h_{\bk}^{\text{BBG}} =
\begin{pmatrix}
\Delta/2     & \phi(\bk)  & \gamma_1   & 0     \\
\phi^*(\bk)  & \Delta/2   & 0          & 0         \\
\gamma_1     & 0          & -\Delta/2  & \phi^*(\bk)   \\
0            & 0          & \phi(\bk)  & -\Delta/2
\end{pmatrix}
\!,
\end{align}
where $ \gamma_1 $ represents the interlayer hopping and $ \Delta $ the 
difference in potential energy in the two layers induced by the external field. 
The dispersion function $\phi(\bk)$ is the same that appears in Eq.
\ref{eq:Ek-2band}. The Hamiltonian \eqref{eq:bilayer:Hamils:graph} is 
represented in the basis $\lbrace A_1,\, B_1, \, B_2, \, A_2 \rbrace$.

The largest gaps obtained by field effect with top and bottom-gated devices 
have not so far exceeded 0.5\,eV \cite{Castro2007, Szafranek2011, Velasco2012, 
Pachoud2010, Britnell2012, Stabile2015}. Hence, to be specific, we shall 
analyze the photocurrent in BBG with $\Delta \teq 200 \, \mathrm{meV}$ in the 
parametrization of \eqref{eq:sigma2:full}.

The band structure of BBG can accommodate different types of vertical 
transitions in the low energy regime, thereby increasing the richness of the 
interaction with light. Having now one pair of conduction and another of 
valence bands means that transitions such as $\epsilon_3 \leftrightarrow 
\epsilon_1 $ or $ \epsilon_2 \leftrightarrow \epsilon_4 $ become important at 
low frequencies because the separation between these bands (set by the 
interlayer hopping $\gamma_1$) can easily be comparable to the bias-induced 
band-gap. 
What is more, chemical potential and bias (hence gap) can be controlled 
independently in experiments, allowing for a selective suppression of different 
types of transitions. 

%
\begin{figure}
  \includegraphics[width=1.\columnwidth]{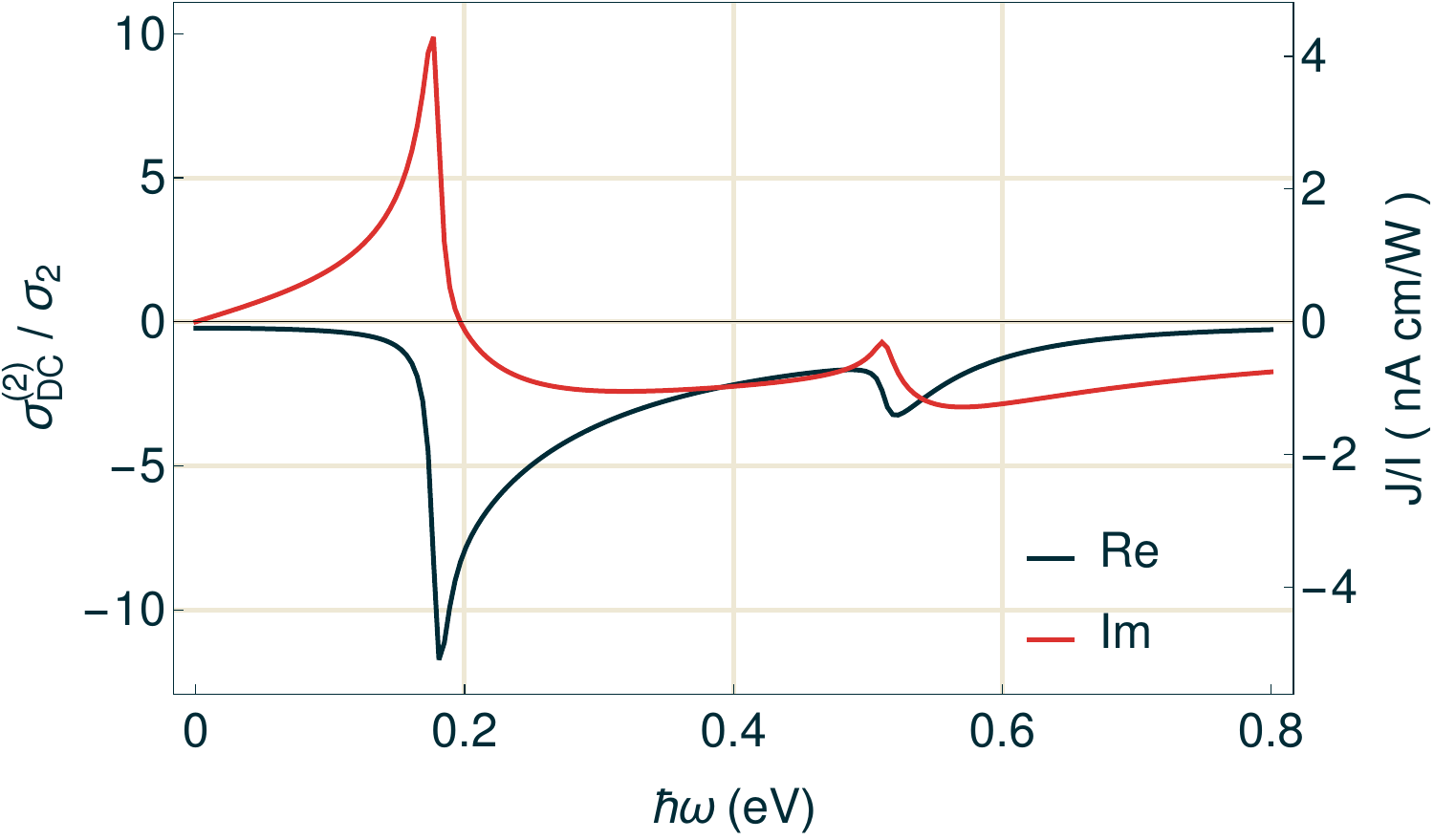}%
  \caption{\label{fig:graph:AB}
  Photoconductivity of BBG at charge neutrality in units of $ \sigma_2 \teq 
  2.88 \times 10^{-15} \, \mathrm{S\, m/V}$.
  Right scale measures the electric current per laser intensity $( J/I 
  )$\cite{EndNote-1}.
  We consider the  TB model \eqref{eq:bilayer:Hamils:graph} 
  $ ( \gamma_0 \teq 3 \, \mathrm{eV}$, \,
  $ \gamma_1 \teq 0.4 \mathrm{eV}$, \,
  $ \Delta \teq 0.2 \, \mathrm{eV}$, \,
  $ \Gamma \teq 5 \mathrm{meV}$, \,
  $ \mu \teq 0 \, \mathrm{eV} )$.
} 
\end{figure}%

\begin{figure}
\includegraphics[width=\columnwidth]{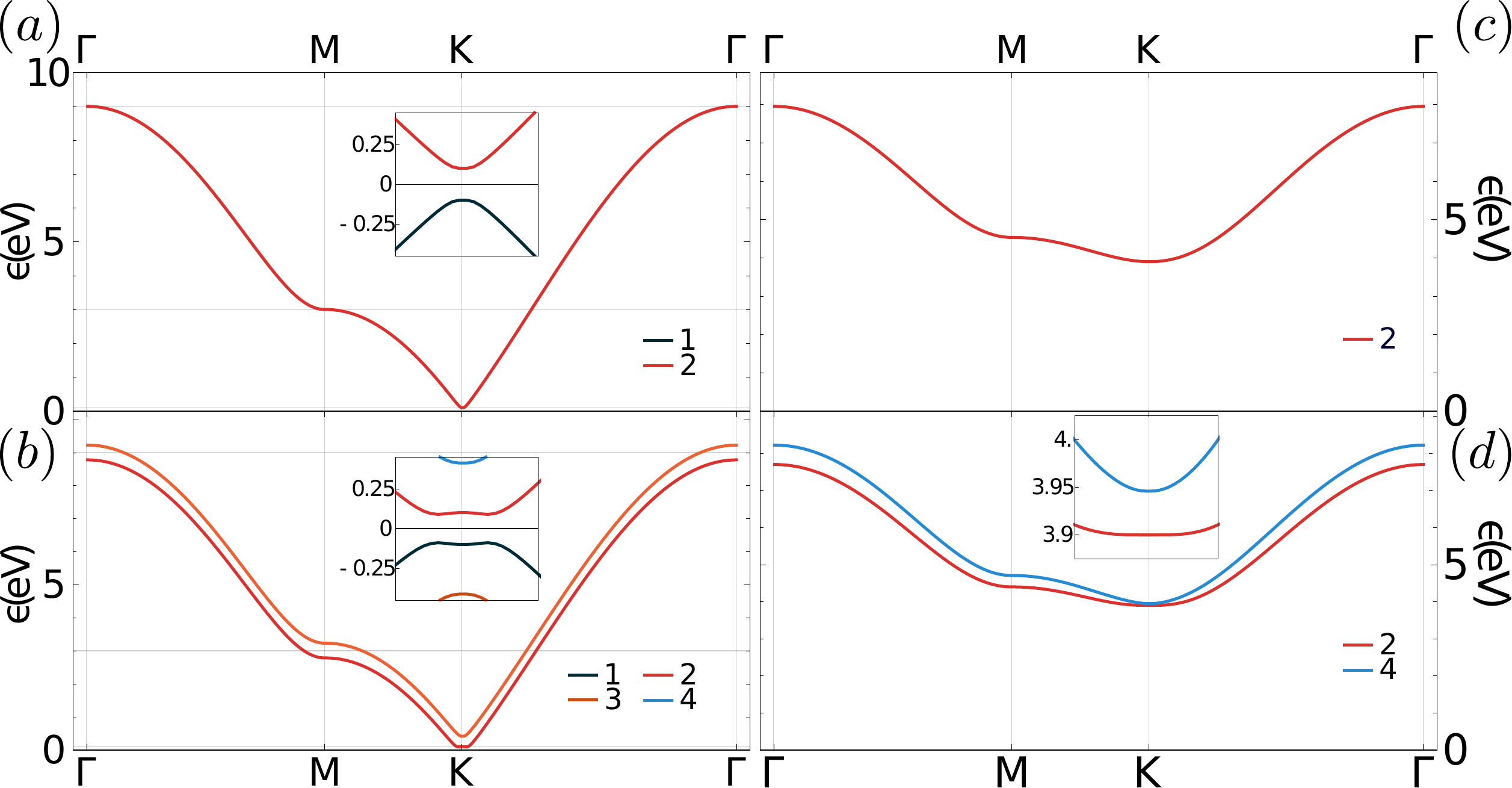}
\caption{Conduction band structures along the high symmetry path $ 
\bs{\Gamma} \to \mathbf{M} \to \mathbf{K} \to \bs{\Gamma} $ using 
a TB Hamiltonian (given that all systems under consideration have particle-hole 
symmetry, we only display the conduction bands).
(a) Monolayer graphene $ ( \gamma_0 \teq 3 \, \mathrm{eV}, \, \Delta \teq 
0.2 \, \mathrm{eV} ) $.
(b) BBG $ ( \gamma_0 \teq 3 \, \mathrm{eV}, \, \gamma_1 \teq 0.4, \, 
\mathrm{eV} \, \Delta \teq 0.2 \, \mathrm{eV}  ) $.
(c) Monolayer hBN $ ( \gamma_0 \teq 2.33 \, \mathrm{eV}, \, \Delta \teq 7.8 
\, \mathrm{eV} ) $.
(d) Bilayer $AB$ hBN $ ( \gamma_0 = 2.33 \, \mathrm{eV}, \, \gamma_1 \teq 
0.6, \, \mathrm{eV} \, \Delta \teq 7.8 \, \mathrm{eV} ) $.
Insets in (b) and (d) show the band structure in the vicinity of the $K$ point.
}
\label{fig:edisp}%
\end{figure}

Figure \ref{fig:graph:AB} shows the intrinsic photoconductivity expected in 
charge-neutral BBG.
At low temperature, the response exhibits two key features associated with 
the onset of the virtual transitions $ \epsilon_1 \leftrightarrow \epsilon_2 $ 
and $ \epsilon_1 \leftrightarrow \epsilon_4 $ (see \Fref{fig:edisp} for 
details of the band labeling). As expected, no features appear 
related to transitions involving bands $\epsilon_3 \leftrightarrow \epsilon_1$ 
or $ \epsilon_2 \leftrightarrow \epsilon_4 $  due to Pauli blocking.

\begin{figure}
\subfigure[][]{%
 \includegraphics[width=\columnwidth]{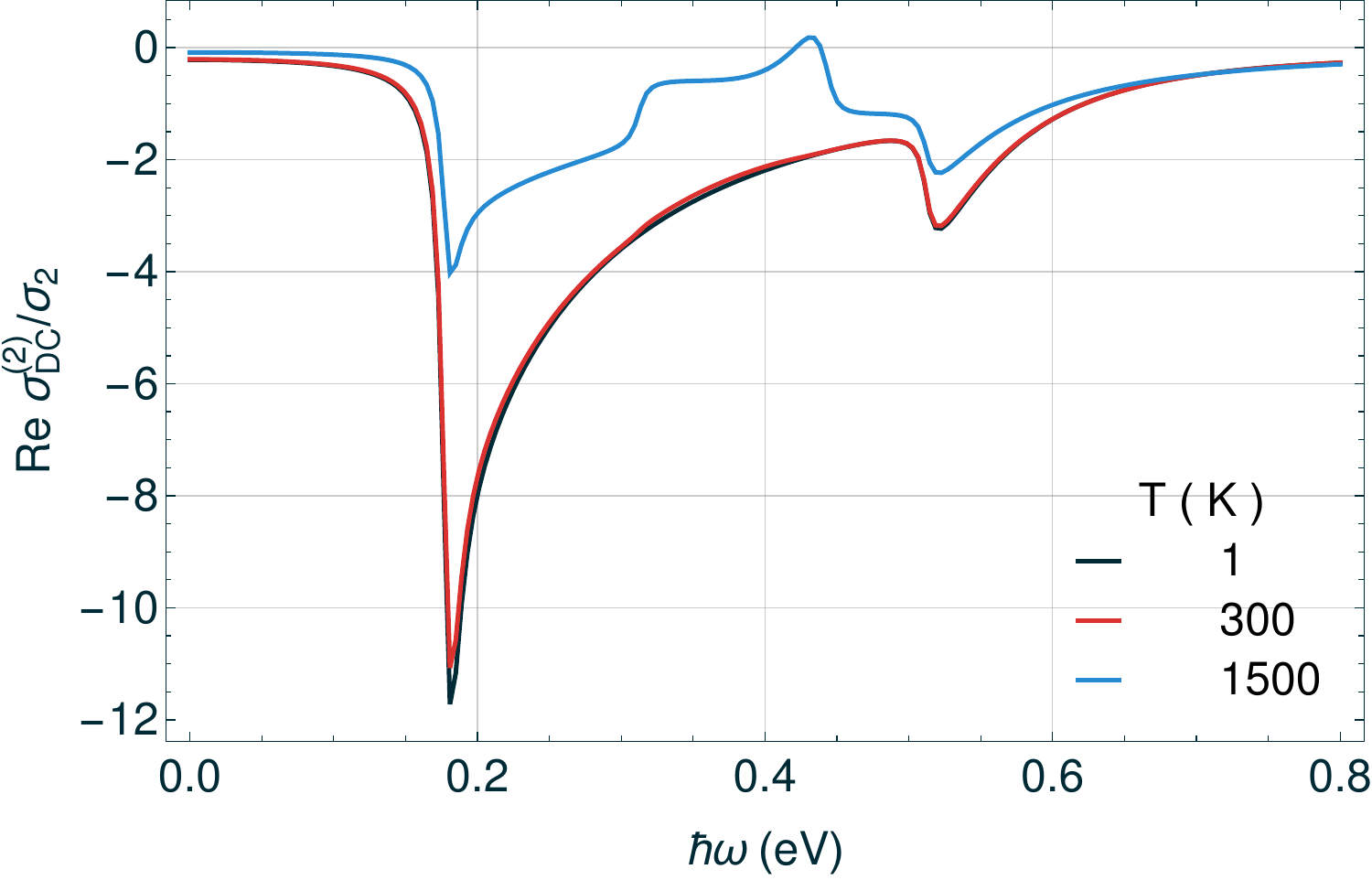}%
 \label{fig:graph:AB:temp}%
}
\hfill
\subfigure[][]{%
 \includegraphics[width=\columnwidth]{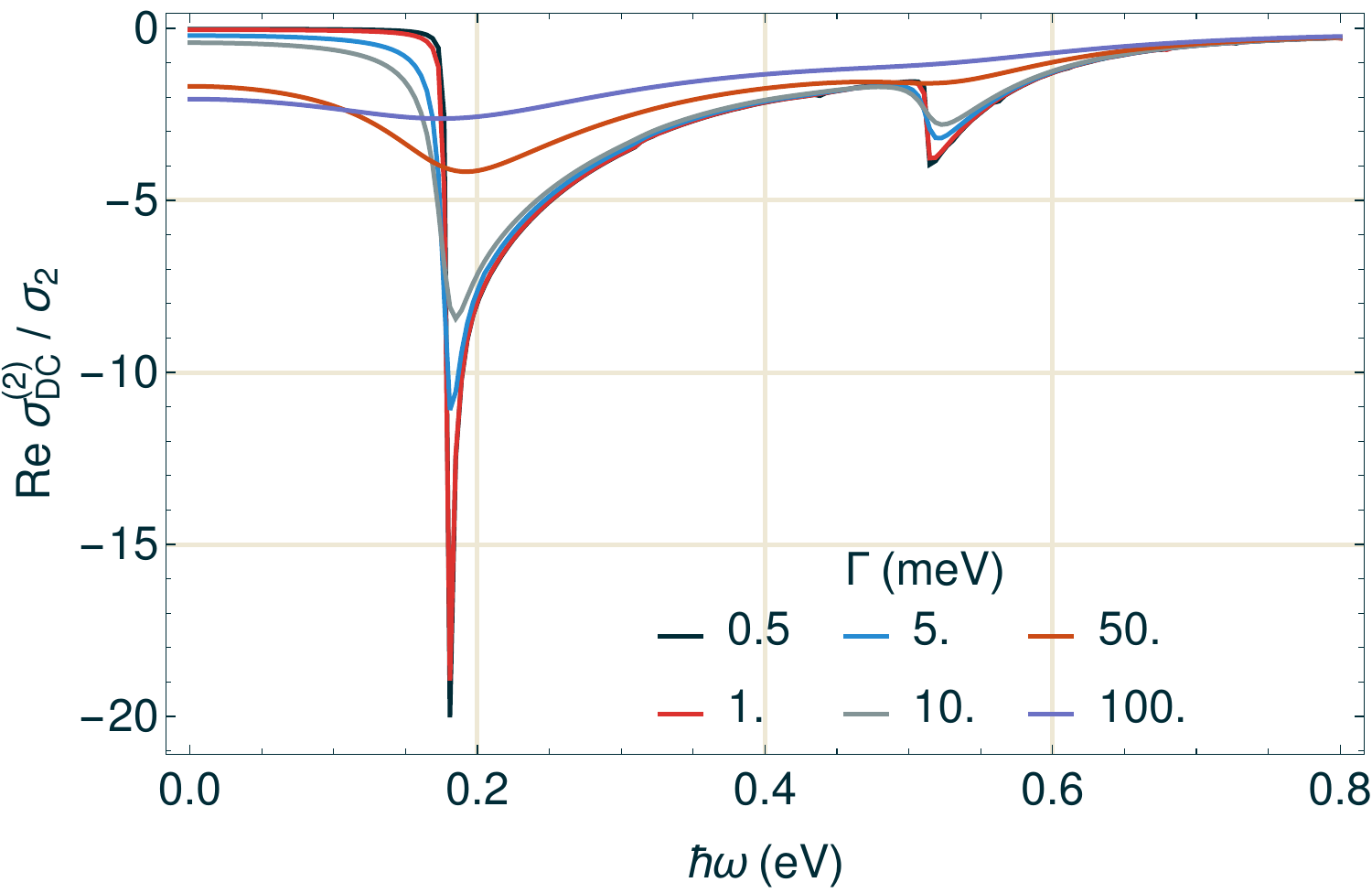}%
 \label{fig:graph:AB:eta}%
}
\hfill
\subfigure[][]{%
  \includegraphics[width=\columnwidth]{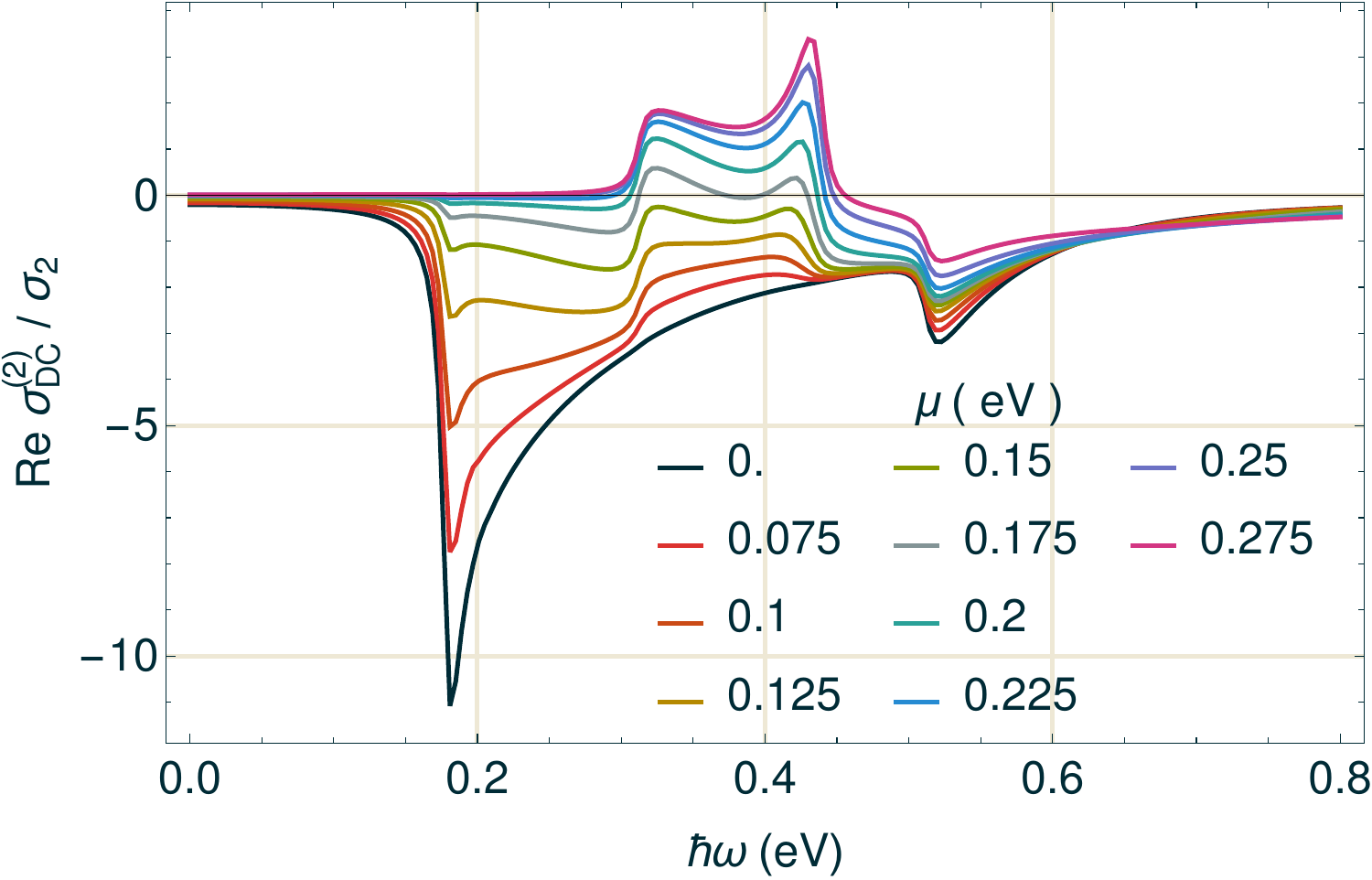}%
  \label{fig:graph:AB:mu}%
}
\caption{%
  Photoconductivity of BBG at charge neutrality in units of $ \sigma_2 \teq 
  2.88 \times 10^{-15} \, \mathrm{S\, m/V}$.
  In \subref{fig:graph:AB:temp} we plot the variation of the photoconductivity 
  with temperature. Panels \subref{fig:graph:AB:eta} and
  \subref{fig:graph:AB:mu} are calculated at $T \teq 300$\,K and show,
  respectively, the dependence on the broadening parameter $\Gamma$ (meV) and 
  chemical potential $\mu$. 
  Other parameters are as in \Fref{fig:graph:AB}.
} 
 \label{fig:graph:AB:2}
\end{figure}%

\subsection{Finite temperatures, doping and broadening}

If the temperature is high enough, we can see in 
\Fref{fig:graph:AB:2}\subref{fig:graph:AB:temp} that the frequency response 
becomes much richer, and $\sigmadc$ can even change sign in the range 
$0.3\lesssim \hbar\omega \lesssim 0.5$ that is dictated by the magnitude of the 
inter-layer hopping $ \gamma_1 $.
This is caused by the suppression of Pauli blocking with increasing 
temperature, leading to the emergence of spectral features associated with 
transitions from the bands $ \epsilon_3 \leftrightarrow \epsilon_1 $ and $ 
\epsilon_2 \leftrightarrow \epsilon_4 $ that are not effectively possible at 
room temperature.
It is easy to conceive a direct application of this characteristic: since 
experiments that probe the non-linear response must be frequently done under 
relatively high power, the observation of spectral features expected to be 
Pauli blocked could be used as an indirect thermometer to estimate the local 
temperature of the electron gas in the system under illumination.

Since in this paper we are concerned with the generation of a dc electrical 
current through optical means, it is natural to wonder what effect the 
scattering of charge carriers by impurities might have in the strength and 
nature of the photocurrents. Although we do not explicitly include disorder in 
our calculations, and to the best of our knowledge, disorder has not been 
addressed in microscopic calculations of the non-linear optical response, we 
can draw insight from the effects that weak disorder has in the linear optical 
conductivity. In this context, several types of disorder have been studied and, 
in addition to the fact that dilute scatterers lead to perturbations 
proportional to their concentration, the leading qualitative effect of both 
short-ranged and Coulomb impurities is the broadening/smearing of the line 
shape characteristic of the pristine crystals\cite{Peres2008a, Stauber2008b, 
Stauber2008a, Abergel2012, Yuan2011}.
Correspondingly, we expect the main effect of weak disorder in the non-linear 
response to be captured qualitatively by making the adiabatic parameter $\Gamma$ 
in our calculations explicitly finite, with a magnitude that reflects a 
phenomenological scattering rate. That would be equivalent to assuming that 
self-energy corrections arising from disorder are featureless in momentum and 
frequency, which is a reasonable approximation for these cases.
As an illustration, in Fig. \ref{fig:graph:AB:2}\subref{fig:graph:AB:eta}, we 
study the sensitivity of the photocurrent to that scattering rate for the BBG 
at room temperature. The main features remain identifiable up to $\Gamma 
\gtrsim 50$\,meV.

Notwithstanding the impact that disorder might have in the dc currents 
generated by the photogalvanic effect, the continuous progress in the 
production of high quality samples, such as encapsulating graphene between hBN 
crystals \cite{Gannett2011, Chari2015, Banszerus2016, Ni2016} and other 
techniques \cite{Pallecchi2014}, has delivered procedures to achieve 
graphene-based electronic devices of progressively higher mobility and mean 
free paths up to $\ell_f \tsim 23 \, \mu\mathrm{m}$ \cite{Banszerus2016}.
This corresponds to a typical lifetime for ballistic transport of $ \tau \teq 
\ell_f / v_F \tsim 2.3$\,ps or a scattering rate $ \Gamma{\,\sim\,}\hbar 
/2\Delta t \teq 0.14 $\,meV.
According to the data in Fig. \ref{fig:graph:AB:2}\subref{fig:graph:AB:eta}, 
this means that, even though effects such as photon drag might compete with the 
photogalvanic effect at the quadratic order, it is not unrealistic to 
anticipate a class of systems where contributions to the photocurrent arising 
from impurity-assisted processes are minimal, and it is mostly determined by 
the intrinsic photogalvanic effect discussed here.

Finally, the results presented in \Fref{fig:graph:AB:2}\subref{fig:graph:AB:mu} 
at different $\mu$ demonstrate that the ability to experimentally vary the 
chemical potential on demand through simple gating might allow external control 
over the polarity of the induced photocurrents within target frequency ranges. 
The sign of $\sigmadc$ directly translates into the sign of the dc current in 
the material and, as we can see in this figure, the photocurrent can be made to 
switch from positive to negative at frequencies that are controlled by $\mu$.

\subsection{Bilayer boron nitride}
\label{sec:BNab}

As discussed earlier, several types of hBN bilayers can arise from distinct 
stacking arrangements, but only two are non-centrosymmetric (cf. 
Table~\ref{tab:sym:threefold}) and thus relevant in the context of quadratic 
response. We will consider the $AB$ bilayer in the following discussion, which 
is the counterpart of the Bernal graphene bilayer that results when two 
superimposed hBN monolayers are displaced so that the N atom ($A_1$) lies above 
the B ($B_2$) in the layer underneath.
The corresponding Hamiltonian in the basis $ \lbrace A_1, \, B_1, \, B_2, \, 
A_2 \rbrace $ is
\begin{equation}
  \label{eq:bilayer:Hamils:BN:AB}
  h_{\bk}^{\text{AB}} = 
  \begin{pmatrix}
  -\Delta/2   & \phi(\bk) & \gamma_1  & 0           \\
  \phi^*(\bk) & \Delta/2  & 0         & 0           \\
  \gamma_1    & 0         & \Delta/2  & \phi^*(\bk) \\
  0           & 0         & \phi(\bk) & -\Delta/2
  \end{pmatrix}
  \!,
\end{equation}
where, analogously to Eq. \ref{eq:bilayer:Hamils:graph}, $ \Delta $ is the 
interlayer bias parameter and $\gamma_1$ the interlayer hopping.
For consistency with the calculations done earlier in the monolayer, we 
consider the same in-plane tight-binding parameters $ \gamma_0 \teq 2.33 \, 
\mathrm{eV} $, $ \Delta \teq 7.80 \, \mathrm{eV}$, and 
$ a \teq 1.45 \times 10^{-10} \, \mathrm{m} $.
In reference \onlinecite{Ribeiro2011}, a fit to a first-principles calculation 
of the bandstructure of $AB$-BN finds $\gamma_1 \teq 0.60 \, \mathrm{eV} $ and 
we use this value.
Although a finite inter-layer bias voltage can be added similarly to BBG and is 
expected to modify the gap \cite{Yang2010}, we consider only unbiased hBN 
bilayers ($\Delta \teq 0$).
%

\begin{figure}
 \includegraphics[width=\columnwidth]{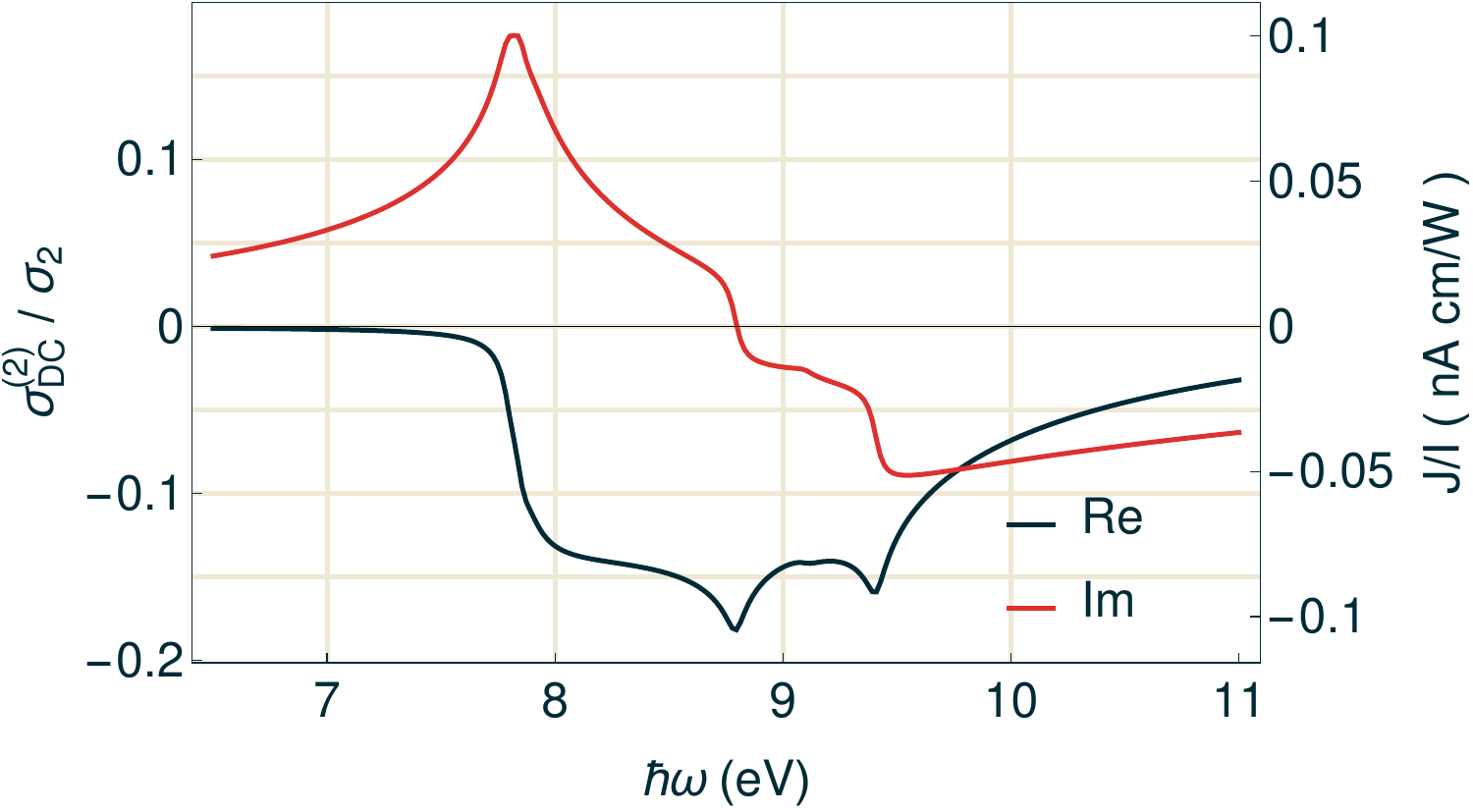}%
\caption{%
  Photoconductivity of an $AB$-hBN bilayer in units of $ \sigma_2 \teq 3.79 
  \times 10^{-15} \, \mathrm{S\, m/V}$.
  Right scale measures the electric current per laser intensity $( J/I   
  )$\cite{EndNote-1}.
  We consider the  TB model \eqref{eq:bilayer:Hamils:BN:AB}
  $ ( \gamma_0 \teq 2.33 \, \mathrm{eV}$, \,
  $ \gamma_1 \teq 0.6\, \mathrm{eV}$, \,
  $ \Delta \teq   7.8 \,   \mathrm{eV}$, \,
  $ \Gamma \teq 0.03 \,\mathrm{eV}$, \,
  $ \mu \teq 0 \,   \mathrm{eV}$,\,
  $ T \teq 1 \, \mathrm{K} )$.
} 
\label{fig:BN:AB}
\end{figure}%

Figure \ref{fig:BN:AB} displays the resulting photoconductivity at low 
temperature for an undoped ($\mu\teq0$) hBN bilayer. Since $\Delta$ is by far 
the largest energy scale in hBN, it is no surprise that the shape of $\sigmadc$ 
seen here is almost entirely similar to that of the monolayer (cf. 
\Fref{fig:BN:SL}), except for a factor of 2 enhancement in the case of the 
bilayer on account of the system now having twice as many layers.
The one noticeable difference appears associated with the van Hove 
singularities at the $M$ points of the BZ because, at these saddle points in the 
energy dispersion, the separation of the pair of valence (and conduction) bands 
is large ($\approx 307 \,\mathrm{meV}$), leading to a sizable separation 
($\approx 614\,\mathrm{meV}$) of the two possible vertical transitions.

\begin{figure}
\subfigure[][]{%
 \includegraphics[width=\columnwidth]{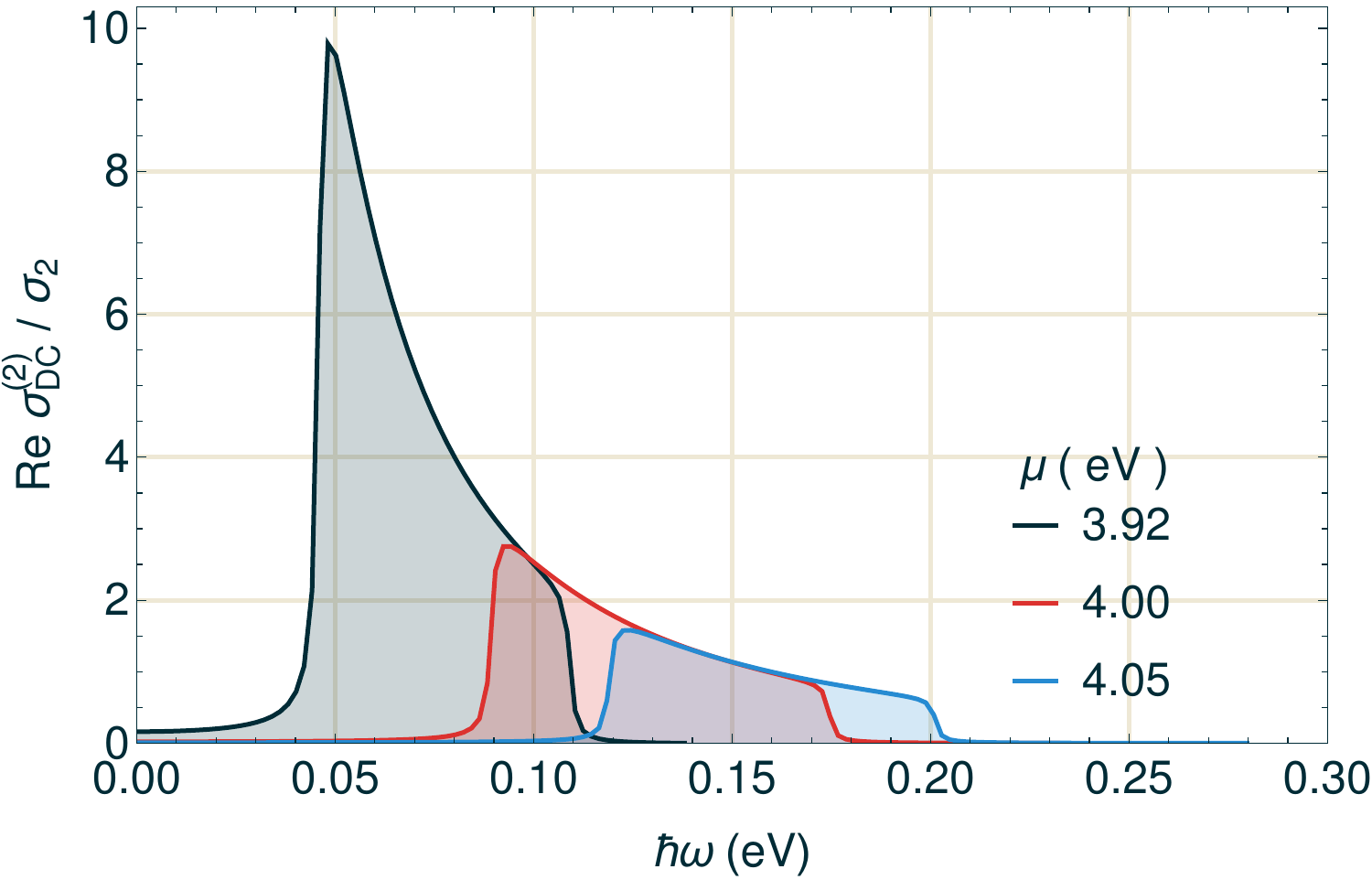}%
  \label{fig:BN:AB:mu:low}%
}
\hfill
\subfigure[][]{%
 \includegraphics[width=\columnwidth]{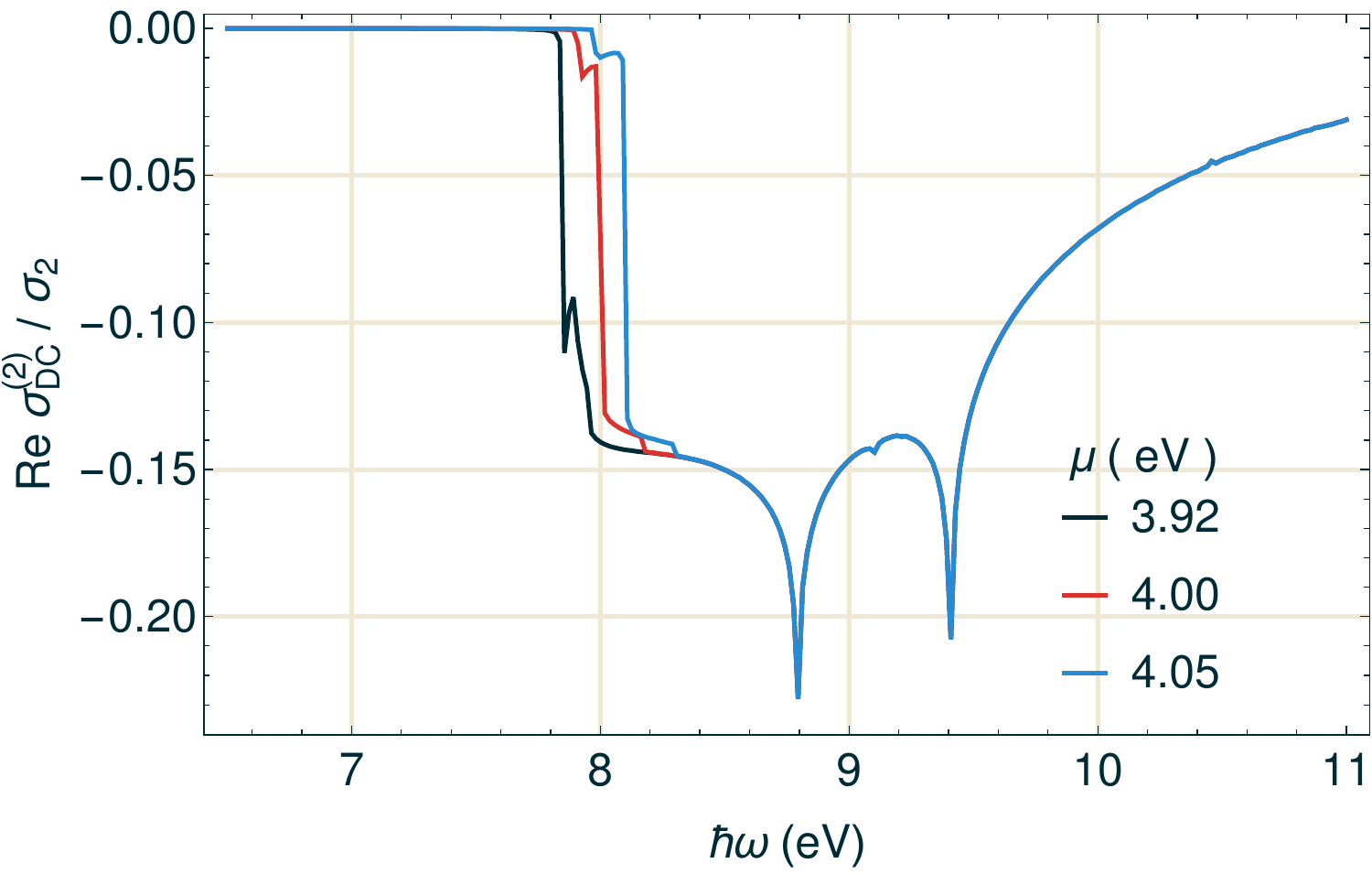}%
 \label{fig:BN:AB:mu:high}%
}
\caption{\label{fig:BN:AB:mu}%
  Real part of $\sigmadc$ in a bilayer of $AB$-stacked hBN at finite doping 
  and $ T \teq 1 \, \mathrm{K}$.
  Only the low frequency range is shown in panel \subref{fig:BN:AB:mu:low}, 
  while \subref{fig:BN:AB:mu:high} displays the high-frequency response. All 
  frequencies in between show zero $\rm{Re}\,\sigmadc$.
  Note that the energy range in \subref{fig:BN:AB:mu:low} is much smaller than 
  in \subref{fig:BN:AB:mu:high}.
  Other parameters are as in \Fref{fig:BN:AB}.
}
\label{fig:BN:AB:2}
\end{figure}%

An interesting scenario is possible if the bilayer is slightly doped (possibly 
intentionally by impurities) in order to place the chemical potential at or 
slightly above the bottom of the lowest conduction band
(the same happens if $\mu$ is slightly below the top of the highest valence 
band in a slightly hole-doped scenario)\cite{Terrones2007, Wei2011}.
Consider the results shown in \Fref{fig:BN:AB:2}(a).
At the $K$ point the separation between bands 2 and 4 [cf. \Fref{fig:edisp}(d)] 
is
\begin{align}
\epsilon_{42} = \frac{ \Delta }{2} \Big( \sqrt{ 1 + \tfrac{4 
\gamma^2}{\Delta^2} } -1 \Big)
\approx \gamma_1^2 / \Delta = 46 \, \mathrm{meV} \, .
\end{align}
Since $ \gamma_1 / \Delta \ll 1$ in hBN, this separation is rather small when 
compared with $ \epsilon_2-\epsilon_1 \approx  \epsilon_4-\epsilon_1 \approx 
\Delta $ to leading order in the gap.
As a result, transitions between bands $\epsilon_2 \leftrightarrow \epsilon_4 $ 
open a channel for second order response at a very low energy ($\sim \gamma_1^2 
/\Delta$) when compared with typical valence-conduction transitions.
Consequently, when hBN is slightly electron-doped so that $\mu$ straddles the 
bottom of the two conduction bands one obtains: (i) a strong response 
at frequencies $\hbar\omega \approx  46$\,meV that can easily be an order 
of magnitude higher than the response characteristic of the valence-conduction 
transitions; (ii) an inversion of the sign of the induced current for low 
stimulating frequencies in comparison with that for frequencies above the 
fundamental band gap. 
The high intensity seen in the low frequency window shown 
in \Fref{fig:BN:AB:2}(a) stems from the fact that these processes arise
from interband transitions between bands having approximately the same 
curvature [see \Fref{fig:edisp}(d)]. As a result, there is a much larger 
underlying joint density of states than at higher energies, where the 
transitions always connect states in bands with opposite curvature. The 
frequency band showing strong infrared response is controlled by the position 
of the chemical potential, suggesting that the effect can be manipulated by 
tailoring the doping level.

%
\section{\label{sec:excitons}The role of excitons in \lowercase{h}BN}

Excitons are not a crucial element in elementary descriptions of the optical 
response of graphene. Even though they do lead to quantitative changes in the 
position of the van Hove singularities \cite{Yang2009} and, therefore, should 
be properly accounted for in quantitative comparisons with experimental data 
\cite{Kravets2010,Mak2014}, their presence doesn't introduce significant 
\emph{qualitative} changes to the frequency dependence of the response 
functions \cite{Kravets2010}. 
In particular, the low-frequency behavior is not visibly sensitive to the 
inclusion of excitonic corrections due mostly to the fact that graphene has no 
band-gap. In BBG, the band-gap allows for the absorption spectrum to be modified 
in the gap region due to excitons, but the screening from substrates is enough 
to make these corrections relatively small in comparison with a single particle 
description \cite{Park2010}.

Boron nitride is different. First principles calculations indicate that 
the linear optical conductivity of hBN is strongly renormalized by excitonic 
corrections \cite{Wirtz2006}. This, although a fact common to all 
two-dimensional insulator crystals on account of the reduced screening of 
Coulomb interactions, leads to a particularly strong correction in hBN on 
because of its very large band-gap.
The second-harmonic susceptibility of hBN studied in Ref. 
\onlinecite{Pedersen2015} is entirely dominated by the two-particle spectrum 
and we expect the photoconductivity, $\sigmadc$, to be likewise strongly 
modified.

\begin{figure}[t]
\centering
\includegraphics[width=\columnwidth]{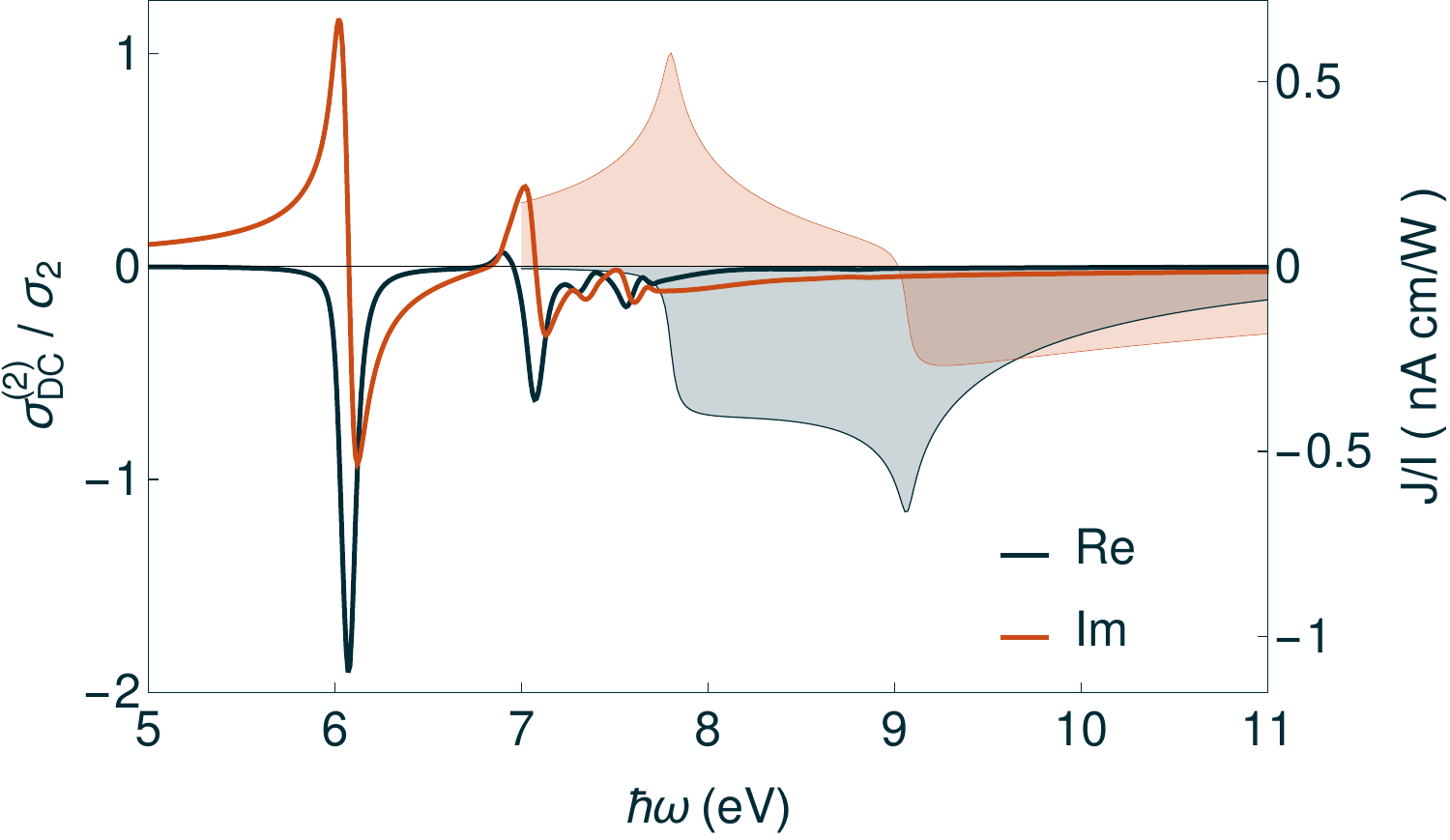}
\caption{
Photoconductivity of a hBN monolayer with excitonic corrections. 
Model parameters are identical to those in \Fref{fig:BN:SL}.
The shadowed curves reproduce the independent-particle results shown in 
\Fref{fig:BN:SL}, vertically scaled up by a factor of 10. 
}
  \label{fig:graph:Excitons}
\end{figure}

To that end, we have applied the two-band model \cite{Pedersen2015} of the 
second-order intraband response to the non-linear photoconductivity in hBN. The 
band structure is based on the parameters applied above, i.e. $\gamma_0 \teq 
2.33 \, \mathrm{eV}, \, \Delta \teq 7.8 \, \mathrm{eV}$ and screening is 
implemented as in reference \onlinecite{Pedersen2015}. The resulting exciton 
spectrum is shown in \Fref{fig:graph:Excitons}, which should be 
contrasted with the independent-electron result in \Fref{fig:BN:SL}(a). For a 
direct comparison, the latter is reproduced as the shaded curves in 
\Fref{fig:graph:Excitons}. 
It is apparent that Coulomb effects cause a marked red-shift by nearly 2\,eV of 
the onset of the photoconductive response: the prominent band-gap feature 
characteristic of the independent-electron response at 7.8\,eV 
is down-shifted to the fundamental exciton energy around 6.1\,eV. It 
is also noted that the magnitude of the response at frequencies associated with 
well isolated exciton levels is dramatically increased (in excess of tenfold) 
in comparison with the non-interacting case. At the same time, whereas the 
curve of $\sigmadc$ is markedly structured in the frequency range containing 
well defined excitons, it becomes quite featureless above the non-interacting 
band-gap. This is a consequence of the large spectral weight carried by the 
excitonic peaks that implies a depletion in the response at frequencies above 
the band-gap. This is analogous to the behavior observed in the 
(linear) optical absorption spectrum of this material \cite{Wirtz2006}.
The colossal excitonic effects derive mainly from the poor screening 
in large-gap two-dimensional insulators, of which hBN is clearly an extreme 
case. 
In view of the close similarity between the photoconductivity of hBN monolayers 
and bilayers seen at the independent-particle level [cf. Figs. 
\ref{fig:BN:SL}(a) and \Fref{fig:BN:AB}], we expect an equally strong 
renormalization of the response in the undoped bilayers. In general, less 
pronounced modifications are expected for low-gap systems such as bilayer 
graphene or other realizations of the more general ``gapped graphene'' model 
discussed above, as well as in the lightly doped scenarios discussed before.

%
\section{\label{sec:concl}Conclusion}

The photoconductivity provides a direct measure of the ability to directly 
inject dc currents in a system by purely optical means. Harnessing and being 
able to tailor this effect can lead to improved optoelectronic device concepts 
and functionalities. Two-dimensional crystals are excellent materials to 
explore towards this end due to the intrinsic ease of integration in flat 
heterostructures, the possibility of controlling the electronic density through 
field effect, or the ability to modify their electronic structure by various 
types of surface modification.
We have explored the photoconductivity of honeycomb-based electronic systems in 
monolayer and bilayer form and resorted to the examples of graphene and hBN for 
a definite illustration of its behavior in the cases of small and large 
band-gap. 

Our calculations were done in the length-gauge within the framework originally 
discussed by Sipe and collaborators \cite{Aversa1995}. In the systems derived 
from graphene, an independent-particle approach should provide a good 
qualitative and quantitative characterization of the response. We therefore 
trust that our results for BBG are entirely realistic, and similarly for those 
pertaining to the ``gapped graphene'' model, provided the gaps are kept small. 
In particular, our results in Sec. \ref{sec:graphAB} show that the ability 
to independently tune both the gap and density in BBG can lead to a very rich 
and density dependent photogalvanic response.
On the other hand, those structures based on undoped hBN necessarily require an 
explicit treatment of the Coulomb interactions between particle-hole pairs. 
By solving the Bethe-Salpeter equation and computing the resulting 
photogalvanic response, we have shown in Sec. \ref{sec:excitons} that the 
reduced screening in two-dimensions leads to robust excitons with large binding 
energies that, not only strongly renormalize the onset of the response to much 
lower energies, but, importantly, concentrate most of the spectral weight for 
$\sigmadc$. Therefore, similarly to their crucial impact in the absorption 
spectrum, interaction effects are clearly unavoidable in an accurate model of 
the photoconductivity for hBN. In a lightly doped scenario, however, the 
enhanced metallic screening is expected to significantly suppress the Coulomb 
interaction and bring the system's response closer to that of an 
independent-particle description.
Experiments reporting doping by carbon substitution have been reported in hBN 
films, nanotubes and nanoribbons \cite{Terrones2007, Wei2011}.
It has also been predicted that intercalation or adsorption with alkali elements 
can produce shallow donor states with minimal impact in the underlying 
band structure \cite{Oba2010}.
In this case, the features discussed in Sec. \ref{sec:BNab} should hold, and 
one expects a strong photoconductivity in a narrow frequency band in the 
infra-red. The existence and width of this band are controlled by the position 
of the chemical potential within the two conduction sub-bands, and suggests the 
possibility of generating photogalvanic currents in hBN with frequencies much 
smaller than the fundamental band-gap.

\acknowledgments

F.H. expresses thanks to A. H. Castro Neto for his support and discussions 
throughout this project.
This work was supported by the National Research Foundation Singapore, under 
its Medium Sized Centre Programme and the CRP Award No. NRF-CRP6-2010-05. 
T.G.P. is financially supported by the CNG center under the Danish National 
Research Foundation, Project No. DNRF103, and the QUSCOPE center sponsored by 
the Villum foundation.
The numerical computations were carried out at the HPC facilities of the Centre 
for Advanced 2D Materials at the National University of Singapore.

\bibliography{pchoneycomb}

\begin{thebibliography}{69}%
\makeatletter
\providecommand \@ifxundefined [1]{%
 \@ifx{#1\undefined}
}%
\providecommand \@ifnum [1]{%
 \ifnum #1\expandafter \@firstoftwo
 \else \expandafter \@secondoftwo
 \fi
}%
\providecommand \@ifx [1]{%
 \ifx #1\expandafter \@firstoftwo
 \else \expandafter \@secondoftwo
 \fi
}%
\providecommand \natexlab [1]{#1}%
\providecommand \enquote  [1]{``#1''}%
\providecommand \bibnamefont  [1]{#1}%
\providecommand \bibfnamefont [1]{#1}%
\providecommand \citenamefont [1]{#1}%
\providecommand \href@noop [0]{\@secondoftwo}%
\providecommand \href [0]{\begingroup \@sanitize@url \@href}%
\providecommand \@href[1]{\@@startlink{#1}\@@href}%
\providecommand \@@href[1]{\endgroup#1\@@endlink}%
\providecommand \@sanitize@url [0]{\catcode `\\12\catcode `\$12\catcode
  `\&12\catcode `\#12\catcode `\^12\catcode `\_12\catcode `\%12\relax}%
\providecommand \@@startlink[1]{}%
\providecommand \@@endlink[0]{}%
\providecommand \url  [0]{\begingroup\@sanitize@url \@url }%
\providecommand \@url [1]{\endgroup\@href {#1}{\urlprefix }}%
\providecommand \urlprefix  [0]{URL }%
\providecommand \Eprint [0]{\href }%
\providecommand \doibase [0]{http://dx.doi.org/}%
\providecommand \selectlanguage [0]{\@gobble}%
\providecommand \bibinfo  [0]{\@secondoftwo}%
\providecommand \bibfield  [0]{\@secondoftwo}%
\providecommand \translation [1]{[#1]}%
\providecommand \BibitemOpen [0]{}%
\providecommand \bibitemStop [0]{}%
\providecommand \bibitemNoStop [0]{.\EOS\space}%
\providecommand \EOS [0]{\spacefactor3000\relax}%
\providecommand \BibitemShut  [1]{\csname bibitem#1\endcsname}%
\let\auto@bib@innerbib\@empty
\bibitem [{\citenamefont {Bass}\ \emph {et~al.}(1962)\citenamefont {Bass},
  \citenamefont {Franken}, \citenamefont {Ward},\ and\ \citenamefont
  {Weinreich}}]{Bass1962}%
  \BibitemOpen
  \bibfield  {author} {\bibinfo {author} {\bibfnamefont {M.}~\bibnamefont
  {Bass}}, \bibinfo {author} {\bibfnamefont {P.~A.}\ \bibnamefont {Franken}},
  \bibinfo {author} {\bibfnamefont {F.}~\bibnamefont {Ward}}, \ and\ \bibinfo
  {author} {\bibfnamefont {G.}~\bibnamefont {Weinreich}},\ }\href {\doibase
  10.1103/PhysRevLett.9.446} {\bibfield  {journal} {\bibinfo  {journal} {Phys.
  Rev. Lett.}\ }\textbf {\bibinfo {volume} {9}},\ \bibinfo {pages} {446}
  (\bibinfo {year} {1962})}\BibitemShut {NoStop}%
\bibitem [{\citenamefont {Ivchenko}(2005)}]{Ivchenko2005}%
  \BibitemOpen
  \bibfield  {author} {\bibinfo {author} {\bibfnamefont {E.~L.}\ \bibnamefont
  {Ivchenko}},\ }\href@noop {} {\emph {\bibinfo {title} {{Optical Spectroscopy
  of Semiconductor Nanostructures}}}}\ (\bibinfo  {publisher} {Alpha Science
  International, Ltd},\ \bibinfo {year} {2005})\ p.\ \bibinfo {pages}
  {350}\BibitemShut {NoStop}%
\bibitem [{\citenamefont {Ghosh}\ and\ \citenamefont
  {Mandel}(1987)}]{Ghosh1987}%
  \BibitemOpen
  \bibfield  {author} {\bibinfo {author} {\bibfnamefont {R.}~\bibnamefont
  {Ghosh}}\ and\ \bibinfo {author} {\bibfnamefont {L.}~\bibnamefont {Mandel}},\
  }\href {\doibase 10.1103/PhysRevLett.59.1903} {\bibfield  {journal} {\bibinfo
   {journal} {Phys. Rev. Lett.}\ }\textbf {\bibinfo {volume} {59}},\ \bibinfo
  {pages} {1903} (\bibinfo {year} {1987})}\BibitemShut {NoStop}%
\bibitem [{\citenamefont {Shih}\ and\ \citenamefont {Alley}(1988)}]{Shih1988}%
  \BibitemOpen
  \bibfield  {author} {\bibinfo {author} {\bibfnamefont {Y.~H.}\ \bibnamefont
  {Shih}}\ and\ \bibinfo {author} {\bibfnamefont {C.~O.}\ \bibnamefont
  {Alley}},\ }\href {\doibase 10.1103/PhysRevLett.61.2921} {\bibfield
  {journal} {\bibinfo  {journal} {Phys. Rev. Lett.}\ }\textbf {\bibinfo
  {volume} {61}},\ \bibinfo {pages} {2921} (\bibinfo {year}
  {1988})}\BibitemShut {NoStop}%
\bibitem [{\citenamefont {Boyd}(2008)}]{Boyd2008}%
  \BibitemOpen
  \bibfield  {author} {\bibinfo {author} {\bibfnamefont {R.~W.}\ \bibnamefont
  {Boyd}},\ }\href
  {http://store.elsevier.com/product.jsp?isbn=9780123694706&_requestid=659054}
  {\emph {\bibinfo {title} {{Nonlinear Optics}}}},\ \bibinfo {edition} {3rd}\
  ed.\ (\bibinfo  {publisher} {Elsevier Science Publishing Co Inc},\ \bibinfo
  {year} {2008})\BibitemShut {NoStop}%
\bibitem [{\citenamefont {Shen}(2002)}]{Shen2002}%
  \BibitemOpen
  \bibfield  {author} {\bibinfo {author} {\bibfnamefont {Y.~R.}\ \bibnamefont
  {Shen}},\ }\href
  {http://as.wiley.com/WileyCDA/WileyTitle/productCd-0471430803.html} {\emph
  {\bibinfo {title} {{The Principles of Nonlinear Optics}}}}\ (\bibinfo {year}
  {2002})\BibitemShut {NoStop}%
\bibitem [{\citenamefont {Hauss{\"u}hl}(2007)}]{Haussuhl2007}%
  \BibitemOpen
  \bibfield  {author} {\bibinfo {author} {\bibfnamefont {S.}~\bibnamefont
  {Hauss{\"u}hl}},\ }\href {\doibase 10.1002/9783527621156} {\emph {\bibinfo
  {title} {{Physical Properties of Crystals}}}},\ edited by\ \bibinfo {editor}
  {\bibfnamefont {S.}~\bibnamefont {Hausshl}}\ (\bibinfo  {publisher}
  {Wiley-VCH Verlag GmbH},\ \bibinfo {address} {Weinheim, Germany},\ \bibinfo
  {year} {2007})\BibitemShut {NoStop}%
\bibitem [{\citenamefont {Zhang}\ \emph {et~al.}(2009)\citenamefont {Zhang},
  \citenamefont {Brar}, \citenamefont {Girit}, \citenamefont {Zettl},\ and\
  \citenamefont {Crommie}}]{Zhang2009}%
  \BibitemOpen
  \bibfield  {author} {\bibinfo {author} {\bibfnamefont {Y.}~\bibnamefont
  {Zhang}}, \bibinfo {author} {\bibfnamefont {V.~W.}\ \bibnamefont {Brar}},
  \bibinfo {author} {\bibfnamefont {C.}~\bibnamefont {Girit}}, \bibinfo
  {author} {\bibfnamefont {A.}~\bibnamefont {Zettl}}, \ and\ \bibinfo {author}
  {\bibfnamefont {M.~F.}\ \bibnamefont {Crommie}},\ }\href {\doibase
  10.1038/nphys1365} {\bibfield  {journal} {\bibinfo  {journal} {Nat. Phys.}\
  }\textbf {\bibinfo {volume} {5}},\ \bibinfo {pages} {722} (\bibinfo {year}
  {2009})}\BibitemShut {NoStop}%
\bibitem [{\citenamefont {Castro}\ \emph {et~al.}(2007)\citenamefont {Castro},
  \citenamefont {Novoselov}, \citenamefont {Morozov}, \citenamefont {Peres},
  \citenamefont {dos Santos}, \citenamefont {Nilsson}, \citenamefont {Guinea},
  \citenamefont {Geim},\ and\ \citenamefont {Neto}}]{Castro2007}%
  \BibitemOpen
  \bibfield  {author} {\bibinfo {author} {\bibfnamefont {E.~V.}\ \bibnamefont
  {Castro}}, \bibinfo {author} {\bibfnamefont {K.~S.}\ \bibnamefont
  {Novoselov}}, \bibinfo {author} {\bibfnamefont {S.~V.}\ \bibnamefont
  {Morozov}}, \bibinfo {author} {\bibfnamefont {N.~M.~R.}\ \bibnamefont
  {Peres}}, \bibinfo {author} {\bibfnamefont {J.~M. B.~L.}\ \bibnamefont {dos
  Santos}}, \bibinfo {author} {\bibfnamefont {J.}~\bibnamefont {Nilsson}},
  \bibinfo {author} {\bibfnamefont {F.}~\bibnamefont {Guinea}}, \bibinfo
  {author} {\bibfnamefont {A.~K.}\ \bibnamefont {Geim}}, \ and\ \bibinfo
  {author} {\bibfnamefont {A.~H.~C.}\ \bibnamefont {Neto}},\ }\href {\doibase
  10.1103/PhysRevLett.99.216802} {\bibfield  {journal} {\bibinfo  {journal}
  {Phys. Rev. Lett.}\ }\textbf {\bibinfo {volume} {99}},\ \bibinfo {pages}
  {216802} (\bibinfo {year} {2007})}\BibitemShut {NoStop}%
\bibitem [{\citenamefont {Das}\ \emph {et~al.}(2008)\citenamefont {Das},
  \citenamefont {Pisana}, \citenamefont {Chakraborty}, \citenamefont
  {Piscanec}, \citenamefont {Saha}, \citenamefont {Waghmare}, \citenamefont
  {K.~S.}, \citenamefont {Krishnamurthy}, \citenamefont {Geim}, \citenamefont
  {Ferrari},\ and\ \citenamefont {Sood}}]{Das2008}%
  \BibitemOpen
  \bibfield  {author} {\bibinfo {author} {\bibfnamefont {A.}~\bibnamefont
  {Das}}, \bibinfo {author} {\bibfnamefont {S.}~\bibnamefont {Pisana}},
  \bibinfo {author} {\bibfnamefont {B.}~\bibnamefont {Chakraborty}}, \bibinfo
  {author} {\bibfnamefont {S.}~\bibnamefont {Piscanec}}, \bibinfo {author}
  {\bibfnamefont {S.~K.}\ \bibnamefont {Saha}}, \bibinfo {author}
  {\bibfnamefont {U.~V.}\ \bibnamefont {Waghmare}}, \bibinfo {author}
  {\bibfnamefont {N.}~\bibnamefont {K.~S.}}, \bibinfo {author} {\bibfnamefont
  {H.~R.}\ \bibnamefont {Krishnamurthy}}, \bibinfo {author} {\bibfnamefont
  {A.~K.}\ \bibnamefont {Geim}}, \bibinfo {author} {\bibfnamefont {A.~C.}\
  \bibnamefont {Ferrari}}, \ and\ \bibinfo {author} {\bibfnamefont {A.~K.}\
  \bibnamefont {Sood}},\ }\href
  {http://www.nature.com/nnano/journal/v3/n4/full/nnano.2008.67.html}
  {\bibfield  {journal} {\bibinfo  {journal} {Nat. Nano.}\ }\textbf {\bibinfo
  {volume} {3}},\ \bibinfo {pages} {210} (\bibinfo {year} {2008})}\BibitemShut
  {NoStop}%
\bibitem [{\citenamefont {Aversa}\ and\ \citenamefont
  {Sipe}(1995)}]{Aversa1995}%
  \BibitemOpen
  \bibfield  {author} {\bibinfo {author} {\bibfnamefont {C.}~\bibnamefont
  {Aversa}}\ and\ \bibinfo {author} {\bibfnamefont {J.~E.}\ \bibnamefont
  {Sipe}},\ }\href {\doibase 10.1103/PhysRevB.52.14636} {\bibfield  {journal}
  {\bibinfo  {journal} {Phys. Rev. B}\ }\textbf {\bibinfo {volume} {52}},\
  \bibinfo {pages} {14636} (\bibinfo {year} {1995})}\BibitemShut {NoStop}%
\bibitem [{\citenamefont {Brun}\ and\ \citenamefont
  {Pedersen}(2015)}]{Brun2015}%
  \BibitemOpen
  \bibfield  {author} {\bibinfo {author} {\bibfnamefont {S.~J.}\ \bibnamefont
  {Brun}}\ and\ \bibinfo {author} {\bibfnamefont {T.~G.}\ \bibnamefont
  {Pedersen}},\ }\href {\doibase 10.1103/PhysRevB.91.205405} {\bibfield
  {journal} {\bibinfo  {journal} {Phys. Rev. B}\ }\textbf {\bibinfo {volume}
  {91}},\ \bibinfo {pages} {205405} (\bibinfo {year} {2015})}\BibitemShut
  {NoStop}%
\bibitem [{\citenamefont {Trolle}\ \emph {et~al.}(2014)\citenamefont {Trolle},
  \citenamefont {Seifert},\ and\ \citenamefont {Pedersen}}]{Trolle2013}%
  \BibitemOpen
  \bibfield  {author} {\bibinfo {author} {\bibfnamefont {M.~L.}\ \bibnamefont
  {Trolle}}, \bibinfo {author} {\bibfnamefont {G.}~\bibnamefont {Seifert}}, \
  and\ \bibinfo {author} {\bibfnamefont {T.~G.}\ \bibnamefont {Pedersen}},\
  }\href {\doibase 10.1103/PhysRevB.89.235410} {\bibfield  {journal} {\bibinfo
  {journal} {Phys. Rev. B}\ }\textbf {\bibinfo {volume} {89}},\ \bibinfo
  {pages} {235410} (\bibinfo {year} {2014})}\BibitemShut {NoStop}%
\bibitem [{\citenamefont {Pedersen}(2015)}]{Pedersen2015}%
  \BibitemOpen
  \bibfield  {author} {\bibinfo {author} {\bibfnamefont {T.~G.}\ \bibnamefont
  {Pedersen}},\ }\href {\doibase 10.1103/PhysRevB.92.235432} {\bibfield
  {journal} {\bibinfo  {journal} {Phys. Rev. B}\ }\textbf {\bibinfo {volume}
  {92}},\ \bibinfo {pages} {235432} (\bibinfo {year} {2015})}\BibitemShut
  {NoStop}%
\bibitem [{\citenamefont {Dresselhaus}\ \emph {et~al.}(2008)\citenamefont
  {Dresselhaus}, \citenamefont {Dresselhaus},\ and\ \citenamefont
  {Jorio}}]{Dresselhaus2008}%
  \BibitemOpen
  \bibfield  {author} {\bibinfo {author} {\bibfnamefont {M.~S.}\ \bibnamefont
  {Dresselhaus}}, \bibinfo {author} {\bibfnamefont {G.}~\bibnamefont
  {Dresselhaus}}, \ and\ \bibinfo {author} {\bibfnamefont {A.}~\bibnamefont
  {Jorio}},\ }\href {\doibase 10.1007/978-3-540-32899-5} {\emph {\bibinfo
  {title} {{Group Theory}}}}\ (\bibinfo  {publisher} {Springer Berlin
  Heidelberg},\ \bibinfo {address} {Berlin, Heidelberg},\ \bibinfo {year}
  {2008})\BibitemShut {NoStop}%
\bibitem [{\citenamefont {Ribeiro}\ and\ \citenamefont
  {Peres}(2011)}]{Ribeiro2011}%
  \BibitemOpen
  \bibfield  {author} {\bibinfo {author} {\bibfnamefont {R.~M.}\ \bibnamefont
  {Ribeiro}}\ and\ \bibinfo {author} {\bibfnamefont {N.~M.~R.}\ \bibnamefont
  {Peres}},\ }\href {\doibase 10.1103/PhysRevB.83.235312} {\bibfield  {journal}
  {\bibinfo  {journal} {Phys. Rev. B}\ }\textbf {\bibinfo {volume} {83}},\
  \bibinfo {pages} {235312} (\bibinfo {year} {2011})}\BibitemShut {NoStop}%
\bibitem [{\citenamefont {Sakurai}(1967)}]{Sakurai}%
  \BibitemOpen
  \bibfield  {author} {\bibinfo {author} {\bibfnamefont {J.~J.}\ \bibnamefont
  {Sakurai}},\ }\href@noop {} {\emph {\bibinfo {title} {{Advanced Quantum
  Physics}}}},\ \bibinfo {edition} {1st}\ ed.\ (\bibinfo  {publisher}
  {Addison-Wesley},\ \bibinfo {address} {Berlin, Heidelberg},\ \bibinfo {year}
  {1967})\BibitemShut {NoStop}%
\bibitem [{\citenamefont {Blount}(1962)}]{Blount1962}%
  \BibitemOpen
  \bibfield  {author} {\bibinfo {author} {\bibfnamefont {E.}~\bibnamefont
  {Blount}},\ }in\ \href {\doibase 10.1016/S0081-1947(08)60459-2} {\emph
  {\bibinfo {booktitle} {{Solid State Phys.}}}},\ Vol.~\bibinfo {volume} {13}\
  (\bibinfo {year} {1962})\ pp.\ \bibinfo {pages} {305--373}\BibitemShut
  {NoStop}%
\bibitem [{\citenamefont {Moss}\ \emph {et~al.}(1990)\citenamefont {Moss},
  \citenamefont {Ghahramani}, \citenamefont {Sipe},\ and\ \citenamefont {van
  Driel}}]{Moss1990}%
  \BibitemOpen
  \bibfield  {author} {\bibinfo {author} {\bibfnamefont {D.~J.}\ \bibnamefont
  {Moss}}, \bibinfo {author} {\bibfnamefont {E.}~\bibnamefont {Ghahramani}},
  \bibinfo {author} {\bibfnamefont {J.~E.}\ \bibnamefont {Sipe}}, \ and\
  \bibinfo {author} {\bibfnamefont {H.~M.}\ \bibnamefont {van Driel}},\ }\href
  {\doibase 10.1103/PhysRevB.41.1542} {\bibfield  {journal} {\bibinfo
  {journal} {Phys. Rev. B}\ }\textbf {\bibinfo {volume} {41}},\ \bibinfo
  {pages} {1542} (\bibinfo {year} {1990})}\BibitemShut {NoStop}%
\bibitem [{\citenamefont {Berry}(1984)}]{Berry1984}%
  \BibitemOpen
  \bibfield  {author} {\bibinfo {author} {\bibfnamefont {M.~V.}\ \bibnamefont
  {Berry}},\ }\href {\doibase 10.1098/rspa.1984.0023} {\bibfield  {journal}
  {\bibinfo  {journal} {Proc. R. Soc. A}\ }\textbf {\bibinfo {volume} {392}},\
  \bibinfo {pages} {45} (\bibinfo {year} {1984})}\BibitemShut {NoStop}%
\bibitem [{\citenamefont {Xiao}\ \emph {et~al.}(2010)\citenamefont {Xiao},
  \citenamefont {Chang},\ and\ \citenamefont {Niu}}]{Xiao2010}%
  \BibitemOpen
  \bibfield  {author} {\bibinfo {author} {\bibfnamefont {D.}~\bibnamefont
  {Xiao}}, \bibinfo {author} {\bibfnamefont {M.-C.}\ \bibnamefont {Chang}}, \
  and\ \bibinfo {author} {\bibfnamefont {Q.}~\bibnamefont {Niu}},\ }\href
  {\doibase 10.1103/RevModPhys.82.1959} {\bibfield  {journal} {\bibinfo
  {journal} {Rev. Mod. Phys.}\ }\textbf {\bibinfo {volume} {82}},\ \bibinfo
  {pages} {1959} (\bibinfo {year} {2010})}\BibitemShut {NoStop}%
\bibitem [{\citenamefont {Gusynin}\ and\ \citenamefont
  {Sharapov}(2006)}]{Gusynin2006}%
  \BibitemOpen
  \bibfield  {author} {\bibinfo {author} {\bibfnamefont {V.~P.}\ \bibnamefont
  {Gusynin}}\ and\ \bibinfo {author} {\bibfnamefont {S.~G.}\ \bibnamefont
  {Sharapov}},\ }\href {\doibase 10.1103/PhysRevB.73.245411} {\bibfield
  {journal} {\bibinfo  {journal} {Phys. Rev. B}\ }\textbf {\bibinfo {volume}
  {73}},\ \bibinfo {pages} {245411} (\bibinfo {year} {2006})}\BibitemShut
  {NoStop}%
\bibitem [{\citenamefont {Peres}\ \emph {et~al.}(2006)\citenamefont {Peres},
  \citenamefont {Guinea},\ and\ \citenamefont {{Castro Neto}}}]{Peres2006a}%
  \BibitemOpen
  \bibfield  {author} {\bibinfo {author} {\bibfnamefont {N.~M.~R.}\
  \bibnamefont {Peres}}, \bibinfo {author} {\bibfnamefont {F.}~\bibnamefont
  {Guinea}}, \ and\ \bibinfo {author} {\bibfnamefont {A.}~\bibnamefont {{Castro
  Neto}}},\ }\href {\doibase 10.1103/PhysRevB.73.125411} {\bibfield  {journal}
  {\bibinfo  {journal} {Phys. Rev. B}\ }\textbf {\bibinfo {volume} {73}},\
  \bibinfo {pages} {125411} (\bibinfo {year} {2006})}\BibitemShut {NoStop}%
\bibitem [{\citenamefont {Stauber}\ \emph
  {et~al.}(2008{\natexlab{a}})\citenamefont {Stauber}, \citenamefont {Peres},\
  and\ \citenamefont {Geim}}]{Stauber2008}%
  \BibitemOpen
  \bibfield  {author} {\bibinfo {author} {\bibfnamefont {T.}~\bibnamefont
  {Stauber}}, \bibinfo {author} {\bibfnamefont {N.~M.~R.}\ \bibnamefont
  {Peres}}, \ and\ \bibinfo {author} {\bibfnamefont {A.~K.}\ \bibnamefont
  {Geim}},\ }\href {\doibase 10.1103/PhysRevB.78.085432} {\bibfield  {journal}
  {\bibinfo  {journal} {Phys. Rev. B}\ }\textbf {\bibinfo {volume} {78}},\
  \bibinfo {pages} {085432} (\bibinfo {year} {2008}{\natexlab{a}})}\BibitemShut
  {NoStop}%
\bibitem [{\citenamefont {Mak}\ \emph {et~al.}(2008)\citenamefont {Mak},
  \citenamefont {Sfeir}, \citenamefont {Wu}, \citenamefont {Lui}, \citenamefont
  {Misewich},\ and\ \citenamefont {Heinz}}]{Mak2008a}%
  \BibitemOpen
  \bibfield  {author} {\bibinfo {author} {\bibfnamefont {K.~F.}\ \bibnamefont
  {Mak}}, \bibinfo {author} {\bibfnamefont {M.~Y.}\ \bibnamefont {Sfeir}},
  \bibinfo {author} {\bibfnamefont {Y.}~\bibnamefont {Wu}}, \bibinfo {author}
  {\bibfnamefont {C.~H.}\ \bibnamefont {Lui}}, \bibinfo {author} {\bibfnamefont
  {J.~A.}\ \bibnamefont {Misewich}}, \ and\ \bibinfo {author} {\bibfnamefont
  {T.~F.}\ \bibnamefont {Heinz}},\ }\href {\doibase
  10.1103/PhysRevLett.101.196405} {\bibfield  {journal} {\bibinfo  {journal}
  {Phys. Rev. Lett.}\ }\textbf {\bibinfo {volume} {101}},\ \bibinfo {pages}
  {196405} (\bibinfo {year} {2008})}\BibitemShut {NoStop}%
\bibitem [{End()}]{EndNote-1}%
  \BibitemOpen
  \href@noop {} {}\bibinfo {note} {Given a linearly polarized electric field
  along $ \hat{\mathbf{e}}_2 $, the photocurrent depends exclusively on the
  real part of the photoconductivity. Hence, the ratio between the electric
  curent per laser intensity $(J/I)$ to the real part of photoconductivity
  tensor in units of $\sigma_2$ reads $\frac{ J/I }{ \mathrm{Re} \, \sigmadc /
  \sigma_2 } = 4 \times 10^{11} \, \sigma_2 / n \varepsilon_0 c_0
  \,\mathrm{nA\,cm/W}$, where $n$ is refrective index, $\varepsilon_0$ the
  vacuum permitivity and $c_0$ the speed of light in vacuum. For hBN or
  graphene based systems, the ratios are $ \approx \lbrace 0.576\,,\; 0.434
  \rbrace $ respectively.}\BibitemShut {Stop}%
\bibitem [{\citenamefont {Sun}\ \emph {et~al.}(2008)\citenamefont {Sun},
  \citenamefont {Wu}, \citenamefont {Divin}, \citenamefont {Li}, \citenamefont
  {Berger}, \citenamefont {de~Heer}, \citenamefont {First},\ and\ \citenamefont
  {Norris}}]{Sun2008}%
  \BibitemOpen
  \bibfield  {author} {\bibinfo {author} {\bibfnamefont {D.}~\bibnamefont
  {Sun}}, \bibinfo {author} {\bibfnamefont {Z.-K.}\ \bibnamefont {Wu}},
  \bibinfo {author} {\bibfnamefont {C.}~\bibnamefont {Divin}}, \bibinfo
  {author} {\bibfnamefont {X.}~\bibnamefont {Li}}, \bibinfo {author}
  {\bibfnamefont {C.}~\bibnamefont {Berger}}, \bibinfo {author} {\bibfnamefont
  {W.~A.}\ \bibnamefont {de~Heer}}, \bibinfo {author} {\bibfnamefont {P.~N.}\
  \bibnamefont {First}}, \ and\ \bibinfo {author} {\bibfnamefont {T.~B.}\
  \bibnamefont {Norris}},\ }\href {\doibase 10.1103/PhysRevLett.101.157402}
  {\bibfield  {journal} {\bibinfo  {journal} {Phys. Rev. Lett.}\ }\textbf
  {\bibinfo {volume} {101}},\ \bibinfo {pages} {157402} (\bibinfo {year}
  {2008})}\BibitemShut {NoStop}%
\bibitem [{\citenamefont {Ruzicka}\ \emph
  {et~al.}(2010{\natexlab{a}})\citenamefont {Ruzicka}, \citenamefont {Wang},
  \citenamefont {Werake}, \citenamefont {Weintrub}, \citenamefont {Loh},\ and\
  \citenamefont {Zhao}}]{Ruzicka2010}%
  \BibitemOpen
  \bibfield  {author} {\bibinfo {author} {\bibfnamefont {B.~A.}\ \bibnamefont
  {Ruzicka}}, \bibinfo {author} {\bibfnamefont {S.}~\bibnamefont {Wang}},
  \bibinfo {author} {\bibfnamefont {L.~K.}\ \bibnamefont {Werake}}, \bibinfo
  {author} {\bibfnamefont {B.}~\bibnamefont {Weintrub}}, \bibinfo {author}
  {\bibfnamefont {K.~P.}\ \bibnamefont {Loh}}, \ and\ \bibinfo {author}
  {\bibfnamefont {H.}~\bibnamefont {Zhao}},\ }\href {\doibase
  10.1103/PhysRevB.82.195414} {\bibfield  {journal} {\bibinfo  {journal} {Phys.
  Rev. B}\ }\textbf {\bibinfo {volume} {82}},\ \bibinfo {pages} {195414}
  (\bibinfo {year} {2010}{\natexlab{a}})}\BibitemShut {NoStop}%
\bibitem [{\citenamefont {Ruzicka}\ \emph
  {et~al.}(2010{\natexlab{b}})\citenamefont {Ruzicka}, \citenamefont {Werake},
  \citenamefont {Zhao}, \citenamefont {Wang},\ and\ \citenamefont
  {Loh}}]{Ruzicka2010a}%
  \BibitemOpen
  \bibfield  {author} {\bibinfo {author} {\bibfnamefont {B.~A.}\ \bibnamefont
  {Ruzicka}}, \bibinfo {author} {\bibfnamefont {L.~K.}\ \bibnamefont {Werake}},
  \bibinfo {author} {\bibfnamefont {H.}~\bibnamefont {Zhao}}, \bibinfo {author}
  {\bibfnamefont {S.}~\bibnamefont {Wang}}, \ and\ \bibinfo {author}
  {\bibfnamefont {K.~P.}\ \bibnamefont {Loh}},\ }\href {\doibase
  10.1063/1.3421541} {\bibfield  {journal} {\bibinfo  {journal} {App. Phys.
  Lett.}\ }\textbf {\bibinfo {volume} {96}},\ \bibinfo {pages} {173106}
  (\bibinfo {year} {2010}{\natexlab{b}})}\BibitemShut {NoStop}%
\bibitem [{\citenamefont {Lui}\ \emph {et~al.}(2010)\citenamefont {Lui},
  \citenamefont {Mak}, \citenamefont {Shan},\ and\ \citenamefont
  {Heinz}}]{Lui2010}%
  \BibitemOpen
  \bibfield  {author} {\bibinfo {author} {\bibfnamefont {C.~H.}\ \bibnamefont
  {Lui}}, \bibinfo {author} {\bibfnamefont {K.~F.}\ \bibnamefont {Mak}},
  \bibinfo {author} {\bibfnamefont {J.}~\bibnamefont {Shan}}, \ and\ \bibinfo
  {author} {\bibfnamefont {T.~F.}\ \bibnamefont {Heinz}},\ }\href {\doibase
  10.1103/PhysRevLett.105.127404} {\bibfield  {journal} {\bibinfo  {journal}
  {Phys. Rev. Lett.}\ }\textbf {\bibinfo {volume} {105}},\ \bibinfo {pages}
  {127404} (\bibinfo {year} {2010})}\BibitemShut {NoStop}%
\bibitem [{\citenamefont {Sun}\ \emph {et~al.}(2012)\citenamefont {Sun},
  \citenamefont {Aivazian}, \citenamefont {Jones}, \citenamefont {Ross},
  \citenamefont {Yao}, \citenamefont {Cobden},\ and\ \citenamefont
  {Xu}}]{Sun2012}%
  \BibitemOpen
  \bibfield  {author} {\bibinfo {author} {\bibfnamefont {D.}~\bibnamefont
  {Sun}}, \bibinfo {author} {\bibfnamefont {G.}~\bibnamefont {Aivazian}},
  \bibinfo {author} {\bibfnamefont {A.~M.}\ \bibnamefont {Jones}}, \bibinfo
  {author} {\bibfnamefont {J.~S.}\ \bibnamefont {Ross}}, \bibinfo {author}
  {\bibfnamefont {W.}~\bibnamefont {Yao}}, \bibinfo {author} {\bibfnamefont
  {D.}~\bibnamefont {Cobden}}, \ and\ \bibinfo {author} {\bibfnamefont
  {X.}~\bibnamefont {Xu}},\ }\href {\doibase 10.1038/nnano.2011.243} {\bibfield
   {journal} {\bibinfo  {journal} {Nat. Nanotechnol.}\ }\textbf {\bibinfo
  {volume} {7}},\ \bibinfo {pages} {114} (\bibinfo {year} {2012})}\BibitemShut
  {NoStop}%
\bibitem [{\citenamefont {Tielrooij}\ \emph {et~al.}(2013)\citenamefont
  {Tielrooij}, \citenamefont {Song}, \citenamefont {Jensen}, \citenamefont
  {Centeno}, \citenamefont {Pesquera}, \citenamefont {{Zurutuza Elorza}},
  \citenamefont {Bonn}, \citenamefont {Levitov},\ and\ \citenamefont
  {Koppens}}]{Tielrooij2013}%
  \BibitemOpen
  \bibfield  {author} {\bibinfo {author} {\bibfnamefont {K.~J.}\ \bibnamefont
  {Tielrooij}}, \bibinfo {author} {\bibfnamefont {J.~C.~W.}\ \bibnamefont
  {Song}}, \bibinfo {author} {\bibfnamefont {S.~A.}\ \bibnamefont {Jensen}},
  \bibinfo {author} {\bibfnamefont {A.}~\bibnamefont {Centeno}}, \bibinfo
  {author} {\bibfnamefont {A.}~\bibnamefont {Pesquera}}, \bibinfo {author}
  {\bibfnamefont {A.}~\bibnamefont {{Zurutuza Elorza}}}, \bibinfo {author}
  {\bibfnamefont {M.}~\bibnamefont {Bonn}}, \bibinfo {author} {\bibfnamefont
  {L.~S.}\ \bibnamefont {Levitov}}, \ and\ \bibinfo {author} {\bibfnamefont
  {F.~H.~L.}\ \bibnamefont {Koppens}},\ }\href {\doibase 10.1038/nphys2564}
  {\bibfield  {journal} {\bibinfo  {journal} {Nat. Phys.}\ }\textbf {\bibinfo
  {volume} {9}},\ \bibinfo {pages} {248} (\bibinfo {year} {2013})}\BibitemShut
  {NoStop}%
\bibitem [{\citenamefont {Jorio}\ \emph {et~al.}(2011)\citenamefont {Jorio},
  \citenamefont {Saito}, \citenamefont {Dresselhaus},\ and\ \citenamefont
  {Dresselhaus}}]{Jorio2011a}%
  \BibitemOpen
  \bibfield  {author} {\bibinfo {author} {\bibfnamefont {A.}~\bibnamefont
  {Jorio}}, \bibinfo {author} {\bibfnamefont {R.}~\bibnamefont {Saito}},
  \bibinfo {author} {\bibfnamefont {G.}~\bibnamefont {Dresselhaus}}, \ and\
  \bibinfo {author} {\bibfnamefont {M.~S.}\ \bibnamefont {Dresselhaus}},\
  }\href {\doibase 10.1002/9783527632695} {\emph {\bibinfo {title} {{Raman
  Spectroscopy in Graphene Related Systems}}}}\ (\bibinfo  {publisher}
  {Wiley-VCH Verlag GmbH \& Co. KGaA},\ \bibinfo {address} {Weinheim,
  Germany},\ \bibinfo {year} {2011})\BibitemShut {NoStop}%
\bibitem [{\citenamefont {Giovannetti}\ \emph {et~al.}(2007)\citenamefont
  {Giovannetti}, \citenamefont {Khomyakov}, \citenamefont {Brocks},
  \citenamefont {Kelly},\ and\ \citenamefont {van~den
  Brink}}]{Giovannetti2007}%
  \BibitemOpen
  \bibfield  {author} {\bibinfo {author} {\bibfnamefont {G.}~\bibnamefont
  {Giovannetti}}, \bibinfo {author} {\bibfnamefont {P.~A.}\ \bibnamefont
  {Khomyakov}}, \bibinfo {author} {\bibfnamefont {G.}~\bibnamefont {Brocks}},
  \bibinfo {author} {\bibfnamefont {P.~J.}\ \bibnamefont {Kelly}}, \ and\
  \bibinfo {author} {\bibfnamefont {J.}~\bibnamefont {van~den Brink}},\ }\href
  {\doibase 10.1103/PhysRevB.76.073103} {\bibfield  {journal} {\bibinfo
  {journal} {Phys. Rev. B}\ }\textbf {\bibinfo {volume} {76}},\ \bibinfo
  {pages} {073103} (\bibinfo {year} {2007})}\BibitemShut {NoStop}%
\bibitem [{\citenamefont {Zhou}\ \emph {et~al.}(2007)\citenamefont {Zhou},
  \citenamefont {Gweon}, \citenamefont {Fedorov}, \citenamefont {First},
  \citenamefont {de~Heer}, \citenamefont {Lee}, \citenamefont {Guinea},
  \citenamefont {{Castro Neto}},\ and\ \citenamefont {Lanzara}}]{Zhou2007}%
  \BibitemOpen
  \bibfield  {author} {\bibinfo {author} {\bibfnamefont {S.~Y.}\ \bibnamefont
  {Zhou}}, \bibinfo {author} {\bibfnamefont {G.-H.}\ \bibnamefont {Gweon}},
  \bibinfo {author} {\bibfnamefont {a.~V.}\ \bibnamefont {Fedorov}}, \bibinfo
  {author} {\bibfnamefont {P.~N.}\ \bibnamefont {First}}, \bibinfo {author}
  {\bibfnamefont {W.~A.}\ \bibnamefont {de~Heer}}, \bibinfo {author}
  {\bibfnamefont {D.-H.}\ \bibnamefont {Lee}}, \bibinfo {author} {\bibfnamefont
  {F.}~\bibnamefont {Guinea}}, \bibinfo {author} {\bibfnamefont {A.~H.}\
  \bibnamefont {{Castro Neto}}}, \ and\ \bibinfo {author} {\bibfnamefont
  {A.}~\bibnamefont {Lanzara}},\ }\href {\doibase 10.1038/nmat2056} {\bibfield
  {journal} {\bibinfo  {journal} {Nat. Mater.}\ }\textbf {\bibinfo {volume}
  {6}},\ \bibinfo {pages} {916} (\bibinfo {year} {2007})}\BibitemShut {NoStop}%
\bibitem [{\citenamefont {S{\l}awi{\'n}ska}\ \emph {et~al.}(2010)\citenamefont
  {S{\l}awi{\'n}ska}, \citenamefont {Zasada},\ and\ \citenamefont
  {Klusek}}]{Slawinska2010}%
  \BibitemOpen
  \bibfield  {author} {\bibinfo {author} {\bibfnamefont {J.}~\bibnamefont
  {S{\l}awi{\'n}ska}}, \bibinfo {author} {\bibfnamefont {I.}~\bibnamefont
  {Zasada}}, \ and\ \bibinfo {author} {\bibfnamefont {Z.}~\bibnamefont
  {Klusek}},\ }\href {\doibase 10.1103/PhysRevB.81.155433} {\bibfield
  {journal} {\bibinfo  {journal} {Phys. Rev. B}\ }\textbf {\bibinfo {volume}
  {81}},\ \bibinfo {pages} {155433} (\bibinfo {year} {2010})}\BibitemShut
  {NoStop}%
\bibitem [{\citenamefont {Dean}\ \emph {et~al.}(2010)\citenamefont {Dean},
  \citenamefont {Young}, \citenamefont {Meric}, \citenamefont {Lee},
  \citenamefont {Wang}, \citenamefont {Sorgenfrei}, \citenamefont {Watanabe},
  \citenamefont {Taniguchi}, \citenamefont {Kim}, \citenamefont {Shepard},\
  and\ \citenamefont {Hone}}]{Dean2010}%
  \BibitemOpen
  \bibfield  {author} {\bibinfo {author} {\bibfnamefont {C.~R.}\ \bibnamefont
  {Dean}}, \bibinfo {author} {\bibfnamefont {A.~F.}\ \bibnamefont {Young}},
  \bibinfo {author} {\bibfnamefont {I.}~\bibnamefont {Meric}}, \bibinfo
  {author} {\bibfnamefont {C.}~\bibnamefont {Lee}}, \bibinfo {author}
  {\bibfnamefont {L.}~\bibnamefont {Wang}}, \bibinfo {author} {\bibfnamefont
  {S.}~\bibnamefont {Sorgenfrei}}, \bibinfo {author} {\bibfnamefont
  {K.}~\bibnamefont {Watanabe}}, \bibinfo {author} {\bibfnamefont
  {T.}~\bibnamefont {Taniguchi}}, \bibinfo {author} {\bibfnamefont
  {P.}~\bibnamefont {Kim}}, \bibinfo {author} {\bibfnamefont {K.~L.}\
  \bibnamefont {Shepard}}, \ and\ \bibinfo {author} {\bibfnamefont
  {J.}~\bibnamefont {Hone}},\ }\href {\doibase 10.1038/nnano.2010.172}
  {\bibfield  {journal} {\bibinfo  {journal} {Nat. Nanotechnol.}\ }\textbf
  {\bibinfo {volume} {5}},\ \bibinfo {pages} {722} (\bibinfo {year}
  {2010})}\BibitemShut {NoStop}%
\bibitem [{\citenamefont {Ortix}\ \emph {et~al.}(2012)\citenamefont {Ortix},
  \citenamefont {Yang},\ and\ \citenamefont {van~den Brink}}]{Otrix2012}%
  \BibitemOpen
  \bibfield  {author} {\bibinfo {author} {\bibfnamefont {C.}~\bibnamefont
  {Ortix}}, \bibinfo {author} {\bibfnamefont {L.}~\bibnamefont {Yang}}, \ and\
  \bibinfo {author} {\bibfnamefont {J.}~\bibnamefont {van~den Brink}},\ }\href
  {http://link.aps.org/doi/10.1103/PhysRevB.86.081405} {\bibfield  {journal}
  {\bibinfo  {journal} {Phys. Rev. B}\ }\textbf {\bibinfo {volume} {86}},\
  \bibinfo {pages} {081405} (\bibinfo {year} {2012})}\BibitemShut {NoStop}%
\bibitem [{\citenamefont {Chen}\ \emph {et~al.}(2014)\citenamefont {Chen},
  \citenamefont {Shi}, \citenamefont {Yang}, \citenamefont {Lu}, \citenamefont
  {Lai}, \citenamefont {Yan}, \citenamefont {Wang}, \citenamefont {Zhang},\
  and\ \citenamefont {Li}}]{ZhiGuo2014}%
  \BibitemOpen
  \bibfield  {author} {\bibinfo {author} {\bibfnamefont {Z.-G.}\ \bibnamefont
  {Chen}}, \bibinfo {author} {\bibfnamefont {Z.}~\bibnamefont {Shi}}, \bibinfo
  {author} {\bibfnamefont {W.}~\bibnamefont {Yang}}, \bibinfo {author}
  {\bibfnamefont {X.}~\bibnamefont {Lu}}, \bibinfo {author} {\bibfnamefont
  {Y.}~\bibnamefont {Lai}}, \bibinfo {author} {\bibfnamefont {H.}~\bibnamefont
  {Yan}}, \bibinfo {author} {\bibfnamefont {F.}~\bibnamefont {Wang}}, \bibinfo
  {author} {\bibfnamefont {G.}~\bibnamefont {Zhang}}, \ and\ \bibinfo {author}
  {\bibfnamefont {Z.}~\bibnamefont {Li}},\ }\href
  {http://dx.doi.org/10.1038/ncomms5461} {\bibfield  {journal} {\bibinfo
  {journal} {Nat. Commun.}\ }\textbf {\bibinfo {volume} {5}},\ \bibinfo {pages}
  {4461} (\bibinfo {year} {2014})}\BibitemShut {NoStop}%
\bibitem [{\citenamefont {Bokdam}\ \emph {et~al.}(2014)\citenamefont {Bokdam},
  \citenamefont {Amlaki}, \citenamefont {Brocks},\ and\ \citenamefont
  {Kelly}}]{Bokdam2014}%
  \BibitemOpen
  \bibfield  {author} {\bibinfo {author} {\bibfnamefont {M.}~\bibnamefont
  {Bokdam}}, \bibinfo {author} {\bibfnamefont {T.}~\bibnamefont {Amlaki}},
  \bibinfo {author} {\bibfnamefont {G.}~\bibnamefont {Brocks}}, \ and\ \bibinfo
  {author} {\bibfnamefont {P.~J.}\ \bibnamefont {Kelly}},\ }\href {\doibase
  10.1103/PhysRevB.89.201404} {\bibfield  {journal} {\bibinfo  {journal} {Phys.
  Rev. B}\ }\textbf {\bibinfo {volume} {89}},\ \bibinfo {pages} {201404}
  (\bibinfo {year} {2014})}\BibitemShut {NoStop}%
\bibitem [{\citenamefont {Woods}\ \emph {et~al.}(2014)\citenamefont {Woods},
  \citenamefont {Britnell}, \citenamefont {Eckmann}, \citenamefont {Ma},
  \citenamefont {Lu}, \citenamefont {Guo}, \citenamefont {Lin}, \citenamefont
  {Yu}, \citenamefont {Cao}, \citenamefont {Gorbachev}, \citenamefont
  {Kretinin}, \citenamefont {Park}, \citenamefont {Ponomarenko}, \citenamefont
  {Katsnelson}, \citenamefont {Gornostyrev}, \citenamefont {Watanabe},
  \citenamefont {Taniguchi}, \citenamefont {Casiraghi}, \citenamefont {Gao},
  \citenamefont {Geim},\ and\ \citenamefont {Novoselov}}]{Woods2014}%
  \BibitemOpen
  \bibfield  {author} {\bibinfo {author} {\bibfnamefont {C.~R.}\ \bibnamefont
  {Woods}}, \bibinfo {author} {\bibfnamefont {L.}~\bibnamefont {Britnell}},
  \bibinfo {author} {\bibfnamefont {A.}~\bibnamefont {Eckmann}}, \bibinfo
  {author} {\bibfnamefont {R.~S.}\ \bibnamefont {Ma}}, \bibinfo {author}
  {\bibfnamefont {J.~C.}\ \bibnamefont {Lu}}, \bibinfo {author} {\bibfnamefont
  {H.~M.}\ \bibnamefont {Guo}}, \bibinfo {author} {\bibfnamefont
  {X.}~\bibnamefont {Lin}}, \bibinfo {author} {\bibfnamefont {G.~L.}\
  \bibnamefont {Yu}}, \bibinfo {author} {\bibfnamefont {Y.}~\bibnamefont
  {Cao}}, \bibinfo {author} {\bibfnamefont {R.~V.}\ \bibnamefont {Gorbachev}},
  \bibinfo {author} {\bibfnamefont {A.~V.}\ \bibnamefont {Kretinin}}, \bibinfo
  {author} {\bibfnamefont {J.}~\bibnamefont {Park}}, \bibinfo {author}
  {\bibfnamefont {L.~A.}\ \bibnamefont {Ponomarenko}}, \bibinfo {author}
  {\bibfnamefont {M.~I.}\ \bibnamefont {Katsnelson}}, \bibinfo {author}
  {\bibfnamefont {Y.~N.}\ \bibnamefont {Gornostyrev}}, \bibinfo {author}
  {\bibfnamefont {K.}~\bibnamefont {Watanabe}}, \bibinfo {author}
  {\bibfnamefont {T.}~\bibnamefont {Taniguchi}}, \bibinfo {author}
  {\bibfnamefont {C.}~\bibnamefont {Casiraghi}}, \bibinfo {author}
  {\bibfnamefont {H.-j.}\ \bibnamefont {Gao}}, \bibinfo {author} {\bibfnamefont
  {A.~K.}\ \bibnamefont {Geim}}, \ and\ \bibinfo {author} {\bibfnamefont
  {K.~S.}\ \bibnamefont {Novoselov}},\ }\href {\doibase 10.1038/nphys2954}
  {\bibfield  {journal} {\bibinfo  {journal} {Nat. Phys.}\ }\textbf {\bibinfo
  {volume} {10}},\ \bibinfo {pages} {451} (\bibinfo {year} {2014})}\BibitemShut
  {NoStop}%
\bibitem [{\citenamefont {Yankowitz}\ \emph {et~al.}(2014)\citenamefont
  {Yankowitz}, \citenamefont {Xue},\ and\ \citenamefont
  {LeRoy}}]{Yankowitz2014}%
  \BibitemOpen
  \bibfield  {author} {\bibinfo {author} {\bibfnamefont {M.}~\bibnamefont
  {Yankowitz}}, \bibinfo {author} {\bibfnamefont {J.}~\bibnamefont {Xue}}, \
  and\ \bibinfo {author} {\bibfnamefont {B.~J.}\ \bibnamefont {LeRoy}},\ }\href
  {\doibase 10.1088/0953-8984/26/30/303201} {\bibfield  {journal} {\bibinfo
  {journal} {J. Phys. Condens. Matter}\ }\textbf {\bibinfo {volume} {26}},\
  \bibinfo {pages} {303201} (\bibinfo {year} {2014})}\BibitemShut {NoStop}%
\bibitem [{\citenamefont {Huang}\ \emph {et~al.}(2014)\citenamefont {Huang},
  \citenamefont {Yue}, \citenamefont {Kang}, \citenamefont {Li},\ and\
  \citenamefont {Jingbo}}]{Huang2014}%
  \BibitemOpen
  \bibfield  {author} {\bibinfo {author} {\bibfnamefont {L.}~\bibnamefont
  {Huang}}, \bibinfo {author} {\bibfnamefont {Q.}~\bibnamefont {Yue}}, \bibinfo
  {author} {\bibfnamefont {J.}~\bibnamefont {Kang}}, \bibinfo {author}
  {\bibfnamefont {Y.}~\bibnamefont {Li}}, \ and\ \bibinfo {author}
  {\bibfnamefont {L.}~\bibnamefont {Jingbo}},\ }\href {\doibase
  10.1088/0953-8984/26/29/295304} {\bibfield  {journal} {\bibinfo  {journal}
  {J. Phys. Condens. Matter}\ }\textbf {\bibinfo {volume} {26}},\ \bibinfo
  {pages} {295304} (\bibinfo {year} {2014})}\BibitemShut {NoStop}%
\bibitem [{\citenamefont {Pedersen}\ and\ \citenamefont
  {Pedersen}(2009)}]{Pedersen2009}%
  \BibitemOpen
  \bibfield  {author} {\bibinfo {author} {\bibfnamefont {T.~G.}\ \bibnamefont
  {Pedersen}}\ and\ \bibinfo {author} {\bibfnamefont {K.}~\bibnamefont
  {Pedersen}},\ }\href {\doibase 10.1103/PhysRevB.79.035422} {\bibfield
  {journal} {\bibinfo  {journal} {Phys. Rev. B}\ }\textbf {\bibinfo {volume}
  {79}},\ \bibinfo {pages} {035422} (\bibinfo {year} {2009})}\BibitemShut
  {NoStop}%
\bibitem [{\citenamefont {McCann}\ and\ \citenamefont
  {Fal'ko}(2006)}]{McCann2006}%
  \BibitemOpen
  \bibfield  {author} {\bibinfo {author} {\bibfnamefont {E.}~\bibnamefont
  {McCann}}\ and\ \bibinfo {author} {\bibfnamefont {V.~I.}\ \bibnamefont
  {Fal'ko}},\ }\href {\doibase 10.1103/PhysRevLett.96.086805} {\bibfield
  {journal} {\bibinfo  {journal} {Phys. Rev. Lett.}\ }\textbf {\bibinfo
  {volume} {96}},\ \bibinfo {pages} {086805} (\bibinfo {year}
  {2006})}\BibitemShut {NoStop}%
\bibitem [{\citenamefont {Szafranek}\ \emph {et~al.}(2011)\citenamefont
  {Szafranek}, \citenamefont {Schall}, \citenamefont {Otto}, \citenamefont
  {Neumaier},\ and\ \citenamefont {Kurz}}]{Szafranek2011}%
  \BibitemOpen
  \bibfield  {author} {\bibinfo {author} {\bibfnamefont {B.~N.}\ \bibnamefont
  {Szafranek}}, \bibinfo {author} {\bibfnamefont {D.}~\bibnamefont {Schall}},
  \bibinfo {author} {\bibfnamefont {M.}~\bibnamefont {Otto}}, \bibinfo {author}
  {\bibfnamefont {D.}~\bibnamefont {Neumaier}}, \ and\ \bibinfo {author}
  {\bibfnamefont {H.}~\bibnamefont {Kurz}},\ }\href {\doibase
  10.1021/nl200631m} {\bibfield  {journal} {\bibinfo  {journal} {Nano Lett.}\
  }\textbf {\bibinfo {volume} {11}},\ \bibinfo {pages} {2640} (\bibinfo {year}
  {2011})}\BibitemShut {NoStop}%
\bibitem [{\citenamefont {Velasco}\ \emph {et~al.}(2012)\citenamefont
  {Velasco}, \citenamefont {Jing}, \citenamefont {Bao}, \citenamefont {Lee},
  \citenamefont {Kratz}, \citenamefont {Aji}, \citenamefont {Bockrath},
  \citenamefont {Lau}, \citenamefont {Varma}, \citenamefont {Stillwell},
  \citenamefont {Smirnov}, \citenamefont {Zhang}, \citenamefont {Jung},\ and\
  \citenamefont {MacDonald}}]{Velasco2012}%
  \BibitemOpen
  \bibfield  {author} {\bibinfo {author} {\bibfnamefont {J.}~\bibnamefont
  {Velasco}}, \bibinfo {author} {\bibfnamefont {L.}~\bibnamefont {Jing}},
  \bibinfo {author} {\bibfnamefont {W.}~\bibnamefont {Bao}}, \bibinfo {author}
  {\bibfnamefont {Y.}~\bibnamefont {Lee}}, \bibinfo {author} {\bibfnamefont
  {P.}~\bibnamefont {Kratz}}, \bibinfo {author} {\bibfnamefont
  {V.}~\bibnamefont {Aji}}, \bibinfo {author} {\bibfnamefont {M.}~\bibnamefont
  {Bockrath}}, \bibinfo {author} {\bibfnamefont {C.~N.}\ \bibnamefont {Lau}},
  \bibinfo {author} {\bibfnamefont {C.}~\bibnamefont {Varma}}, \bibinfo
  {author} {\bibfnamefont {R.}~\bibnamefont {Stillwell}}, \bibinfo {author}
  {\bibfnamefont {D.}~\bibnamefont {Smirnov}}, \bibinfo {author} {\bibfnamefont
  {F.}~\bibnamefont {Zhang}}, \bibinfo {author} {\bibfnamefont
  {J.}~\bibnamefont {Jung}}, \ and\ \bibinfo {author} {\bibfnamefont {A.~H.}\
  \bibnamefont {MacDonald}},\ }\href {\doibase 10.1038/nnano.2011.251}
  {\bibfield  {journal} {\bibinfo  {journal} {Nat. Nanotechnol.}\ }\textbf
  {\bibinfo {volume} {7}},\ \bibinfo {pages} {156} (\bibinfo {year}
  {2012})}\BibitemShut {NoStop}%
\bibitem [{\citenamefont {Pachoud}\ \emph {et~al.}(2010)\citenamefont
  {Pachoud}, \citenamefont {Jaiswal}, \citenamefont {Ang}, \citenamefont
  {Loh},\ and\ \citenamefont {Oezyilmaz}}]{Pachoud2010}%
  \BibitemOpen
  \bibfield  {author} {\bibinfo {author} {\bibfnamefont {A.}~\bibnamefont
  {Pachoud}}, \bibinfo {author} {\bibfnamefont {M.}~\bibnamefont {Jaiswal}},
  \bibinfo {author} {\bibfnamefont {P.~K.}\ \bibnamefont {Ang}}, \bibinfo
  {author} {\bibfnamefont {K.~P.}\ \bibnamefont {Loh}}, \ and\ \bibinfo
  {author} {\bibfnamefont {B.}~\bibnamefont {Oezyilmaz}},\ }\href {\doibase
  10.1209/0295-5075/92/27001} {\bibfield  {journal} {\bibinfo  {journal} {EPL}\
  }\textbf {\bibinfo {volume} {92}},\ \bibinfo {pages} {27001} (\bibinfo {year}
  {2010})}\BibitemShut {NoStop}%
\bibitem [{\citenamefont {Britnell}\ \emph {et~al.}(2012)\citenamefont
  {Britnell}, \citenamefont {Gorbachev}, \citenamefont {Jalil}, \citenamefont
  {Belle}, \citenamefont {Schedin}, \citenamefont {Mishchenko}, \citenamefont
  {Georgiou}, \citenamefont {Katsnelson}, \citenamefont {Eaves}, \citenamefont
  {Morozov}, \citenamefont {Peres}, \citenamefont {Leist}, \citenamefont
  {Geim}, \citenamefont {Novoselov},\ and\ \citenamefont
  {Ponomarenko}}]{Britnell2012}%
  \BibitemOpen
  \bibfield  {author} {\bibinfo {author} {\bibfnamefont {L.}~\bibnamefont
  {Britnell}}, \bibinfo {author} {\bibfnamefont {R.~V.}\ \bibnamefont
  {Gorbachev}}, \bibinfo {author} {\bibfnamefont {R.}~\bibnamefont {Jalil}},
  \bibinfo {author} {\bibfnamefont {B.~D.}\ \bibnamefont {Belle}}, \bibinfo
  {author} {\bibfnamefont {F.}~\bibnamefont {Schedin}}, \bibinfo {author}
  {\bibfnamefont {A.}~\bibnamefont {Mishchenko}}, \bibinfo {author}
  {\bibfnamefont {T.}~\bibnamefont {Georgiou}}, \bibinfo {author}
  {\bibfnamefont {M.~I.}\ \bibnamefont {Katsnelson}}, \bibinfo {author}
  {\bibfnamefont {L.}~\bibnamefont {Eaves}}, \bibinfo {author} {\bibfnamefont
  {S.~V.}\ \bibnamefont {Morozov}}, \bibinfo {author} {\bibfnamefont
  {N.~M.~R.}\ \bibnamefont {Peres}}, \bibinfo {author} {\bibfnamefont
  {J.}~\bibnamefont {Leist}}, \bibinfo {author} {\bibfnamefont {A.~K.}\
  \bibnamefont {Geim}}, \bibinfo {author} {\bibfnamefont {K.~S.}\ \bibnamefont
  {Novoselov}}, \ and\ \bibinfo {author} {\bibfnamefont {L.~A.}\ \bibnamefont
  {Ponomarenko}},\ }\href {\doibase 10.1126/science.1218461} {\bibfield
  {journal} {\bibinfo  {journal} {Sci.}\ }\textbf {\bibinfo {volume} {335}},\
  \bibinfo {pages} {947} (\bibinfo {year} {2012})}\BibitemShut {NoStop}%
\bibitem [{\citenamefont {Stabile}\ \emph {et~al.}(2015)\citenamefont
  {Stabile}, \citenamefont {Ferreira}, \citenamefont {Li}, \citenamefont
  {Peres},\ and\ \citenamefont {Zhu}}]{Stabile2015}%
  \BibitemOpen
  \bibfield  {author} {\bibinfo {author} {\bibfnamefont {A.~A.}\ \bibnamefont
  {Stabile}}, \bibinfo {author} {\bibfnamefont {A.}~\bibnamefont {Ferreira}},
  \bibinfo {author} {\bibfnamefont {J.}~\bibnamefont {Li}}, \bibinfo {author}
  {\bibfnamefont {N.~M.~R.}\ \bibnamefont {Peres}}, \ and\ \bibinfo {author}
  {\bibfnamefont {J.}~\bibnamefont {Zhu}},\ }\href {\doibase
  10.1103/PhysRevB.92.121411} {\bibfield  {journal} {\bibinfo  {journal} {Phys.
  Rev. B}\ }\textbf {\bibinfo {volume} {92}},\ \bibinfo {pages} {121411}
  (\bibinfo {year} {2015})}\BibitemShut {NoStop}%
\bibitem [{\citenamefont {Peres}\ \emph {et~al.}(2008)\citenamefont {Peres},
  \citenamefont {Stauber},\ and\ \citenamefont {{Castro Neto}}}]{Peres2008a}%
  \BibitemOpen
  \bibfield  {author} {\bibinfo {author} {\bibfnamefont {N.~M.~R.}\
  \bibnamefont {Peres}}, \bibinfo {author} {\bibfnamefont {T.}~\bibnamefont
  {Stauber}}, \ and\ \bibinfo {author} {\bibfnamefont {A.}~\bibnamefont
  {{Castro Neto}}},\ }\href {\doibase 10.1209/0295-5075/84/38002} {\bibfield
  {journal} {\bibinfo  {journal} {EPL}\ }\textbf {\bibinfo {volume} {84}},\
  \bibinfo {pages} {38002} (\bibinfo {year} {2008})}\BibitemShut {NoStop}%
\bibitem [{\citenamefont {Stauber}\ \emph
  {et~al.}(2008{\natexlab{b}})\citenamefont {Stauber}, \citenamefont {Peres},\
  and\ \citenamefont {Castro~Neto}}]{Stauber2008b}%
  \BibitemOpen
  \bibfield  {author} {\bibinfo {author} {\bibfnamefont {T.}~\bibnamefont
  {Stauber}}, \bibinfo {author} {\bibfnamefont {N.~M.~R.}\ \bibnamefont
  {Peres}}, \ and\ \bibinfo {author} {\bibfnamefont {A.~H.}\ \bibnamefont
  {Castro~Neto}},\ }\href {\doibase 10.1103/PhysRevB.78.085418} {\bibfield
  {journal} {\bibinfo  {journal} {Phys. Rev. B}\ }\textbf {\bibinfo {volume}
  {78}},\ \bibinfo {pages} {085418} (\bibinfo {year}
  {2008}{\natexlab{b}})}\BibitemShut {NoStop}%
\bibitem [{\citenamefont {Stauber}\ and\ \citenamefont
  {Peres}(2008)}]{Stauber2008a}%
  \BibitemOpen
  \bibfield  {author} {\bibinfo {author} {\bibfnamefont {T.}~\bibnamefont
  {Stauber}}\ and\ \bibinfo {author} {\bibfnamefont {N.~M.~R.}\ \bibnamefont
  {Peres}},\ }\href {\doibase 10.1088/0953-8984/20/5/055002} {\bibfield
  {journal} {\bibinfo  {journal} {J. Phys. Condens. Matter}\ }\textbf {\bibinfo
  {volume} {20}},\ \bibinfo {pages} {055002} (\bibinfo {year}
  {2008})}\BibitemShut {NoStop}%
\bibitem [{\citenamefont {Abergel}\ \emph {et~al.}(2012)\citenamefont
  {Abergel}, \citenamefont {Min}, \citenamefont {Hwang},\ and\ \citenamefont
  {{Das Sarma}}}]{Abergel2012}%
  \BibitemOpen
  \bibfield  {author} {\bibinfo {author} {\bibfnamefont {D.~S.~L.}\
  \bibnamefont {Abergel}}, \bibinfo {author} {\bibfnamefont {H.}~\bibnamefont
  {Min}}, \bibinfo {author} {\bibfnamefont {E.~H.}\ \bibnamefont {Hwang}}, \
  and\ \bibinfo {author} {\bibfnamefont {S.}~\bibnamefont {{Das Sarma}}},\
  }\href {\doibase 10.1103/PhysRevB.85.045411} {\bibfield  {journal} {\bibinfo
  {journal} {Phys. Rev. B}\ }\textbf {\bibinfo {volume} {85}},\ \bibinfo
  {pages} {045411} (\bibinfo {year} {2012})}\BibitemShut {NoStop}%
\bibitem [{\citenamefont {Yuan}\ \emph {et~al.}(2011)\citenamefont {Yuan},
  \citenamefont {Rold{\'a}n}, \citenamefont {{De Raedt}},\ and\ \citenamefont
  {Katsnelson}}]{Yuan2011}%
  \BibitemOpen
  \bibfield  {author} {\bibinfo {author} {\bibfnamefont {S.}~\bibnamefont
  {Yuan}}, \bibinfo {author} {\bibfnamefont {R.}~\bibnamefont {Rold{\'a}n}},
  \bibinfo {author} {\bibfnamefont {H.}~\bibnamefont {{De Raedt}}}, \ and\
  \bibinfo {author} {\bibfnamefont {M.~I.}\ \bibnamefont {Katsnelson}},\ }\href
  {\doibase 10.1103/PhysRevB.84.195418} {\bibfield  {journal} {\bibinfo
  {journal} {Phys. Rev. B}\ }\textbf {\bibinfo {volume} {84}},\ \bibinfo
  {pages} {195418} (\bibinfo {year} {2011})}\BibitemShut {NoStop}%
\bibitem [{\citenamefont {Gannett}\ \emph {et~al.}(2011)\citenamefont
  {Gannett}, \citenamefont {Regan}, \citenamefont {Watanabe}, \citenamefont
  {Taniguchi}, \citenamefont {Crommie},\ and\ \citenamefont
  {Zettl}}]{Gannett2011}%
  \BibitemOpen
  \bibfield  {author} {\bibinfo {author} {\bibfnamefont {W.}~\bibnamefont
  {Gannett}}, \bibinfo {author} {\bibfnamefont {W.}~\bibnamefont {Regan}},
  \bibinfo {author} {\bibfnamefont {K.}~\bibnamefont {Watanabe}}, \bibinfo
  {author} {\bibfnamefont {T.}~\bibnamefont {Taniguchi}}, \bibinfo {author}
  {\bibfnamefont {M.~F.}\ \bibnamefont {Crommie}}, \ and\ \bibinfo {author}
  {\bibfnamefont {A.}~\bibnamefont {Zettl}},\ }\href {\doibase
  10.1063/1.3599708} {\bibfield  {journal} {\bibinfo  {journal} {Appl. Phys.
  Lett.}\ }\textbf {\bibinfo {volume} {98}},\ \bibinfo {pages} {242105}
  (\bibinfo {year} {2011})}\BibitemShut {NoStop}%
\bibitem [{\citenamefont {Chari}\ \emph {et~al.}(2015)\citenamefont {Chari},
  \citenamefont {Meric}, \citenamefont {Dean},\ and\ \citenamefont
  {Shepard}}]{Chari2015}%
  \BibitemOpen
  \bibfield  {author} {\bibinfo {author} {\bibfnamefont {T.}~\bibnamefont
  {Chari}}, \bibinfo {author} {\bibfnamefont {I.}~\bibnamefont {Meric}},
  \bibinfo {author} {\bibfnamefont {C.}~\bibnamefont {Dean}}, \ and\ \bibinfo
  {author} {\bibfnamefont {K.}~\bibnamefont {Shepard}},\ }\href {\doibase
  10.1109/TED.2015.2482823} {\bibfield  {journal} {\bibinfo  {journal} {IEEE
  Trans. Electron Devices}\ }\textbf {\bibinfo {volume} {62}},\ \bibinfo
  {pages} {4322} (\bibinfo {year} {2015})}\BibitemShut {NoStop}%
\bibitem [{\citenamefont {Banszerus}\ \emph {et~al.}(2016)\citenamefont
  {Banszerus}, \citenamefont {Schmitz}, \citenamefont {Engels}, \citenamefont
  {Goldsche}, \citenamefont {Watanabe}, \citenamefont {Taniguchi},
  \citenamefont {Beschoten},\ and\ \citenamefont {Stampfer}}]{Banszerus2016}%
  \BibitemOpen
  \bibfield  {author} {\bibinfo {author} {\bibfnamefont {L.}~\bibnamefont
  {Banszerus}}, \bibinfo {author} {\bibfnamefont {M.}~\bibnamefont {Schmitz}},
  \bibinfo {author} {\bibfnamefont {S.}~\bibnamefont {Engels}}, \bibinfo
  {author} {\bibfnamefont {M.}~\bibnamefont {Goldsche}}, \bibinfo {author}
  {\bibfnamefont {K.}~\bibnamefont {Watanabe}}, \bibinfo {author}
  {\bibfnamefont {T.}~\bibnamefont {Taniguchi}}, \bibinfo {author}
  {\bibfnamefont {B.}~\bibnamefont {Beschoten}}, \ and\ \bibinfo {author}
  {\bibfnamefont {C.}~\bibnamefont {Stampfer}},\ }\href {\doibase
  10.1021/acs.nanolett.5b04840} {\bibfield  {journal} {\bibinfo  {journal}
  {Nano Lett.}\ }\textbf {\bibinfo {volume} {16}},\ \bibinfo {pages} {1387}
  (\bibinfo {year} {2016})}\BibitemShut {NoStop}%
\bibitem [{\citenamefont {Ni}\ \emph {et~al.}(2016)\citenamefont {Ni},
  \citenamefont {Wang}, \citenamefont {Goldflam}, \citenamefont {Wagner},
  \citenamefont {Fei}, \citenamefont {McLeod}, \citenamefont {Liu},
  \citenamefont {Keilmann}, \citenamefont {{\"O}zyilmaz}, \citenamefont
  {{Castro Neto}}, \citenamefont {Hone}, \citenamefont {Fogler},\ and\
  \citenamefont {Basov}}]{Ni2016}%
  \BibitemOpen
  \bibfield  {author} {\bibinfo {author} {\bibfnamefont {G.~X.}\ \bibnamefont
  {Ni}}, \bibinfo {author} {\bibfnamefont {L.}~\bibnamefont {Wang}}, \bibinfo
  {author} {\bibfnamefont {M.~D.}\ \bibnamefont {Goldflam}}, \bibinfo {author}
  {\bibfnamefont {M.}~\bibnamefont {Wagner}}, \bibinfo {author} {\bibfnamefont
  {Z.}~\bibnamefont {Fei}}, \bibinfo {author} {\bibfnamefont {A.~S.}\
  \bibnamefont {McLeod}}, \bibinfo {author} {\bibfnamefont {M.~K.}\
  \bibnamefont {Liu}}, \bibinfo {author} {\bibfnamefont {F.}~\bibnamefont
  {Keilmann}}, \bibinfo {author} {\bibfnamefont {B.}~\bibnamefont
  {{\"O}zyilmaz}}, \bibinfo {author} {\bibfnamefont {A.~H.}\ \bibnamefont
  {{Castro Neto}}}, \bibinfo {author} {\bibfnamefont {J.}~\bibnamefont {Hone}},
  \bibinfo {author} {\bibfnamefont {M.~M.}\ \bibnamefont {Fogler}}, \ and\
  \bibinfo {author} {\bibfnamefont {D.~N.}\ \bibnamefont {Basov}},\ }\href
  {\doibase 10.1038/nphoton.2016.45} {\bibfield  {journal} {\bibinfo  {journal}
  {Nat. Photon.}\ }\textbf {\bibinfo {volume} {10}},\ \bibinfo {pages} {244}
  (\bibinfo {year} {2016})}\BibitemShut {NoStop}%
\bibitem [{\citenamefont {Pallecchi}\ \emph {et~al.}(2014)\citenamefont
  {Pallecchi}, \citenamefont {Lafont}, \citenamefont {Cavaliere}, \citenamefont
  {Schopfer}, \citenamefont {Mailly}, \citenamefont {Poirier},\ and\
  \citenamefont {Ouerghi}}]{Pallecchi2014}%
  \BibitemOpen
  \bibfield  {author} {\bibinfo {author} {\bibfnamefont {E.}~\bibnamefont
  {Pallecchi}}, \bibinfo {author} {\bibfnamefont {F.}~\bibnamefont {Lafont}},
  \bibinfo {author} {\bibfnamefont {V.}~\bibnamefont {Cavaliere}}, \bibinfo
  {author} {\bibfnamefont {F.}~\bibnamefont {Schopfer}}, \bibinfo {author}
  {\bibfnamefont {D.}~\bibnamefont {Mailly}}, \bibinfo {author} {\bibfnamefont
  {W.}~\bibnamefont {Poirier}}, \ and\ \bibinfo {author} {\bibfnamefont
  {A.}~\bibnamefont {Ouerghi}},\ }\href {\doibase 10.1038/srep04558} {\bibfield
   {journal} {\bibinfo  {journal} {Sci. Rep.}\ }\textbf {\bibinfo {volume}
  {4}},\ \bibinfo {pages} {4558} (\bibinfo {year} {2014})}\BibitemShut
  {NoStop}%
\bibitem [{\citenamefont {Yang}\ and\ \citenamefont {Ni}(2010)}]{Yang2010}%
  \BibitemOpen
  \bibfield  {author} {\bibinfo {author} {\bibfnamefont {Z.}~\bibnamefont
  {Yang}}\ and\ \bibinfo {author} {\bibfnamefont {J.}~\bibnamefont {Ni}},\
  }\href {\doibase 10.1063/1.3373571} {\bibfield  {journal} {\bibinfo
  {journal} {J. Appl. Phys.}\ }\textbf {\bibinfo {volume} {107}},\ \bibinfo
  {pages} {104301} (\bibinfo {year} {2010})}\BibitemShut {NoStop}%
\bibitem [{\citenamefont {Terrones}\ \emph {et~al.}(2007)\citenamefont
  {Terrones}, \citenamefont {Romo-Herrera}, \citenamefont {Cruz-Silva},
  \citenamefont {L{\'o}pez-Ur{\'i}as}, \citenamefont {Mu{\~n}oz-Sandoval},
  \citenamefont {Vel{\'a}zquez-Salazar}, \citenamefont {Terrones},
  \citenamefont {Bando},\ and\ \citenamefont {Golberg}}]{Terrones2007}%
  \BibitemOpen
  \bibfield  {author} {\bibinfo {author} {\bibfnamefont {M.}~\bibnamefont
  {Terrones}}, \bibinfo {author} {\bibfnamefont {J.~M.}\ \bibnamefont
  {Romo-Herrera}}, \bibinfo {author} {\bibfnamefont {E.}~\bibnamefont
  {Cruz-Silva}}, \bibinfo {author} {\bibfnamefont {F.}~\bibnamefont
  {L{\'o}pez-Ur{\'i}as}}, \bibinfo {author} {\bibfnamefont {E.}~\bibnamefont
  {Mu{\~n}oz-Sandoval}}, \bibinfo {author} {\bibfnamefont {J.~J.}\ \bibnamefont
  {Vel{\'a}zquez-Salazar}}, \bibinfo {author} {\bibfnamefont {H.}~\bibnamefont
  {Terrones}}, \bibinfo {author} {\bibfnamefont {Y.}~\bibnamefont {Bando}}, \
  and\ \bibinfo {author} {\bibfnamefont {D.}~\bibnamefont {Golberg}},\ }\href
  {\doibase 10.1016/S1369-7021(07)70077-9} {\bibfield  {journal} {\bibinfo
  {journal} {Mater. Today}\ }\textbf {\bibinfo {volume} {10}},\ \bibinfo
  {pages} {30} (\bibinfo {year} {2007})}\BibitemShut {NoStop}%
\bibitem [{\citenamefont {Wei}\ \emph {et~al.}(2011)\citenamefont {Wei},
  \citenamefont {Wang}, \citenamefont {Bando},\ and\ \citenamefont
  {Golberg}}]{Wei2011}%
  \BibitemOpen
  \bibfield  {author} {\bibinfo {author} {\bibfnamefont {X.}~\bibnamefont
  {Wei}}, \bibinfo {author} {\bibfnamefont {M.-S.}\ \bibnamefont {Wang}},
  \bibinfo {author} {\bibfnamefont {Y.}~\bibnamefont {Bando}}, \ and\ \bibinfo
  {author} {\bibfnamefont {D.}~\bibnamefont {Golberg}},\ }\href {\doibase
  10.1021/nn103548r} {\bibfield  {journal} {\bibinfo  {journal} {ACS Nano}\
  }\textbf {\bibinfo {volume} {5}},\ \bibinfo {pages} {2916} (\bibinfo {year}
  {2011})}\BibitemShut {NoStop}%
\bibitem [{\citenamefont {Yang}\ \emph {et~al.}(2009)\citenamefont {Yang},
  \citenamefont {Deslippe}, \citenamefont {Park}, \citenamefont {Cohen},\ and\
  \citenamefont {Louie}}]{Yang2009}%
  \BibitemOpen
  \bibfield  {author} {\bibinfo {author} {\bibfnamefont {L.}~\bibnamefont
  {Yang}}, \bibinfo {author} {\bibfnamefont {J.}~\bibnamefont {Deslippe}},
  \bibinfo {author} {\bibfnamefont {C.~H.}\ \bibnamefont {Park}}, \bibinfo
  {author} {\bibfnamefont {M.~L.}\ \bibnamefont {Cohen}}, \ and\ \bibinfo
  {author} {\bibfnamefont {S.~G.}\ \bibnamefont {Louie}},\ }\href {\doibase
  10.1103/PhysRevLett.103.186802} {\bibfield  {journal} {\bibinfo  {journal}
  {Phys. Rev. Lett.}\ }\textbf {\bibinfo {volume} {103}},\ \bibinfo {pages}
  {186802} (\bibinfo {year} {2009})}\BibitemShut {NoStop}%
\bibitem [{\citenamefont {Kravets}\ \emph {et~al.}(2010)\citenamefont
  {Kravets}, \citenamefont {Grigorenko}, \citenamefont {Nair}, \citenamefont
  {Blake}, \citenamefont {Anissimova}, \citenamefont {Novoselov},\ and\
  \citenamefont {Geim}}]{Kravets2010}%
  \BibitemOpen
  \bibfield  {author} {\bibinfo {author} {\bibfnamefont {V.~G.}\ \bibnamefont
  {Kravets}}, \bibinfo {author} {\bibfnamefont {A.~N.}\ \bibnamefont
  {Grigorenko}}, \bibinfo {author} {\bibfnamefont {R.~R.}\ \bibnamefont
  {Nair}}, \bibinfo {author} {\bibfnamefont {P.}~\bibnamefont {Blake}},
  \bibinfo {author} {\bibfnamefont {S.}~\bibnamefont {Anissimova}}, \bibinfo
  {author} {\bibfnamefont {K.~S.}\ \bibnamefont {Novoselov}}, \ and\ \bibinfo
  {author} {\bibfnamefont {A.~K.}\ \bibnamefont {Geim}},\ }\href {\doibase
  10.1103/PhysRevB.81.155413} {\bibfield  {journal} {\bibinfo  {journal} {Phys.
  Rev. B}\ }\textbf {\bibinfo {volume} {81}},\ \bibinfo {pages} {155413}
  (\bibinfo {year} {2010})}\BibitemShut {NoStop}%
\bibitem [{\citenamefont {Mak}\ \emph {et~al.}(2014)\citenamefont {Mak},
  \citenamefont {da~Jornada}, \citenamefont {He}, \citenamefont {Deslippe},
  \citenamefont {Petrone}, \citenamefont {Hone}, \citenamefont {Shan},
  \citenamefont {Louie},\ and\ \citenamefont {Heinz}}]{Mak2014}%
  \BibitemOpen
  \bibfield  {author} {\bibinfo {author} {\bibfnamefont {K.~F.}\ \bibnamefont
  {Mak}}, \bibinfo {author} {\bibfnamefont {F.~H.}\ \bibnamefont {da~Jornada}},
  \bibinfo {author} {\bibfnamefont {K.}~\bibnamefont {He}}, \bibinfo {author}
  {\bibfnamefont {J.}~\bibnamefont {Deslippe}}, \bibinfo {author}
  {\bibfnamefont {N.}~\bibnamefont {Petrone}}, \bibinfo {author} {\bibfnamefont
  {J.}~\bibnamefont {Hone}}, \bibinfo {author} {\bibfnamefont {J.}~\bibnamefont
  {Shan}}, \bibinfo {author} {\bibfnamefont {S.~G.}\ \bibnamefont {Louie}}, \
  and\ \bibinfo {author} {\bibfnamefont {T.~F.}\ \bibnamefont {Heinz}},\ }\href
  {\doibase 10.1103/PhysRevLett.112.207401} {\bibfield  {journal} {\bibinfo
  {journal} {Phys. Rev. Lett.}\ }\textbf {\bibinfo {volume} {112}},\ \bibinfo
  {pages} {207401} (\bibinfo {year} {2014})}\BibitemShut {NoStop}%
\bibitem [{\citenamefont {Park}\ and\ \citenamefont {Louie}(2010)}]{Park2010}%
  \BibitemOpen
  \bibfield  {author} {\bibinfo {author} {\bibfnamefont {C.~H.}\ \bibnamefont
  {Park}}\ and\ \bibinfo {author} {\bibfnamefont {S.~G.}\ \bibnamefont
  {Louie}},\ }\href {\doibase 10.1021/nl902932k} {\bibfield  {journal}
  {\bibinfo  {journal} {Nano Lett.}\ }\textbf {\bibinfo {volume} {10}},\
  \bibinfo {pages} {426} (\bibinfo {year} {2010})}\BibitemShut {NoStop}%
\bibitem [{\citenamefont {Wirtz}\ \emph {et~al.}(2006)\citenamefont {Wirtz},
  \citenamefont {Marini},\ and\ \citenamefont {Rubio}}]{Wirtz2006}%
  \BibitemOpen
  \bibfield  {author} {\bibinfo {author} {\bibfnamefont {L.}~\bibnamefont
  {Wirtz}}, \bibinfo {author} {\bibfnamefont {A.}~\bibnamefont {Marini}}, \
  and\ \bibinfo {author} {\bibfnamefont {A.}~\bibnamefont {Rubio}},\ }\href
  {\doibase 10.1103/PhysRevLett.96.126104} {\bibfield  {journal} {\bibinfo
  {journal} {Phys. Rev. Lett.}\ }\textbf {\bibinfo {volume} {96}},\ \bibinfo
  {pages} {126104} (\bibinfo {year} {2006})}\BibitemShut {NoStop}%
\bibitem [{\citenamefont {Oba}\ \emph {et~al.}(2010)\citenamefont {Oba},
  \citenamefont {Togo}, \citenamefont {Tanaka}, \citenamefont {Watanabe},\ and\
  \citenamefont {Taniguchi}}]{Oba2010}%
  \BibitemOpen
  \bibfield  {author} {\bibinfo {author} {\bibfnamefont {F.}~\bibnamefont
  {Oba}}, \bibinfo {author} {\bibfnamefont {A.}~\bibnamefont {Togo}}, \bibinfo
  {author} {\bibfnamefont {I.}~\bibnamefont {Tanaka}}, \bibinfo {author}
  {\bibfnamefont {K.}~\bibnamefont {Watanabe}}, \ and\ \bibinfo {author}
  {\bibfnamefont {T.}~\bibnamefont {Taniguchi}},\ }\href {\doibase
  10.1103/PhysRevB.81.075125} {\bibfield  {journal} {\bibinfo  {journal} {Phys.
  Rev. B}\ }\textbf {\bibinfo {volume} {81}},\ \bibinfo {pages} {075125}
  (\bibinfo {year} {2010})}\BibitemShut {NoStop}%
\end{thebibliography}%

\end{document}